\documentclass[prc,twocolumn,pdftex]{revtex4-1}
\usepackage{graphicx}
\usepackage{amsmath}
\usepackage{multirow}
\usepackage{color}
\begin{document}

\title{
Constraining the initial temperature and shear viscosity in a hybrid hydrodynamic model of $\sqrt{s_{NN}}$=200 GeV Au+Au collisions using pion spectra, elliptic flow, and femtoscopic radii} 
\author{R.~A.~Soltz}\email{soltz@llnl.gov}
\author{I.~Garishvili}
\author{M.~Cheng}\altaffiliation[Present address: ]{Boston University, Boston, MA 02215}
\author{B.~Abelev}
\author{A.~Glenn}
\author{J.~Newby}\altaffiliation[Present address: ]{Oak Ridge Nat. Lab., Oak Ridge, TN 37831}
\author{L.~A.~Linden Levy}\altaffiliation[Present address: ]{MobiTV, Emeryville, CA 94608}
\affiliation{Lawrence Livermore National Laboratory\\
Livermore, California 94550}
\author{S.~Pratt}
\affiliation{Department of Physics and Astronomy and National Superconducting Cyclotron Laboratory,
Michigan State University\\
East Lansing, Michigan 48824}
\date{\today}

\begin{abstract}

A new framework for evaluating hydrodynamic models of relativistic heavy ion collisions has been developed.  This framework, a Comprehesive Heavy Ion Model Evaluation and  Reporting Algorithm (CHIMERA) has been implemented by augmenting UVH 2+1D viscous hydrodynamic model with eccentricity fluctuations, pre-equilibrium flow, and the Ultra-relativistic Quantum Molecular Dynamic (UrQMD) hadronic cascade.  A range of initial temperatures and shear viscosity to entropy ratios were evaluated for four initial profiles, $N_{part}$ and $N_{coll}$ scaling with and without pre-equilibrium flow.  The model results were compared to pion spectra, elliptic flow, and femtoscopic radii from 200 GeV Au+Au collisions for the 0--20\% centrality range.Two sets of initial density profiles, $N_{part}$ scaling with pre-equilibrium flow and $N_{coll}$ scaling without were shown to provide a consistent description of all three measurements.
\end{abstract}


\maketitle

\section{Introduction}\label{sec:intro}
During the last decade a remarkable success has been achieved in the modeling and analysis of relativistic heavy ion collisions.  Data from the Relativistic Heavy Ion Collider (RHIC) and more recently from the Large Hadron Collider (LHC) have surpassed the petabyte scale and include many classes of observables.
The models that have been most successful in describing the soft physics observables of particle spectra and collective flow measured at RHIC (see \cite{Arsene:2005kd,Back:2005ck,Adams:2005tm,Adcox:2005iw} for a synopsis of the initial results) are those that incorporate relativistic hydrodynamics~\cite{Huovinen:483397,Kolb:2003tx,Hirano:2001bn} coupled to a microscopic transport such as UrQMD~
\cite{Bass:349359,Bleicher:1999wv}.  The relativistic hydrodynamic stage is used to model the quark gluon plasma phase and its transition to hadronic matter, whereas the transport stage simulates the hadronic cascade which follows.  Subsequent refinements to the hydrodynamic models to improve agreement with measured moments of the momentum anisotropy have incorporated shear viscosity~\cite{Luzum:2008hz,Song:2011fb} and initial energy density fluctuations in the initial conditions~\cite{Kodama:2001qv,Schenke:2010di,Qiu:2011js}.  The success of these models in reproducing the general features of particle spectra and elliptic flow has led to the conclusion that the QCD matter created in relativistic heavy ion collisions behaves very much like a fluid with small shear viscosity to entropy ratio.  

Although the comparisons between models and data are both interesting and compelling, there is not yet a complete and rigorous quantification of the uncertainties in the parameters that would describe the quark gluon plasma created in heavy ion collisions at RHIC and the LHC.  The greatest uncertainty lies in our understanding of the initial, pre-thermalized state of the collision.  Differences in the initial density profile that alter the hydrodynamic evolution can lead to large uncertainties in the shear viscosity and initial temperature.  The Equation of State (EoS) is calculated with increasing accuracy with lattice QCD~\cite{Bazavov:2009ep,Borsanyi:2010gh}, but the interplay between the QCD EoS and the other parameters is yet be explored.  Furthermore, most model comparisons are often limited to a subset of measurements that are deemed most relevant to a specific line of inquiry.  Some success in uncertainty quantification has been achieved by studying the relation between two observables, such as the ratio of elliptic flow to eccentricity and the multiplicity density~\cite{Song:2010mg}.  However, a complete determination of the properties of matter created in heavy ion collisions will require a comprehensive comparison between a hybrid hydrodynamic model and a set of physics observables.

This paper describes a first step towards performing such a comparison, with the ultimate goal to fully constrain the properties of the quark gluon plasma.  We have developed a Comprehesive Heavy Ion Model Evaluation and  Reporting Algorithm (CHIMERA) for systematically comparing a set of hybrid hydrodynamic models for heavy ion collisions spanning a range of initial parameters.  This framework enables one to determine the optimal parameters and associated uncertainties that best describe a set of soft physics measurements, incorporating both statistical and systematic errors.  In the current implementation we use the 2D+1 viscous hydro code VH2~\cite{Luzum:2008hz} augmented with initial state eccentricity fluctuations~\cite{Alver:2008fk} and pre-equilibrium flow~\cite{Vredevoogd:2009jt} to describe the hydrodynamic evolution, and the Ultra-relativistic Quantum Molecular Dynamics hadronic cascade code (UrQMD)~\cite{Bass:349359,Bleicher:1999wv} to describe the hadronic transport.  To compare to data we generate particle spectra and elliptic flow directly from the UrQMD output.  We generate femtoscopic correlation lengths, also referred to as HBT radii~\cite{HanburyBrown:433988,Goldhaber:1960vf}, using the Correlation After-Burner (CRAB) code~\cite{Pratt:1990kx,Brown:2005bl}.  A chi-squared statistic is used to determine the best fit initial state parameters and associated uncertainties for the measured results and errors.  For this paper, we compare to published spectra, elliptic flow, and femtoscopic correlations for pions measured by the STAR and PHENIX collaborations for $\sqrt{s_{NN}}$=200~GeV Au+Au collisions in the 0--20\% centrality region.  

In Section~\ref{sec:model} we provide a detailed description of the various components of the model, and in Section~\ref{sec:eval} we explain the procedures used to generate the model results and to evaluate the agreement with the data.  The results of these evaluations are reported in Section~\ref{sec:results} and potential systematic errors are discussed in Section~\ref{sec:sys}.  In Section~\ref{sec:conc} we give our conclusions and discuss future prospects and improvements.

\section{The Model}
\label{sec:model}
\subsection{Initial Conditions and Hydrodynamic Evolution}

The primary component of the model we employ is the freely available 2+1D viscous hydrodynamic code , VH2, developed by Luzum and Romatschke~\cite{Luzum:2008hz}.  VH2 numerically solves the Muller-Israel-Stewart equations with finite shear viscosity in two dimensions assuming Bjorken boost invariant expansion along the longitudinal (beam) axis.  We have modified the distributed version of VH2 in two ways: 1.) to account for the impact of initial density fluctuations on the average eccentricity for a given centrality, and 2.) to allow the initial state to be prepared with pre-equilibrium flow.

The standard version of VH2 determines the initial density profile by calculating the Glauber overlap integral for the Wood-Saxon density distributions of the colliding nuclei.  The initial conditions are therefore smooth, and the density profile may be set by participant or binary collisions scaling.  We have modified the calculation of the initial conditions to account for initial eccentricity fluctuations, using the TGlauber Monte Carlo method~\cite{Alver:2008wu} with the default parameters $R$=6.38~fm, $a$=0.535, and a nucleon-nucleon cross-section $\sigma_{NN}$=42~mb.  Smooth distributions were then obtained by averaging over 10,000 events, in which each event was rotated to align the participant or binary-collision reaction plane along the X-axis.  An event sample of this size will still produce fluctuations in the tails of the distribution that the VH2 code is not designed to handle.  In order to remove these fluctuations the portion of the distribution below 3\% of the maximum was fit to a 2-D Gaussian in X and Y, which was then used to replace the initial distribution.  For this initial study, we used Glauber initial conditions for both participant and binary collision scaling and calculate the mean eccentricity as appropriate for a model analysis with perfect reaction plane resolution~\cite{Alver:2008fk}.  
In~\cite{Luzum:2008hz} the authors set the initial conditions according to binary collision and KLN scaling~\cite{Kharzeev:2005ku}.  Other parameterizations include Gaussian profiles~\cite{Broniowski:2009hp} and non-smooth initial conditions such as IP-Glasma~\cite{Schenke:2012dk}.  
We have limited the choice of initial conditions to the Glauber distributions because they are simple and smooth, the latter as required for VH2.   A more thorough examination of a larger set of initial condition profiles is left for future investigation.

We have also introduced an option to prepare the initial state with pre-equilibrium flow.  Vredevoogd and Pratt have shown that for a system with a traceless stress energy tensor that obeys Bjorken boost-invariant scaling and for which the spatial component anisotropy is largely independent of the spatial coordinates, the flow can be expressed as a universal function of the energy and time~\cite{Vredevoogd:2009jt},
\begin{equation}
\frac{T_{0x}}{T_{00}} \approx \frac{\partial_x T_{00}}{2 T_{00}}t.
\label{eq:preflow}
\end{equation}
\begin{figure*}[th]
\includegraphics[width=0.49\textwidth]{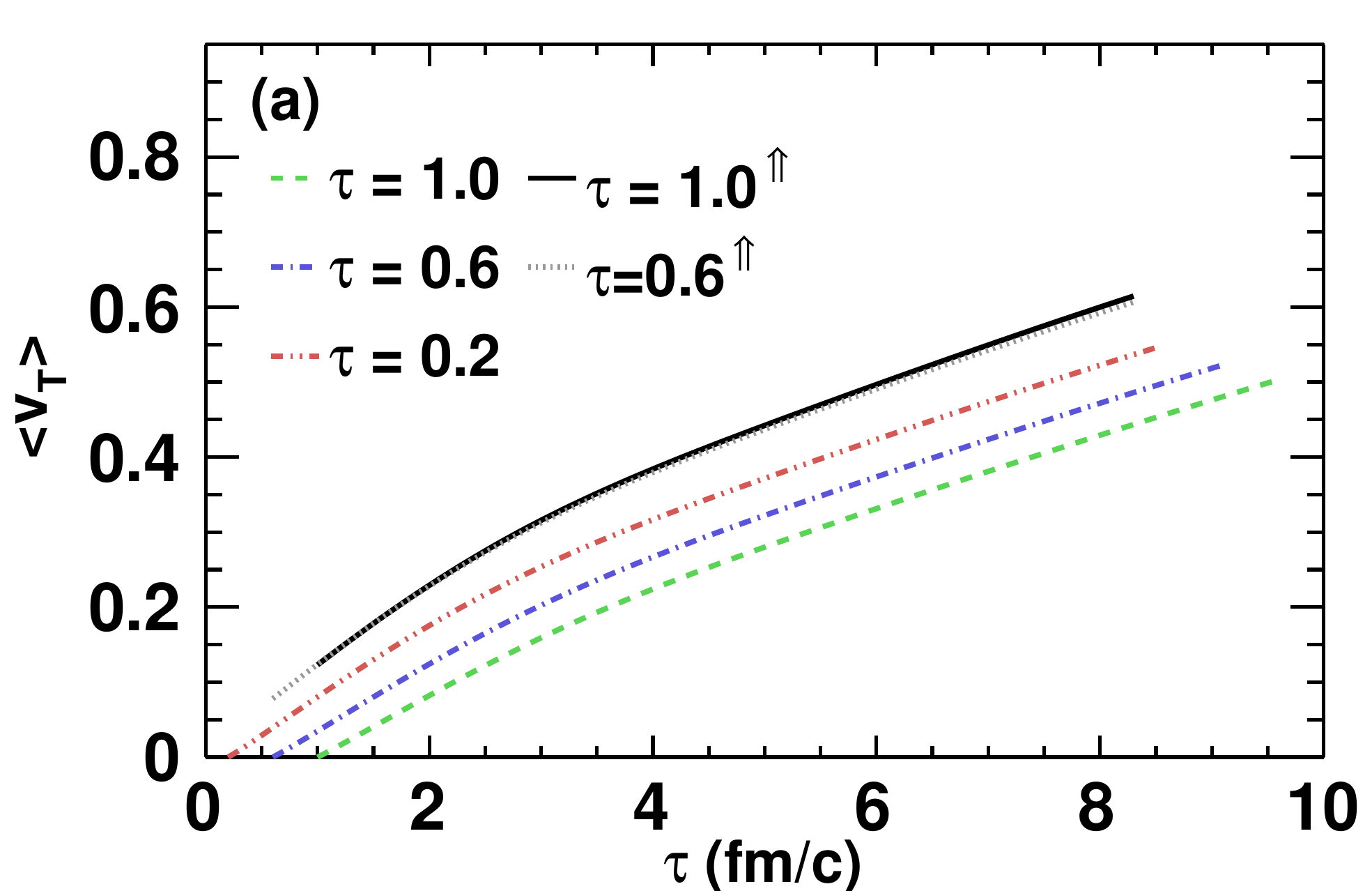}
\includegraphics[width=0.49\textwidth]{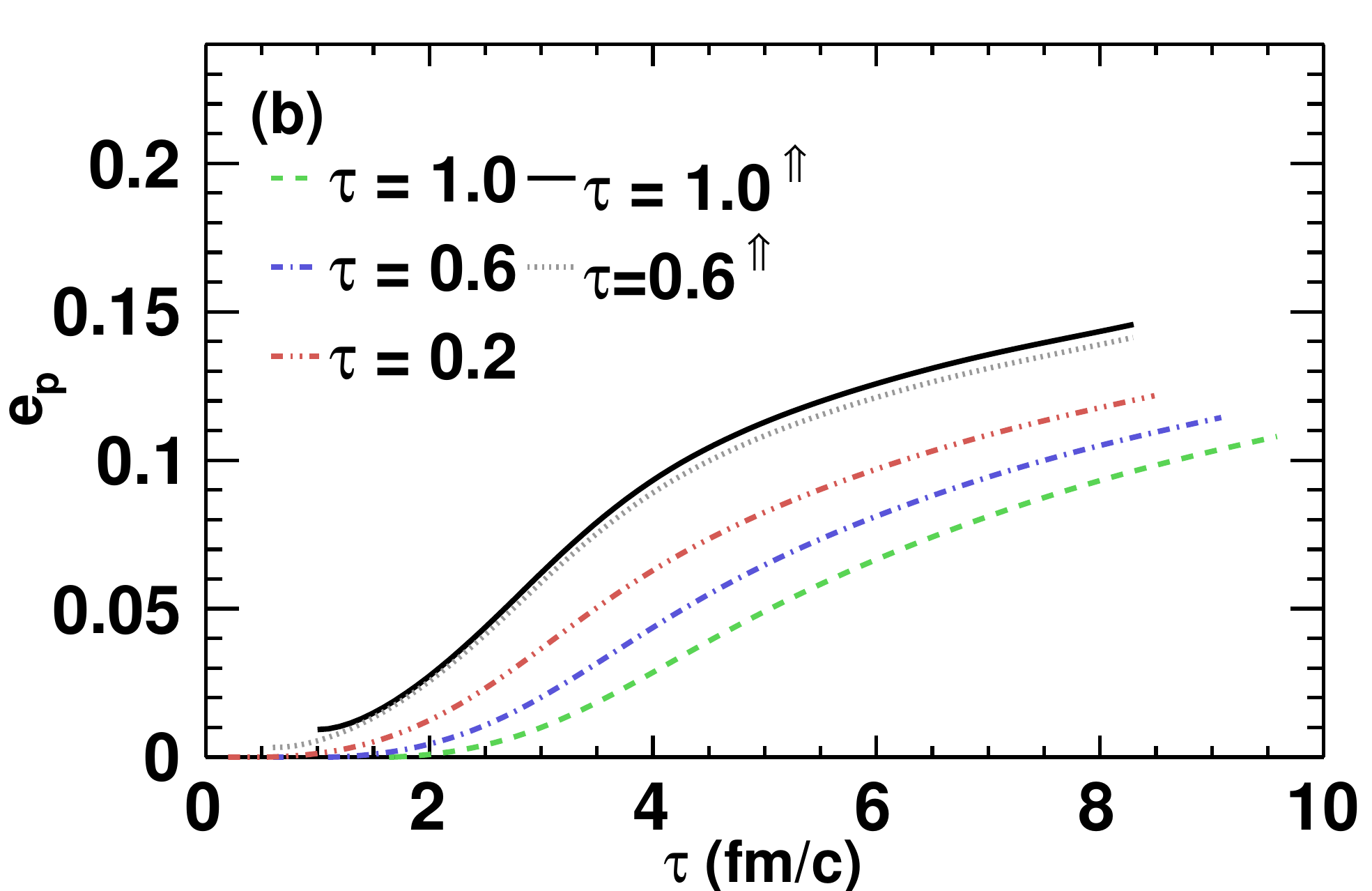}
\includegraphics[width=0.49\textwidth]{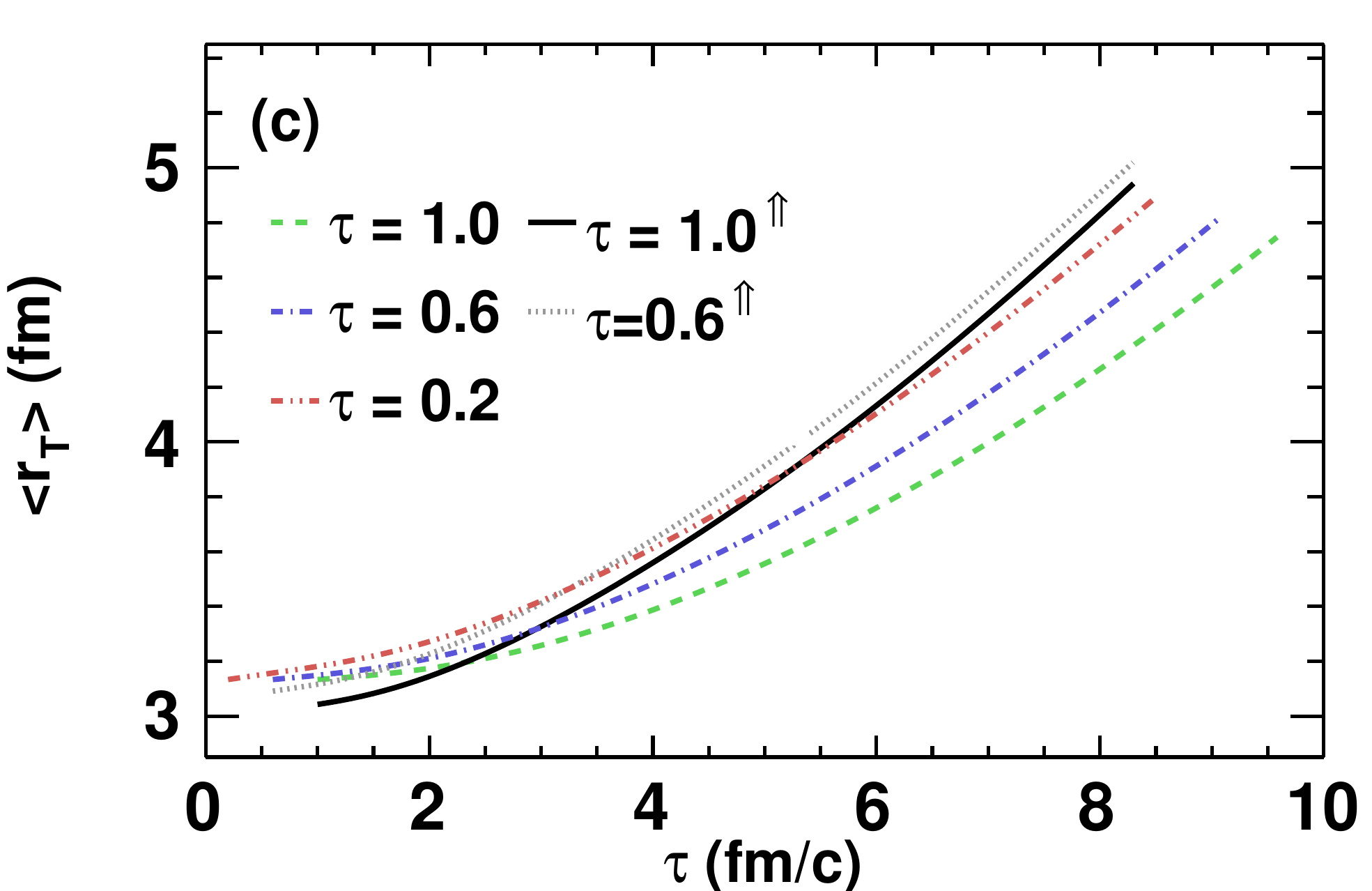}
\includegraphics[width=0.49\textwidth]{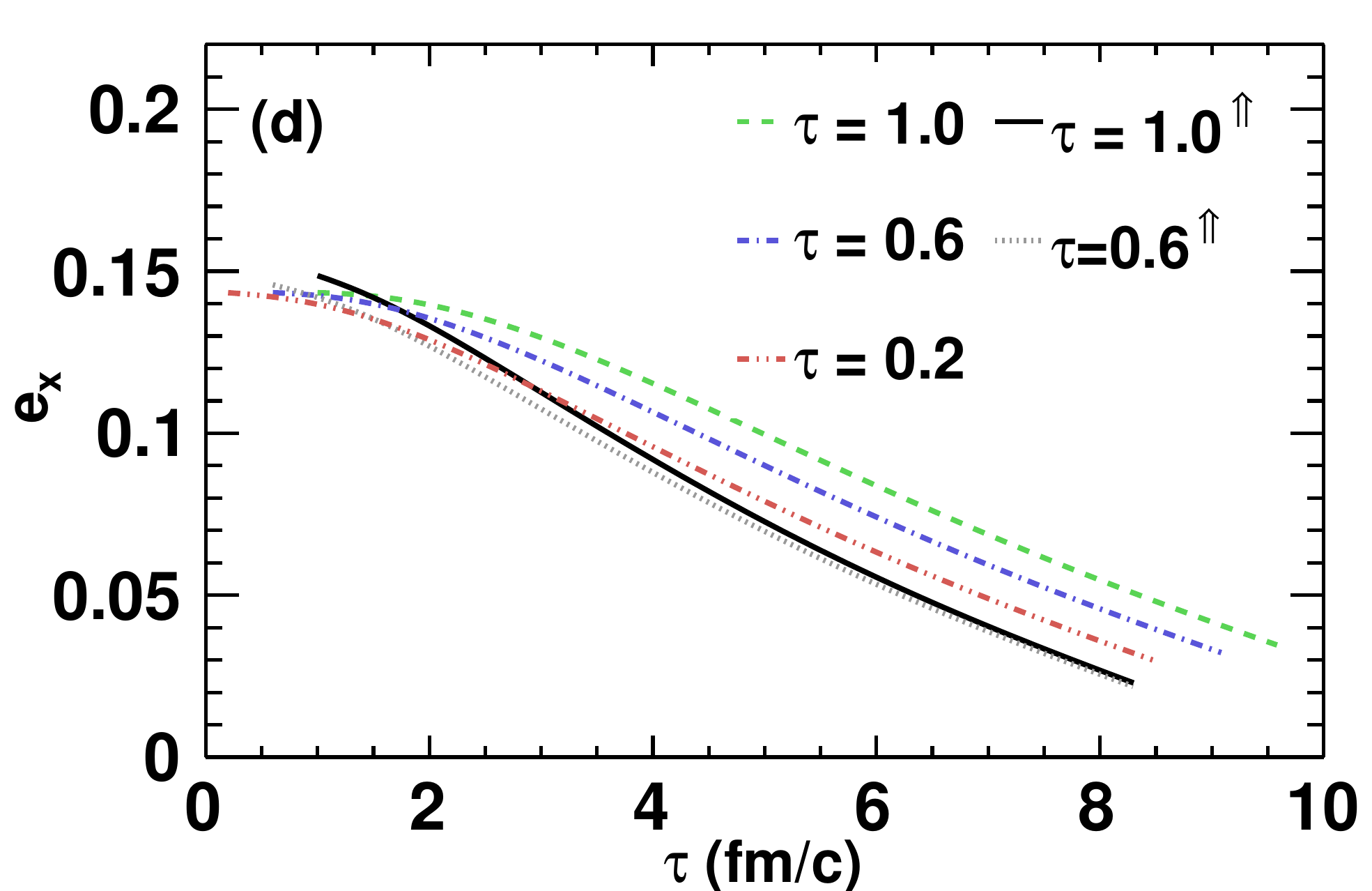}
\caption{(Color online) The proper time evolution of mean transverse radial flow (a) and elliptic flow as defined by the stress tensor asymmetry (b) for VH2 2+1D hydrodynamic evolution for Au+Au collisions with participant Glauber initial conditions for a $b=4.4$~fm impact parameter and corresponding $b=0$ initial temperature of 300~MeV.  Contours are drawn for initial proper start times, $\tau$=0.2, 0.6, and 1.0~fm/c without pre-equilibrium flow and for $\tau$=0.6 and 1.0 with pre-equilibrium flow as indicated in the legend by the $\Uparrow$ symbol.   The mean transverse radial positions are shown in (c) and the eccentricities are plotted in (d).  Note that all means are weighted by energy .}
\label{fig:preflow}
\end{figure*}

As implemented in this work, the individual fluid cells (0.2~GeV$^{-1}$ on a side) are given initial transverse velocities according to Eq.~\ref{eq:preflow} after the initial energy density distribution has been established.  The addition of pre-equilibrium flow is expected to have the most impact on the elliptic flow and transverse radii.  This can be seen in Fig.~\ref{fig:preflow}, where panels (a) and (c) show the proper time evolution of the transverse radial flow, $v_{T}=(u_x^2 + u_y^2)^{1/2}$, and the mean transverse radial position, $r_{T}=(x^2 + y^2)^{1/2}$, and panels (b) and (d) show the evolution of the elliptic flow and eccentricity following the definition in~\cite{Luzum:2008hz},
\begin{eqnarray}
e_x & = & \frac{<x^2-y^2>}{<x^2+y^2>}, \nonumber \\
e_p & = & \frac{<T_{xx}^2-T_{yy}^2>}{<T_{xx}^2+T_{yy}^2>},
\end{eqnarray}
where all averages are weighted by the energy per cell.  All values 
are shown as a function of the proper time for start times of 0.6~fm/c (dotted) and 1.0~fm/c (solid).  These are compared to VH2 run without pre-equilibrium flow for start times of 0.2~fm/c (double-dot-dashed), 0.6~fm/c (dot-dashed), and 1.0~fm/c (dashed).  As the starting time is advanced, the systems that do not include pre-equilibrium flow begin to resemble the systems with pre-equilibrium flow, but they do not reach the same final values.  Advancing the start time without pre-equilibrium flow also advances the freeze-out time by a similar amount.  Note that the two systems prepared with pre-equilibrium flow do not show much dependence on start time, a result that is consistent with the premise of a universal pre-equilibrium flow.  For the comparisons to experimental data that follow, we adopt a start time of 1.0~fm/c with and without inclusion of the pre-equilibrium flow.  This start time is consistent with the choice in~\cite{Luzum:2008hz} for Glauber initial conditions.  

The hydrodynamic evolution then proceeds as with the standard version of the VH2 code, using the QCD-inspired Equation of State based upon the work of Laine and Schroeder~\cite{Laine:2006fj} that interpolates between the hadronic resonance model and the perturbative calculation.  It provides a reasonable albeit imperfect approximation to the cross-over transition that is now calculated non-perturbatively using lattice QCD~\cite{Bazavov:2009ep,Borsanyi:2010gh}.  A thorough study of the implications of various lattice calculations and their equation of state parameterizations is left for future investigation.

\subsection{Freeze-out and Hadronic Cascade}\label{subsec:freezeout}

The hydrodynamic evolution in VH2 ceases when the temperature of a given cell falls below a specified freeze-out value, nominally in the range 140--165~MeV.  The final energy densities are converted to final state particles following the prescription of Cooper and Frye~\cite{Cooper:1974ug} with corrections for the shear viscosity implemented according to the method developed by Pratt and Torrieri~\cite{Pratt:2010kg}.  In this method, the particle number and momentum distributions are determined by Monte Carlo sampling for non-viscous equilibrium distributions, and the momenta are subsequently rescaled according to Eq.~\ref{eq:p_rescale},
\begin{equation}
p_j \rightarrow p_i + \lambda_{ij} p_j.
\label{eq:p_rescale}
\end{equation}
Here $\lambda_{ij}$ is proportional to $\pi_{ij}$, the Israel-Stewart correction to the stress energy tensor.  The constant of proportionality is chosen to reproduce the second order viscous corrections to the final particle distributions for small values of $\pi$.
The list of final state particles is selected to match the complete set of known particles used by the UrQMD code.  For each set of CHIMERA parameters set a set of 5,000  events are generated in the OSCAR-97 format~\cite{OSCAR97} to compare to the measured particle spectra.  For comparisons to elliptic flow and radii, the total number of events is increased to 20,000 in order to achieve smaller statistical errors in the model results prior to fitting.  We use a switching temperature of $T_{\rm sw}$=165~MeV to end the hydrodynamic evolution and generate the particles for the subsequent hadronic cascade stage of the model.  As in~\cite{Song:2011fb}, the value of $T_{\rm sw}$ was chosen to be as close as possible to, but still larger than the transition temperature.  The freeze-out particle times are converted to formation times and the particles are back-propagated to zero time using the standard {\em oscar2u} program provided by the UrQMD developers upon request.

For this paper we use UrQMD v2.3 to model particle interactions that follow the hydrodynamic freeze-out stage.  UrQMD is run with default switches except for modifications needed to read the OSCAR-97 formatted input file.  We also disable the unstable particle decay after the final time step, in order to match the full set of particle spectra measured by the experiments.  The CHIMERA framework can also be used without the hadronic cascade.  In this case, the final state particles are generated from set of stable particles in the 2008 listing of the Particle Data Group~\cite{Amsler:2008kq} and a lower freeze-out temperature $T_{sw}$=140~MeV is used.

\subsection{Post Processing}\label{subsec:postprocess}

The final state particle distributions from UrQMD are then used to construct the observables that can be directly compared to published experimental data.  In this work we restrict our comparison to transverse momentum spectra, the second coefficient of the transverse momentum anisotropy with respect to the event reaction plane $v_2$, and the Bertsch-Pratt femtoscopic radii, $R_{\rm long}$, $R_{\rm side}$, $R_{\rm out}$.  The transverse momentum spectra are calculated as invariant cross-sections using bins of 0.1~GeV/c in the range 0.2--1.5~GeV/c within the rapidity interval of $|y|<0.5$.  The elliptic flow is calculated within the same transverse momentum region.  We use CRAB to generate the three-dimensional femtoscopic correlation functions, including the strong interaction and statistical interference.  The Coulomb interaction is neglected so that a 3D Gaussian can be used to fit the correlation function without corrections.  The correlation functions are binned in 0.2~GeV intervals.  The fits are performed directly on the correlation weights, to avoid the time and cpu-consuming process of constructing an event mixed background for each momentum bin.

The cumulative effect of each modification to the VH2 code is shown in Figures~{\ref{fig:cali_mean_dNdy_pT} and~\ref{fig:cali_mean_v2_radii}.
\begin{figure*}
\includegraphics[width=0.4\textwidth]{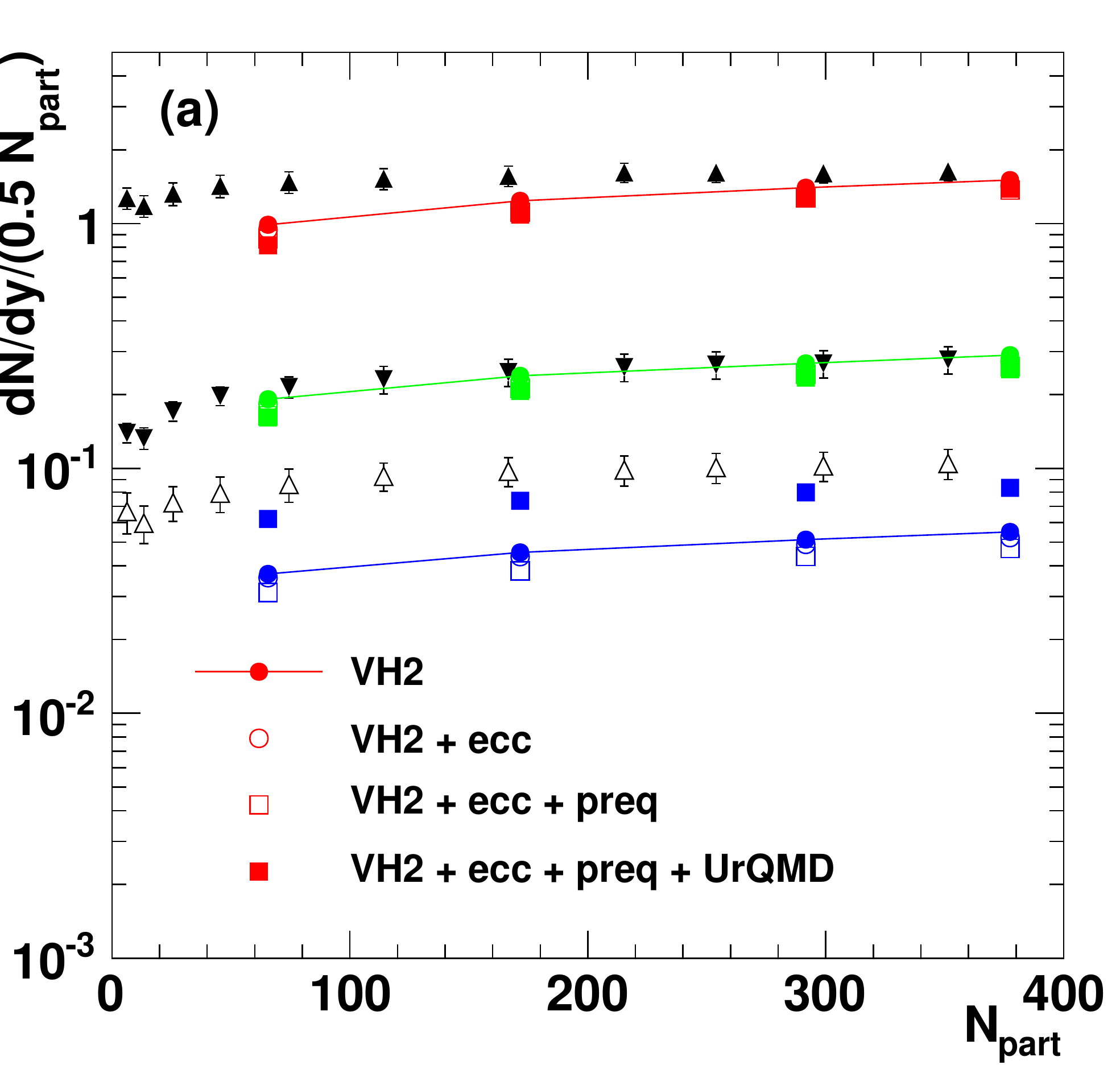}
\includegraphics[width=0.4\textwidth]{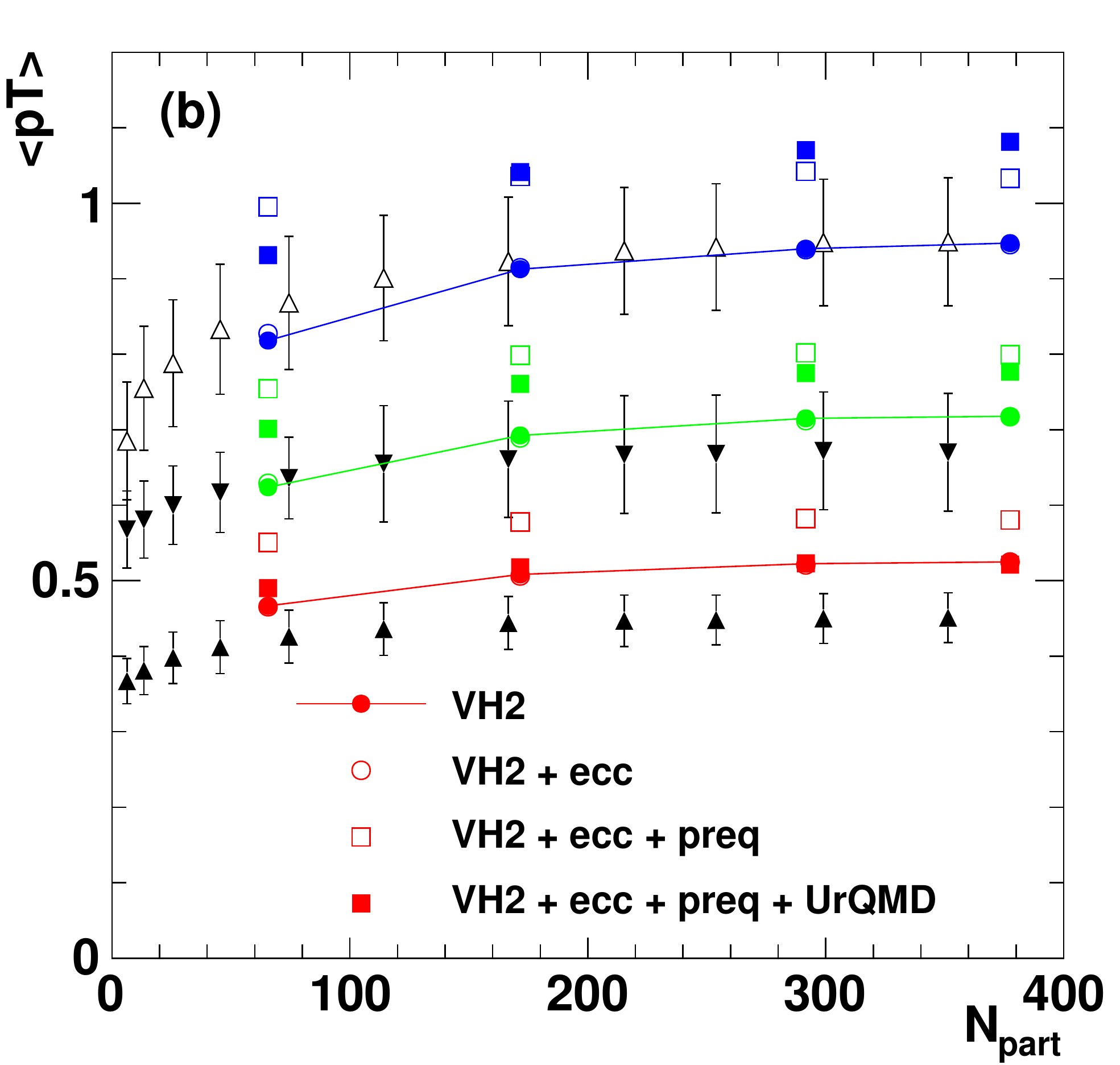}
\caption{(Color online) dN/dy (a) and mean transverse momenta (b) for pions (red), kaons (green), and protons (blue) for the unmodified VH2 (solid connected circles), and for the cumulative effect of adding eccentricity fluctuations (open circles), pre-equilibrium flow (open squares) and switching to the UrQMD hadronic cascade (solid squares) at $T_{sw}=165$~MeV.   Model results are compared to PHENIX measurements for pions (filled triangles), kaons (inverted filled triangles), and protons (open triangles).  In this figure the open-circle symbols for the eccentricity fluctuations are often obscured by filled-circle results from the unmodified VH2 results.}
\label{fig:cali_mean_dNdy_pT}
\end{figure*}
\begin{figure*}
\includegraphics[width=0.4\textwidth]{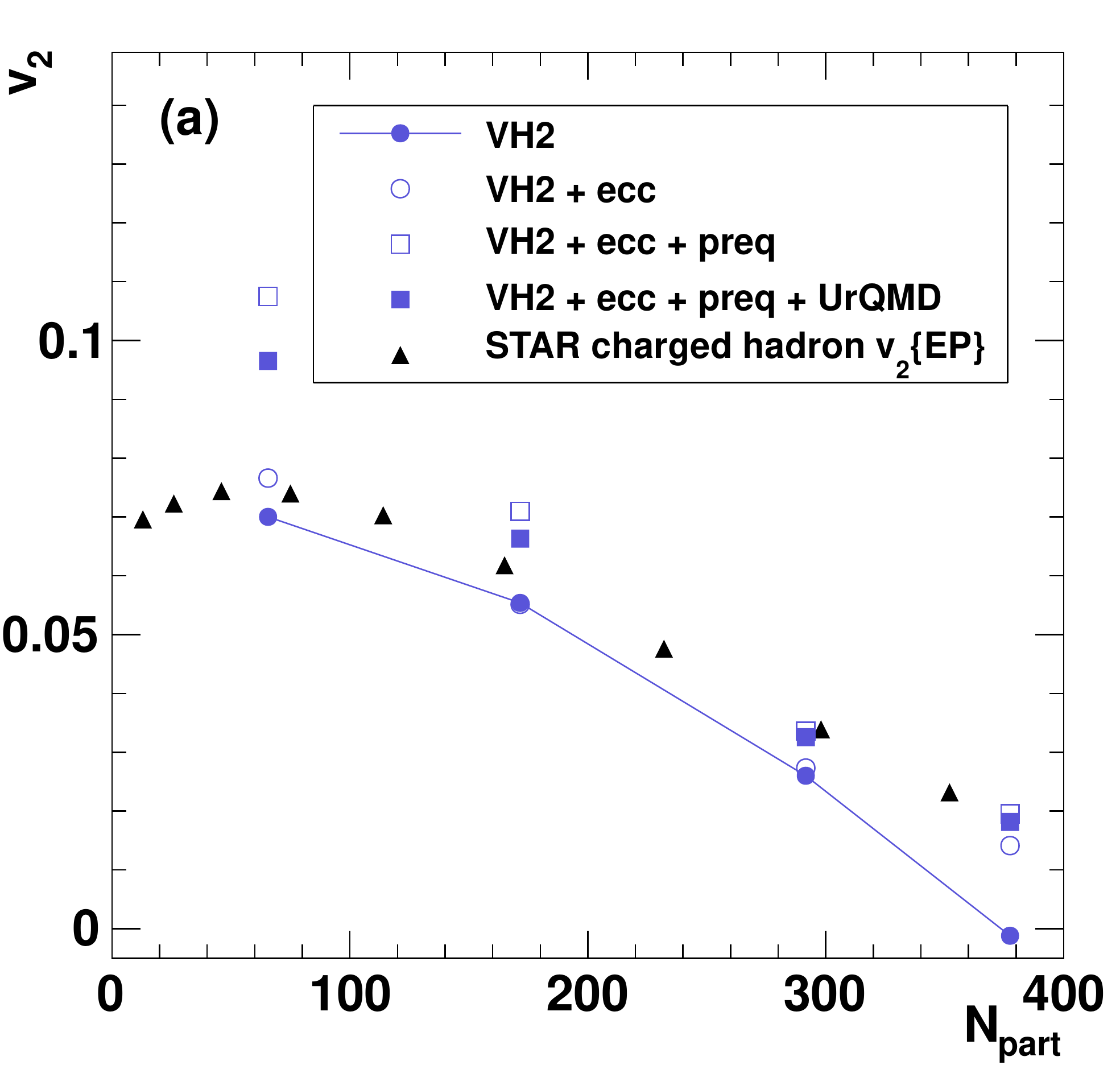}
\includegraphics[width=0.4\textwidth]{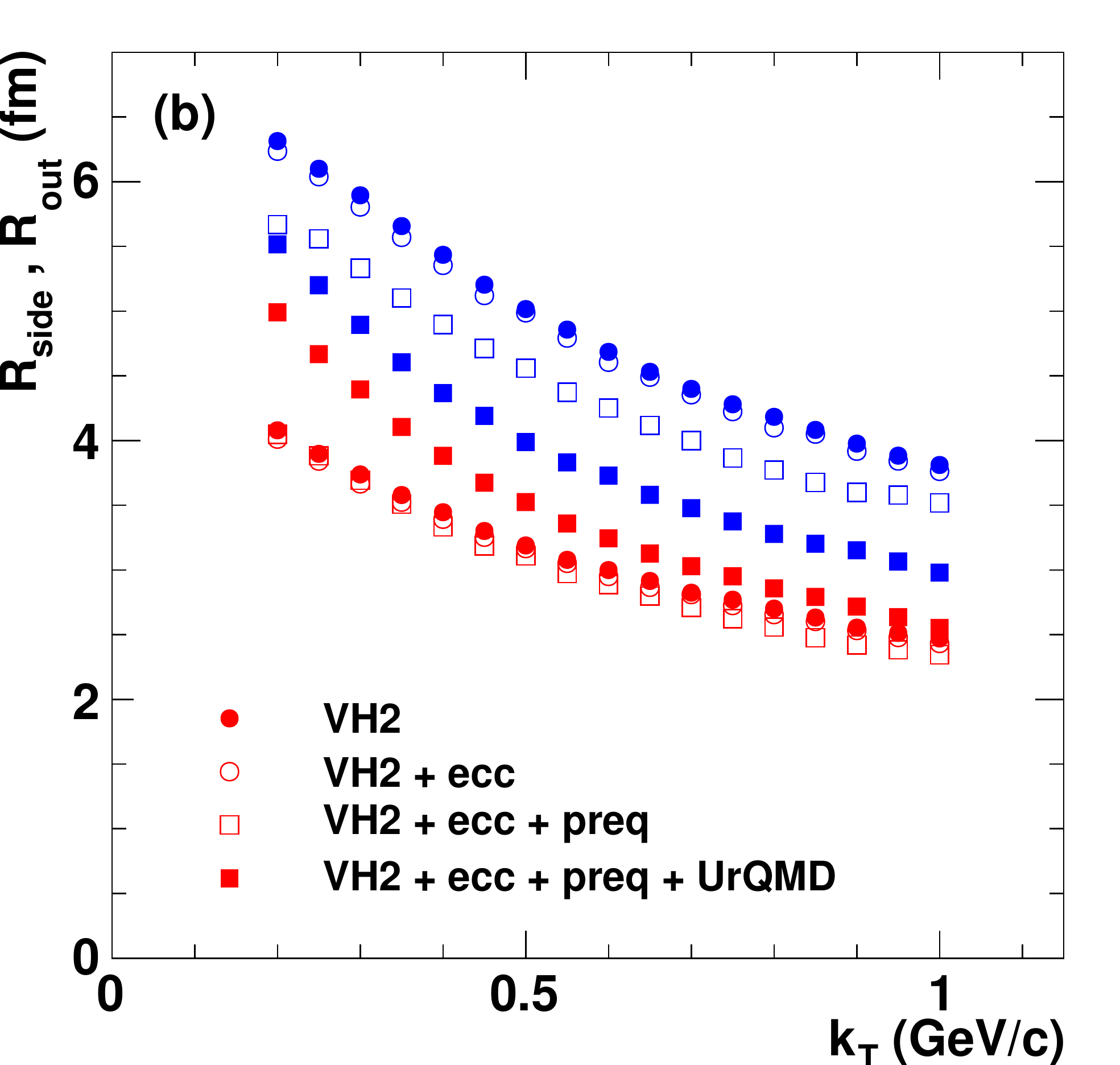}
\caption{(Color online) Mean elliptic flow for VH2, adding eccentricity fluctuations, pre-equilibrium flow, and the UrQMD cascade after-burner (a). $\rm R_{side}$ and $\rm R_{out}$ femtoscopic radii for the same conditions (b).}
\label{fig:cali_mean_v2_radii}
\end{figure*}
Fig.~\ref{fig:cali_mean_dNdy_pT} shows the charged particle yields and mean transverse momentum $\langle p_T \rangle$ for pions (filled triangles), kaons (filled inverse triangles), and protons (open triangles).  The unmodified VH2 results are shown as solid circles with connecting lines, and correspond to a binary scaling optical Glauber model input for an initial central temperature of $T_{cent}=333$~MeV, $\eta/s=0.08$ and a freeze-out temperature of 140~MeV.  These parameters were chosen to match the upper panels of Fig.~7 of~\cite{Luzum:2009bs}.  Here we follow the VH2 convention in reporting the equivalent initial temperature, $T_{cent}$, for a central collision (zero impact parameter).  As in~\cite{Luzum:2009bs}, the model results are compared to PHENIX measurements from~\cite{Adler:2004ki}.  The results in Fig.~\ref{fig:cali_mean_dNdy_pT} make use a different freeze-out routine, described above, that was written to accommodate the need to write OSCAR formatted events for input to CRAB and UrQMD.  Therefore the results may differ slightly from the published VH2 results.  This figure shows the cumulative effect of introducing eccentricity fluctuations (open circles), pre-equilibrium flow (open squares) and UrQMD with a switching temperature of $T_{sw}=165$~MeV (filled squares).  The incorporation of the eccentricity fluctuations has only small impact on the particle yields and $\langle p_T\rangle$, but the addition of pre-equilibrium raises the $\langle p_T \rangle$ by as much as 20\%.  The hadronic cascade phase leads to increased proton yields by allowing for an effective chemical freeze-out as first observed in~\cite{Hirano:2002eh}.  The reduced flow in the cascade relative to full hydrodynamics leads to decreased $\langle p_T \rangle$ values for pions and kaons.  In adding features to the model, we make no attempt to modify the initial parameters to achieve better agreement with the data.  This will be the main focus of the results section.

Similar comparisons for the elliptic flow for charged hadrons and sidewards and outwards radii for pions are shown in Fig.~\ref{fig:cali_mean_v2_radii}.  Each modification to VH2 serves to increase the $v_2$.  For $N_{part}>200$ the model results come closer to the comparison data measured by STAR~\cite{Adams:2005tm}, however for more peripheral collisions the addition of pre-eqiuilibrium flow causes the value of $v_2$ to overshoot the data by a significant margin.  For very peripheral collisions the anisotropy of the stress energy tensor may no longer be independent of the spatial coordinates, thereby violating the conditions required for universal flow~\cite{Vredevoogd:2009jt}.  The hadronic cascade phase leads to a reduction in the value of $v_2$, when compared to a pure hydrodynamic evolution with lower freeze-out temperature.

The right panel of Fig.~\ref{fig:cali_mean_v2_radii} shows the cumulative effects on the transverse radii, $R_{out}$ and $R_{side}$.  The addition of pre-equilibrium flow and the hadronic cascade lead to a reduction in $R_{out}$, and the latter increases $R_{side}$, bringing the two closer to each other and to the ratio observed in the data.  The measurements by STAR and PHENIX for 200 GeV Au+Au are omitted from this figure for clarity in order to better reveal the trends, which are consistent with the one-dimensional relativistic viscous hydro results documented in~\cite{Pratt:2009bk}.

\section{Model Evalulation}\label{sec:eval}

\subsection{Model Parameterizations}\label{subsec:param}

Because the published experimental results do not generally conform to a uniform binning, the model results are fit to a functional form prior to evaluating the overall agreement with the data.  This step also simplifies the treatment of systematic errors in the data.  We selected a purely empirical fitting function in order to reduce any bias in treating the model results prior to comparison to experiment.  The $p_T$ dependence of $v_2$ and the $k_T$ dependence of the femtoscopic radii are both well described by a Chebyshev polynomial.  However, polynomials in general are not easily adapted to the large dynamic range characteristic of spectra.   The spectra require an exponential form to describe the data, and we achieved satisfactory agreement by multiplying a fifth order Chebyshev polynomial by an exponential in transverse mass that has been shown to fit a wide range of particle spectra~\cite{Adler:2004ki}.  The functional form use to fit the model spectra is given by Eq.~\ref{eq:specfit},
 \begin{equation} 
    y(p_{T}) =  \left(\sum_{i=1}^{5} a_{i}T_{i}(p_{T})\right) \cdot \frac
    {\exp{\left[-(m_{T}-m_{0})/T\right]}}
    {2\pi T(T+m_{0})} 
    \label{eq:specfit}	
\end{equation}
We count on the Chebyshev multilpier to cancel any bias that may result from incorporating the slope parameter, $T$, into the fit.  We have verified that these functional forms provide a good description of the model results, with $\chi^2_{ndf}$ close to unity for most systems.  

\subsection{Data Comparisons}\label{subsec:chi2comp}

To evaluate the model parameters we calculate a chi-squared statistic for each combination of model parameterization and data set.  We have selected four different initial states for evaluation, including both participant ($N_{part}$) and binary collision ($N_{coll}$) scaled energy density profiles, each prepared with and without pre-equilibrium flow.  The initial Glauber distributions were generated for a fixed impact parameter of 4.4~fm, corresponding to the 0--20\% centrality bin for 200~GeV Au+Au collisions.  For each of the four initial state distributions we generate a two-dimensional grid of initial temperatures and viscosity to entropy ratios.  The initial viscosity to entropy ratios were sampled in steps of 0.08, from a lower bound of 0.0001 (a non-zero value is required for numerical stability) to an upper bound of 0.48.  The initial temperature was varied in steps of 5~MeV for spectra and 20~MeV for flow and radii, with the ranges adjusted iteratively to reach above and below the optimal parameter values for each set of initial conditions.  The full set of input parameters and data comparison sets are listed in Table~\ref{tab:modelgrid}.  
\begin{table}[th]
\begin{tabular}{|l|c|c|c|}
\hline
Initial Profile                      & $\eta/s$        & $T_{\rm cent}^{\rm spectra}$   & $T_{\rm cent}^{\rm v2-femto}$\\ 
\hline
$N_{part}$                         & 0.0001--0.48  & 280--325 MeV                            &  260--380 MeV \\
\hline
$N_{part,preq}$                   &  0.0001--0.48 & 270--315  MeV                           & 260--380 MeV  \\
\hline
$N_{coll}$                          & 0.0001--0.48  & 330--370 MeV                            &300--420 MeV \\
\hline
$N_{coll,preq}$                    &  0.0001--0.48 & 310--350  MeV                           & 300--420 MeV \\
\hline
\end{tabular}
\caption{(Color online) Initial VH2 density profile and range of initial central temperatures ($T_{cent}$) and $\eta/s$ values.  All runs used a fixed 4.4~fm impact parameter with $\langle N_{part} \rangle$=276, $\tau$=1.0~fm/c, the default QCD-inspired EOS, and $T_{sw}$=165~MeV.}
\label{tab:modelgrid}
\end{table}
A moderate centrality range of 0--20\% was selected to avoid large corrections to the $v_2$ associated with non-flow effects in the data~\cite{Ollitrault:2009gm} and to match existing pion femtoscopy, $v_2$, and spectra.  The pion spectra are compared to the 0--20\% centrality measurements by PHENIX~\cite{Adler:2004ki} and STAR~\cite{Adams:2004dg}.  Centrality matching is performed using the nearest $\langle N_{part} \rangle$ for which data are published.  For this impact parameter we calculate a value of $\langle N_{part} \rangle$=276, whereas PHENIX reports a value of 286 and STAR reports a value of 282 for their respective 0--20\% centrality ranges.  The nearest centrality for pion radii for PHENIX are for the centrality range of 0--30\%, but the presence of two pions in one of the PHENIX central arms imposes a slight centrality bias and $\langle N_{part} \rangle$=281 for these data~\cite{Adcox:2002bo}.  The nearest centrality bin for the STAR pion radii is 5--10\%, which have $\langle N_{part} \rangle$=298~\cite{Adams:2005cx}.  Based on the $N_{part}$ scaling analysis in~\cite{Adler:2004ii}, this centrality difference leads to a difference in radii of less than 2\%, well below the errors of either measurement.  For the $v_2$ we compare to two sets of measurements by PHENIX in this centrality range, a combined analysis of pions and kaons~\cite{Adler:2003gs}, and a recent analysis of pion $v_2$ at higher $p_{T}$~\cite{Adare:2012iv}.  The STAR $v_2$ results for pions are also from the 5--10\% centrality range~\cite{Adams:2005cx}.  From the centrality dependence shown in Fig.~\ref{fig:cali_mean_v2_radii}, a reduction of 26 in $\langle N_{part} \rangle$ can produce an increase in $v_{2}$ $\sim$8\%, which is comparable to the reported errors and may be large enough to influence the results.  The impact of this discrepancy is examined further in Section~\ref{sec:sys}.

To incorporate the systematic errors when evaluating the chi-squared statistic, we follow the procedure defined in~\cite{Adare:2008cs}.  Because it is generally not feasible to calculate a full co-variance matrix for the systematic errors, the authors of~\cite{Adare:2008cs} make a set of simplifying assumptions that correspond to different types of systematic errors.  For this analysis we assume all systematic errors assumed to be of type B, in which the systematic errors are assumed to be fully correlated within a single experimental analysis.   With this assumption, the modified $\chi^2$ formula reduces to,
\begin{equation}
{\chi}^2(\epsilon_b,{p})={\left[ \left( \sum_{i=1}^{n}
{{(y_i+\sigma_{b_i} -\mu_i({p}))^2}  \over \sigma_{i}^2 \left( 1 + {\epsilon_b }\sigma_{b_i}/y_i \right)^2
} \right) + \epsilon_b^2\right]},
\label{eq:chi2B}
\end{equation}
where $\sigma_{b_i}$ represents the systematic error, and $\epsilon_b$ represents the fraction of the error by which the set of correlated measurements move in tandem, either up, down, or around a tipping point.  The full procedure for evaluating the modified $\chi^2$ and the derivation are given in the appendix of~\cite{Adare:2008cs}.  Although hydrodynamic model results have been compared to experimental data at higher values of $\langle p_{T}\rangle$, we restrict the range of comparison to $\langle p_{T}\rangle < 1.5~GeV$ in order to reduce any potential bias from hard processes.  However, within this range all particles are given equal statistical weight.

\section{Results}\label{sec:results}

\begin{figure}
\includegraphics[width=0.35\textwidth]{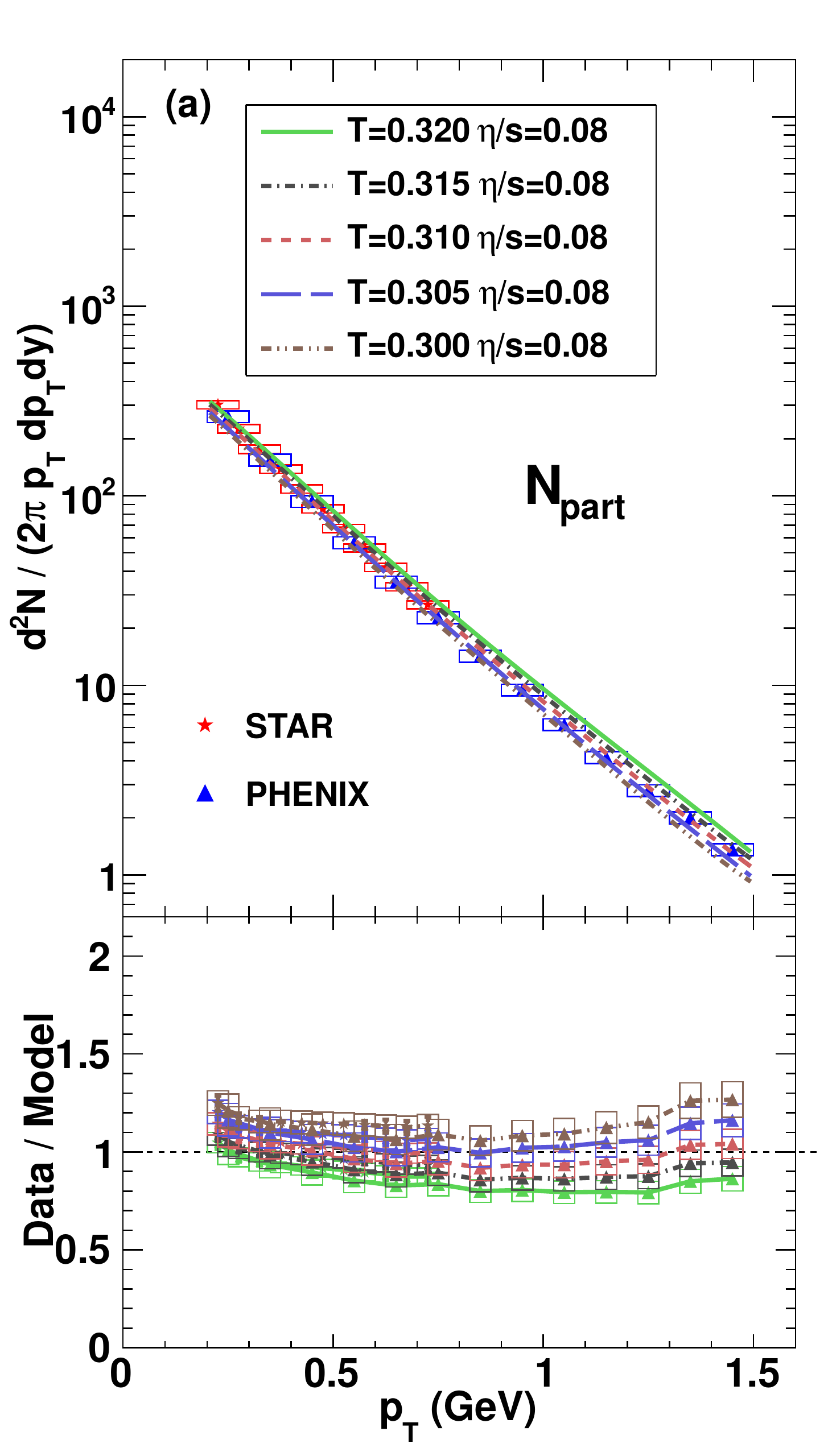}
\includegraphics[width=0.35\textwidth]{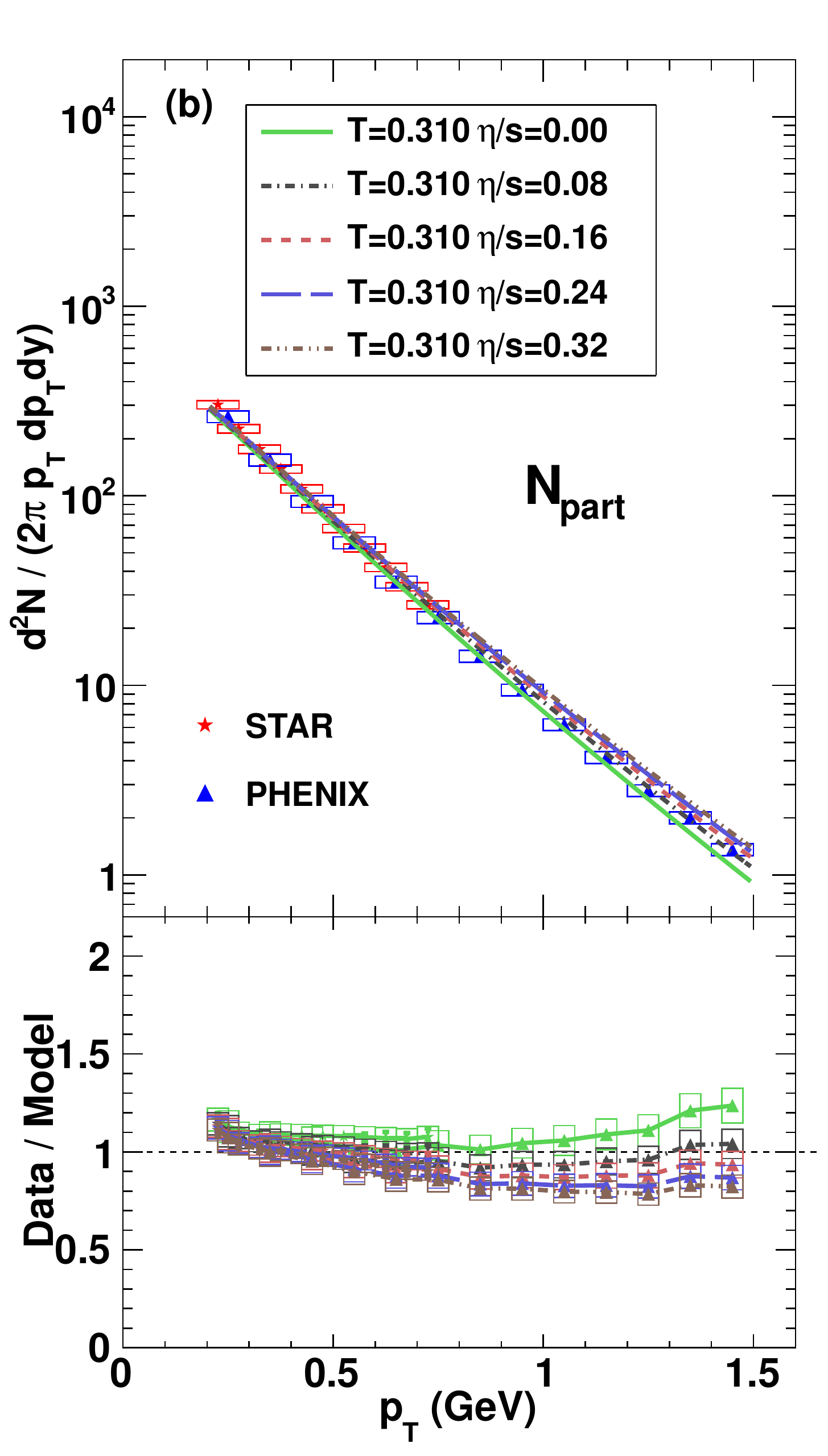}
\caption{(Color online) Model evaluation of pion spectra with $N_{part}$ scaling for fixed $\eta/s$ (a), and fixed $T_{cent}$ (b).}
\label{fig:spec_npart}
\end{figure}
\begin{figure}
\includegraphics[width=0.35\textwidth]{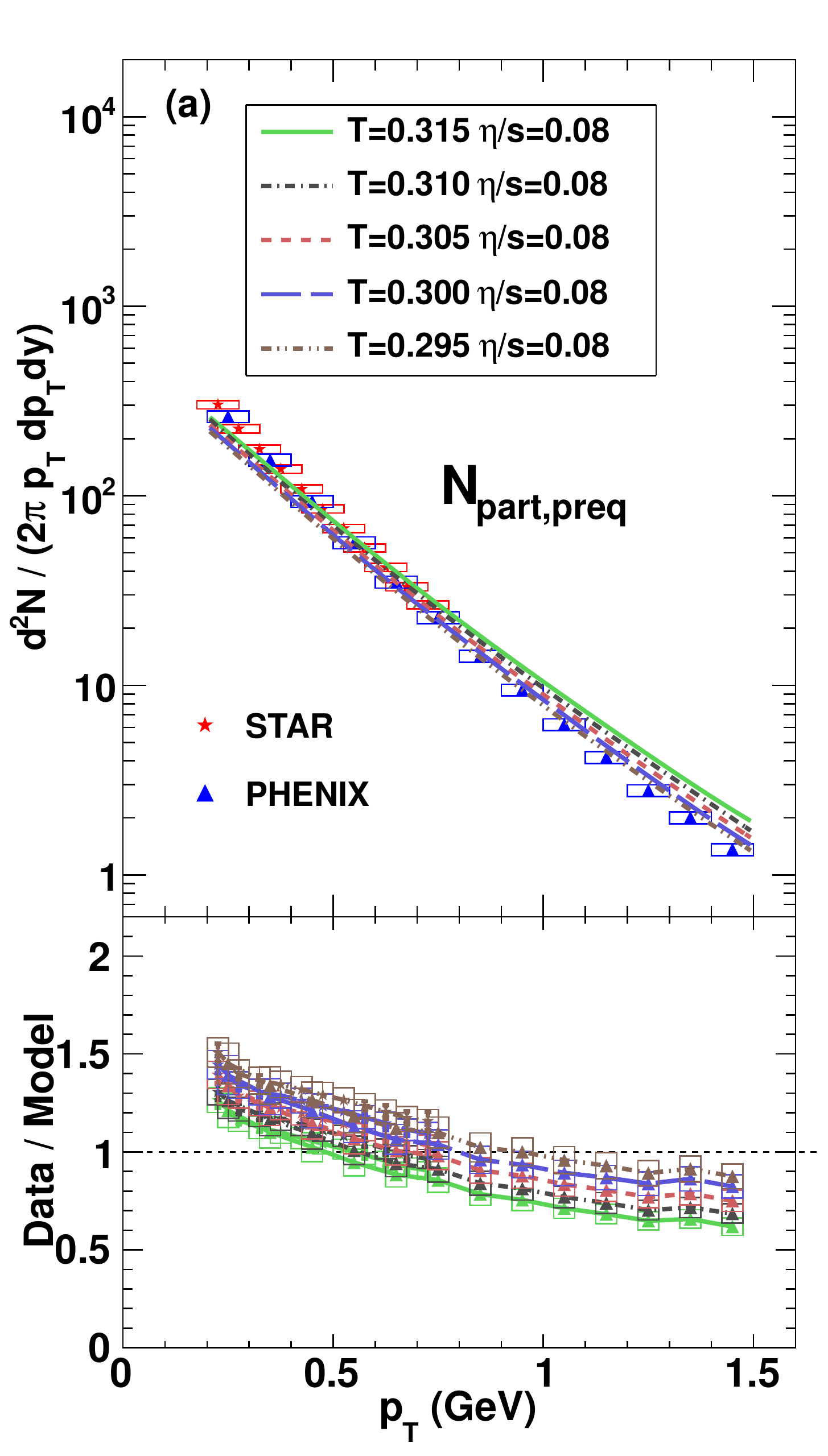}
\includegraphics[width=0.35\textwidth]{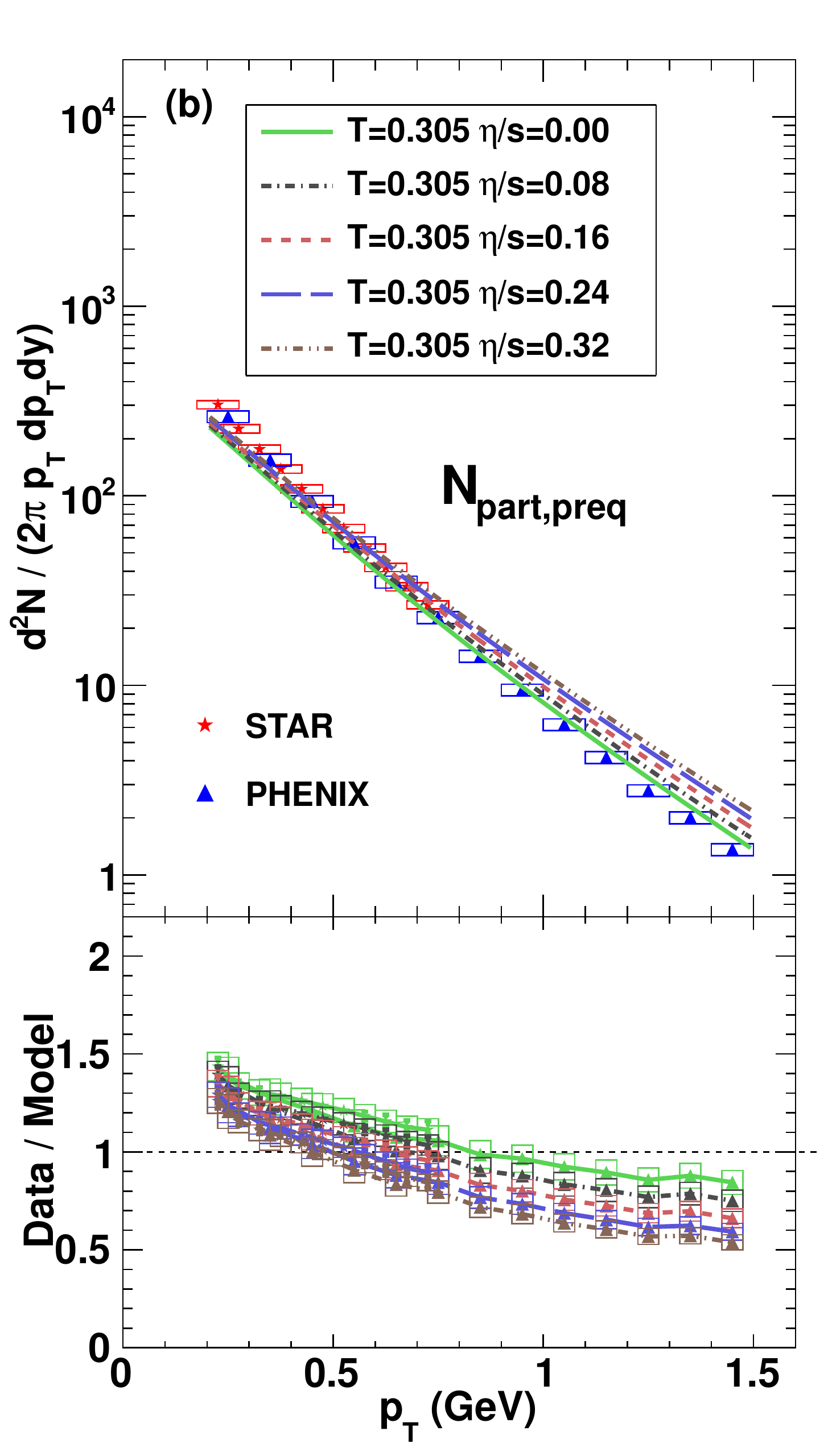}
\caption{(Color online) Model evaluation of pion spectra with $N_{part}$ scaling with pre-equilibrium flow for fixed $\eta/s$ (a), and fixed $T_{cent}$ (b).}
\label{fig:spec_npart_preq}
\end{figure}
We first examine the pion spectra for $N_{part}$ scaling without pre-equilibrium flow in Fig.~\ref{fig:spec_npart}.  Panel (a) shows the variation with initial temperature for a central collision ($T_{cent}$) for fixed $\eta/s$=0.08, and panel (b) shows the variation with $\eta/s$ for fixed $T_{cent}$=0.310~GeV.  The ratios of data to model are plotted in the bottom portion of each panel.  From the figures, it appears that the initial parameters $T_{cent}$=0.310~GeV, $\eta/s$=0.08 provide the best agreement with the data.  Evaluations for pre-equilibrium flow are shown in Fig.~\ref{fig:spec_npart_preq}.  For these figures, the high momentum spectra fall less steeply than the data, and the overall agreement is not as good.  However, the primary motivation for the CHIMERA framework is to evaluate the $\chi^{2}_{ndf}$ for each parameter set.  These $\chi^2$ evaluations for the comparisons shown in Fig.~\ref{fig:spec_npart} are given in Table~\ref{tab:ch2_spec_npart_etafix} for fixed $\eta/s$ and in Table~\ref{tab:ch2_spec_npart_Tfix} for fixed $T_{cent}$.  Because the systematic errors are independent for each experiment, the $\chi^2_{ndf}$ values are reported separately for PHENIX and for STAR.  For the fixed $\eta/s$ comparison, the minimum $\chi^2_{ndf}$ occurs at an initial temperature of 0.310~GeV for both experiments when there is no pre-equilibrium flow, and it occurs a lower initial temperature 0.300--0.305~GeV when pre-equilibrium flow is included.  Although the model spectra with pre-equilibrium flow are noticeably steeper, the independent systematic errors for PHENIX and STAR permit a reasonable $\chi^2_{ndf}$ to be achieved.  Table~\ref{tab:ch2_spec_npart_Tfix} shows that lowest $\chi^2_{ndf}$ values are achieved for $\eta/s$ near the conjectured minimum of 0.08.
\begin{table}[ht]
\begin{tabular}{|c|r r|r r|}
\hline
$T_{\rm cent}$ &
 \multicolumn{2}{c|}{$\chi^2_{ndf}\ N_{part}$} &
 \multicolumn{2}{c|}{$\chi^2_{ndf}\ N_{part,preq}$} \\
(GeV)             & PHNX & STAR & PHNX & STAR \\ 
\hline
0.320 & 21.73 & 1.41 & 0.00 & 0.00 \\
\hline
0.315 & 15.69 & 0.25 & 98.77 & 14.69 \\
\hline
0.310 &   9.27 & 0.60 & 45.56 & 2.98 \\
\hline
0.305 & 20.32 & 1.82 & 7.84 & 2.46 \\
\hline
0.300 & 18.33 & 1.47 & 2.90 & 10.94 \\
\hline
0.295 &  0.00 & 0.00 & 179.12 & 10.57 \\
\hline
\end{tabular}
\caption{$\chi^2_{ndf}$ for evaluation of pion spectra with fixed
  $\eta/s$=0.08 for $N_{part}$ scaling with and without
  pre-equilibrium flow.}
\label{tab:ch2_spec_npart_etafix}
\end{table}
\begin{table}[ht]
\begin{tabular}{|c|r r|r r|}
\hline
& \multicolumn{2}{c|}{$\chi^2_{ndf}\ N_{part}$}
& \multicolumn{2}{c|}{$\chi^2_{ndf}\ N_{part,preq}$} \\
$\eta/s$  & PHNX & STAR & PHNX & STAR \\ 
\hline
0.32 & 14.59 & 0.68 & 231.09 & 9.55 \\
\hline
0.24 & 16.11 & 0.47 & 156.79 & 2.46 \\
\hline
0.16 & 16.64 & 0.49 & 53.23& 3.16 \\
\hline
0.08 &   9.27 & 0.60 & 7.84& 1.83 \\
\hline
$10^{-4}$ & 17.58 & 0.81 & 3.87 & 9.55 \\
\hline
\end{tabular}
\caption{$\chi^2_{ndf}$ for evaluation of pion spectra with fixed
$T_{\rm cent}$=0.310 GeV for $N_{part}$ scaling and 
$T_{\rm cent}$= 0.305 GeV for $N_{part,preq}$ scaling.}
\label{tab:ch2_spec_npart_Tfix}
\end{table}

Because we are interested in combining the evaluations from several measurements, it is also important to study the shape of the $\chi^2$ sum divided by the total number of degrees of freedom for the full set of $T_{cent}$ and $\eta/s$ parameters.  The $\chi^2_{ndf}$ distributions over the full parameter space are shown graphically in Fig.~\ref{fig:ch2_spec_npart} for $N_{part}$ scaling and in Fig.~\ref{fig:ch2_spec_npart_preq} for $N_{part}$ scaling with pre-equilibrium flow.  In each figure the full distribution is shown in panel (a) and a paraboloid fit used to determine the location and curvature near the minimum is shown in panel (b).  The striking feature of these figures is the strong correlation that exists between $T_{cent}$ and $\eta/s$.  Most recent evaluations of hydrodynamic models use the spectra comparisons only to establish the initial temperature.  These figures demonstrate the value of a simultaneous comparison for the temperature and viscosity to entropy ratio.
\begin{figure}
=\includegraphics[width=0.29\textwidth]{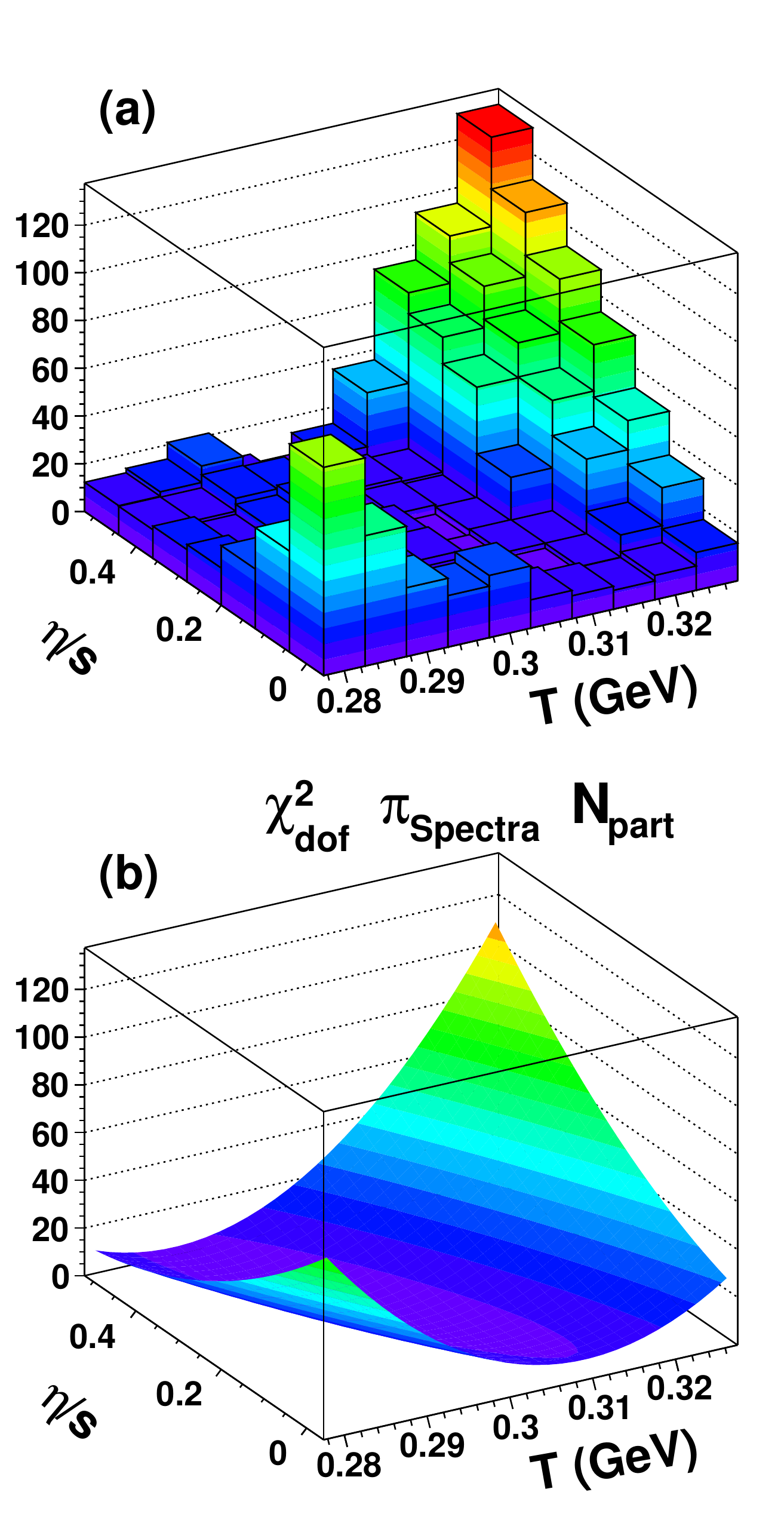}
\caption{(Color online) Model evaluation of pion spectra total $\chi^2_{ndf}$ distribution (a) with paraboloid fit (b) for $N_{part}$ scaling.}
\label{fig:ch2_spec_npart}
\end{figure}
\begin{figure}
\includegraphics[width=0.29\textwidth]{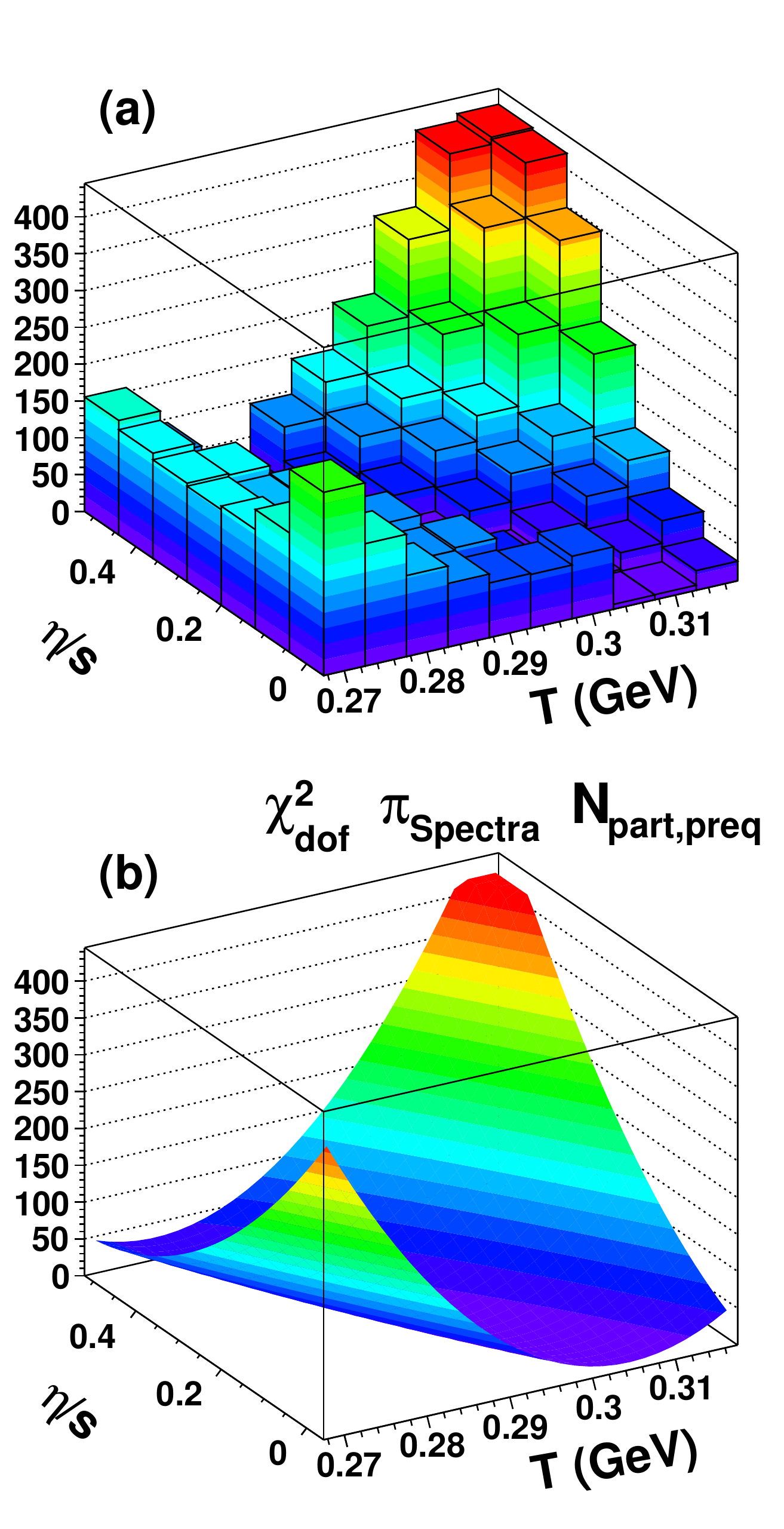}
\caption{(Color online) Model evaluation of pion spectra total $\chi^2_{ndf}$ distribution (a) with paraboloid fit (b) for $N_{part}$ scaling with pre-equilibrium flow.}
\label{fig:ch2_spec_npart_preq}
\end{figure}

\begin{figure}
\includegraphics[width=0.35\textwidth]{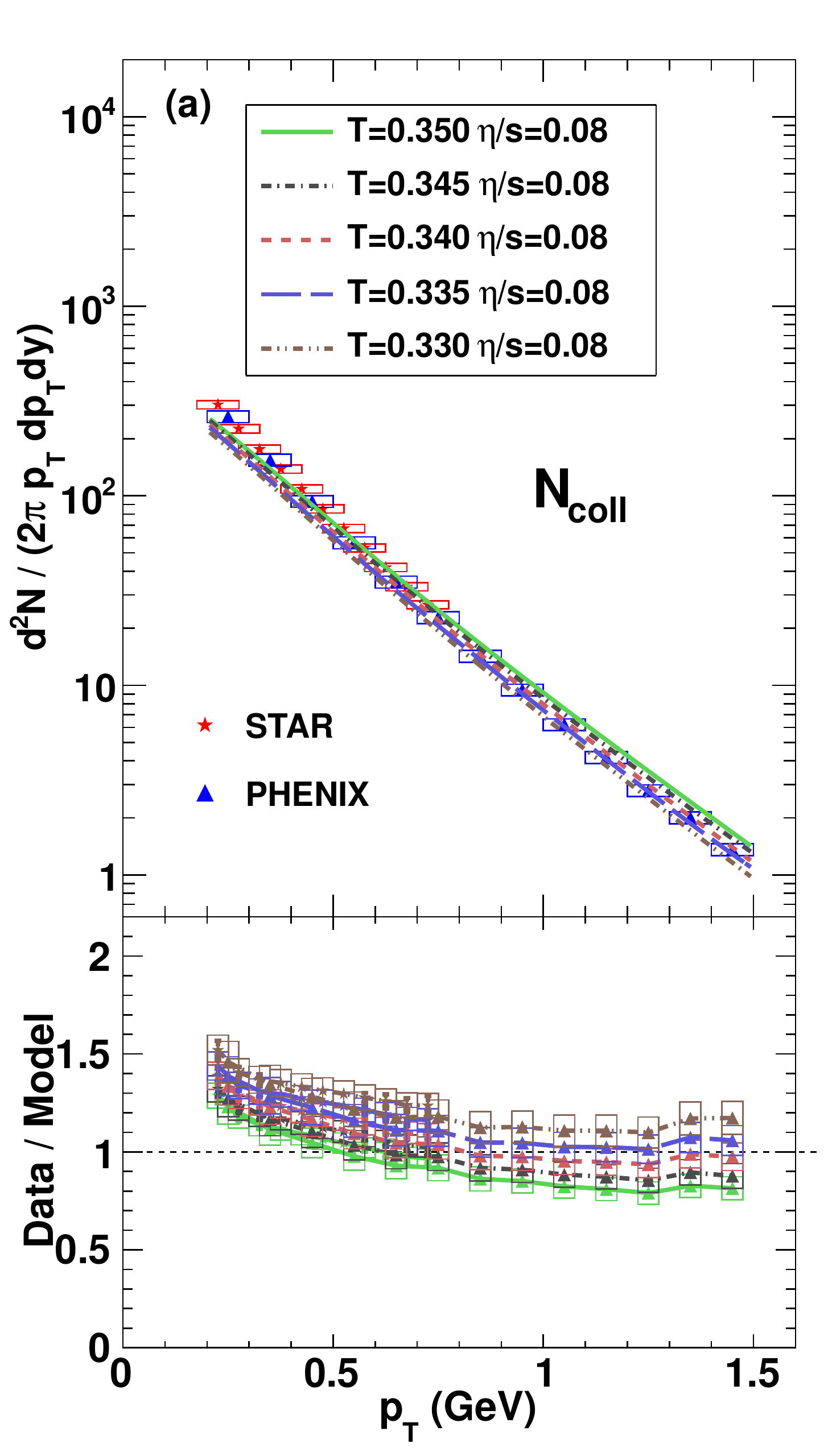}
\includegraphics[width=0.35\textwidth]{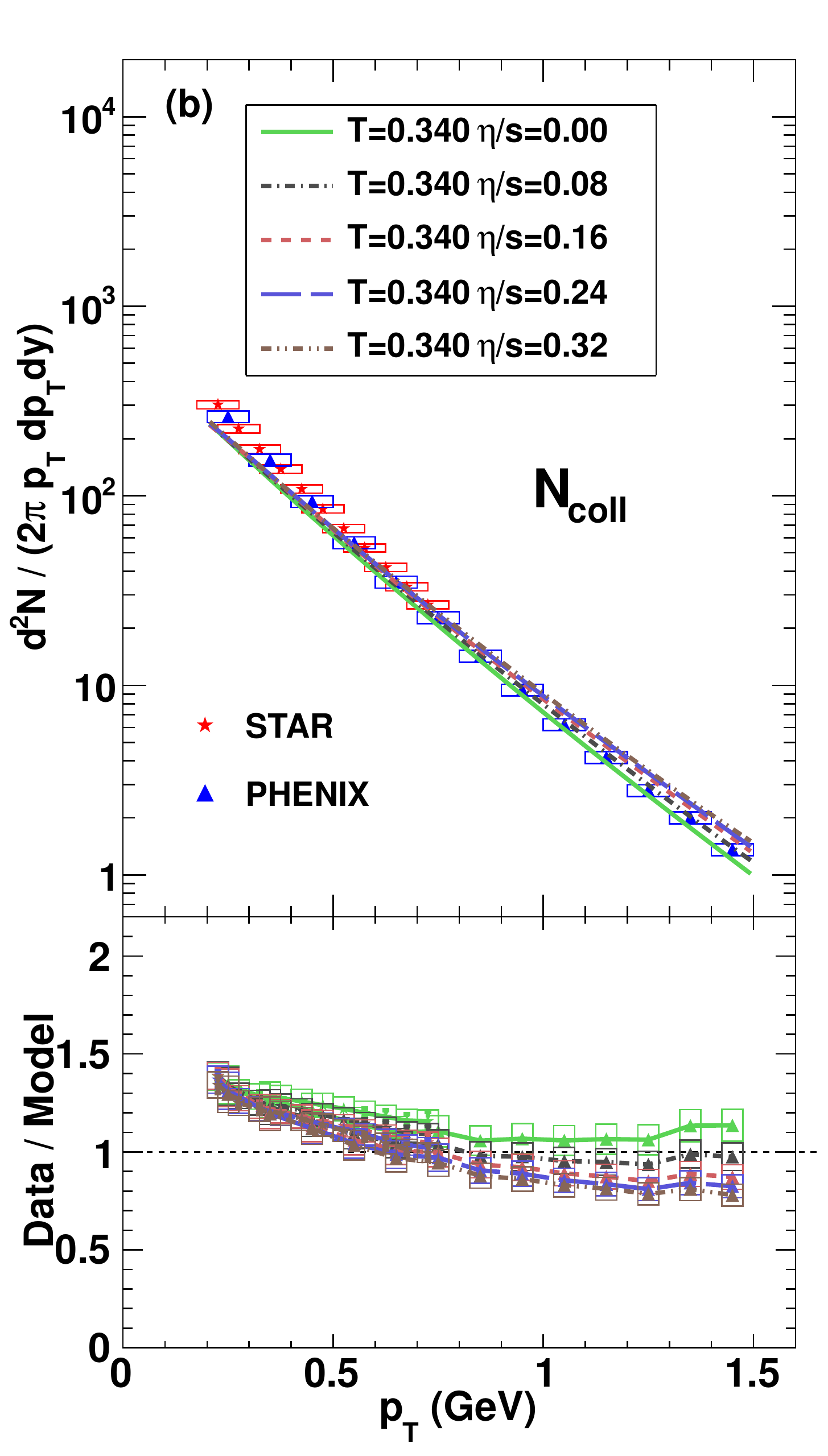}
\caption{(Color online) Model evaluation of pion spectra with $N_{coll}$ scaling for fixed $\eta/s$ (a), and fixed $T_{cent}$ (b).}
\label{fig:spec_ncoll}
\end{figure}
\begin{figure}
\includegraphics[width=0.35\textwidth]{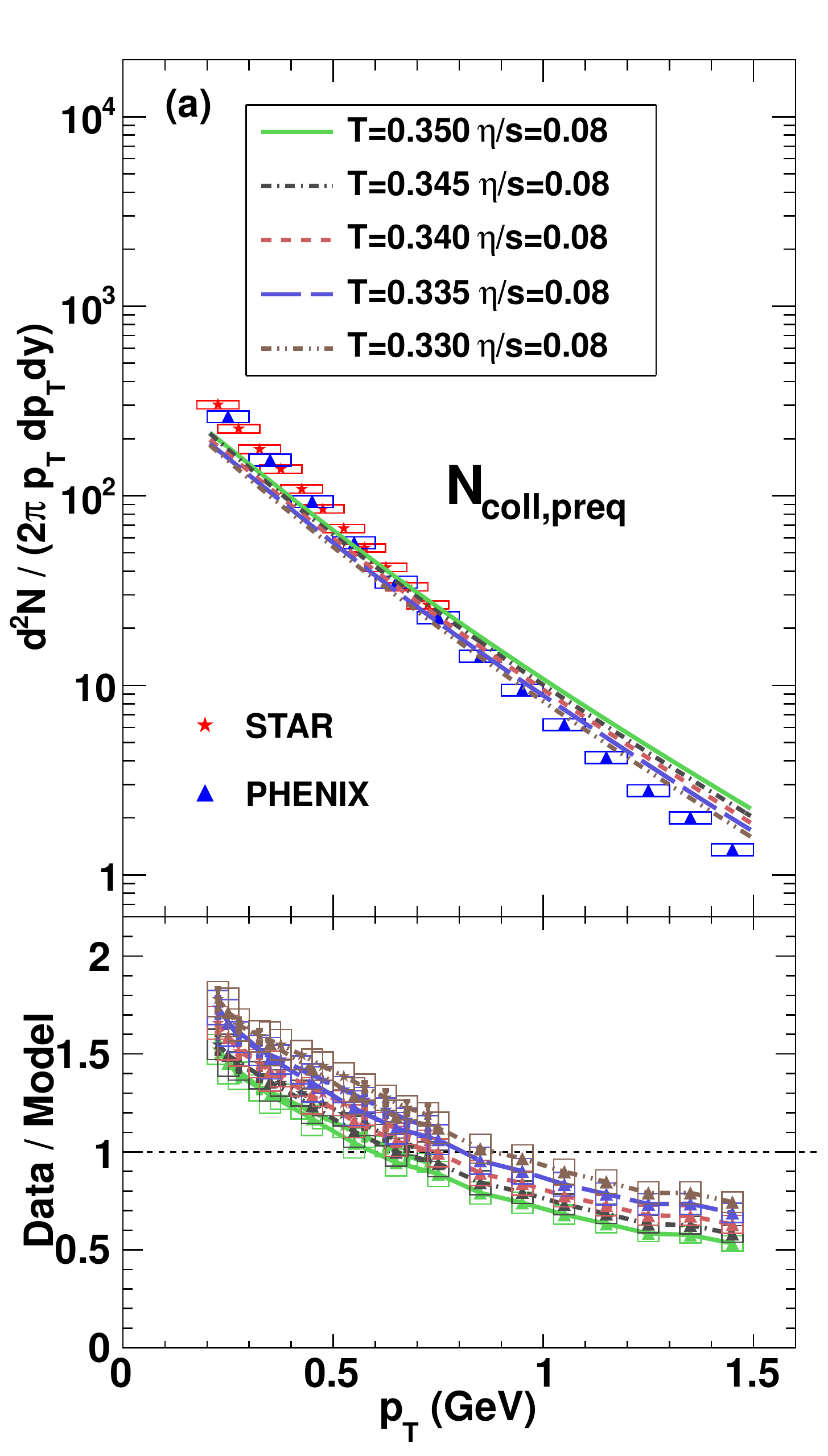}
\includegraphics[width=0.35\textwidth]{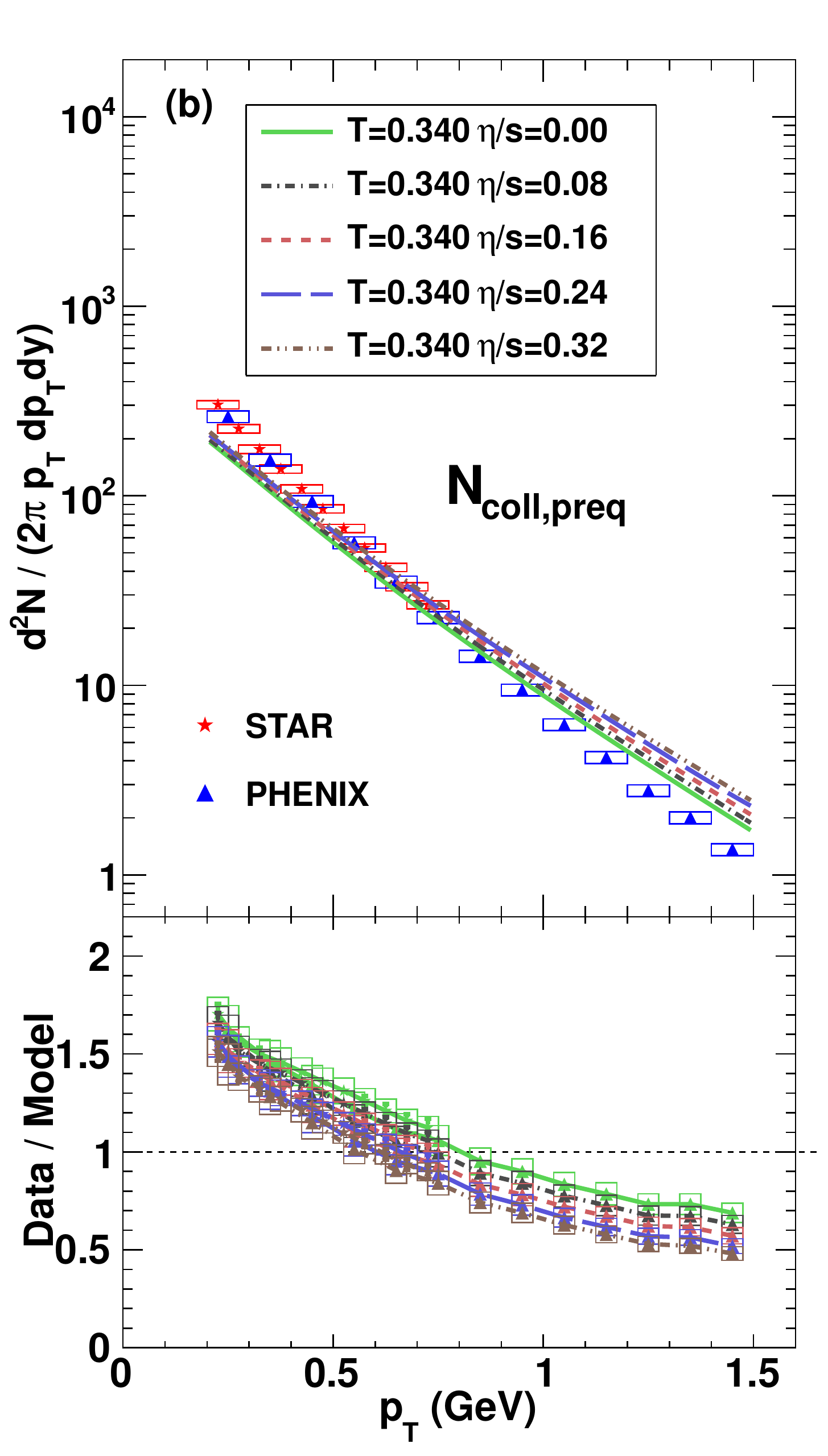}
\caption{(Color online) Model evaluation of pion spectra with $N_{coll}$ scaling with pre-equilibrium flow for fixed $\eta/s$ (a), and fixed $T_{cent}$ (b).}
\label{fig:spec_ncoll_preq}
\end{figure}

The model comparisons for $N_{coll}$ scaling are shown in Fig.~\ref{fig:spec_ncoll} for the case without pre-equilibrium flow and in Fig.~\ref{fig:spec_ncoll_preq} for the case when it is included.  The case of $N_{coll}$ scaling looks similar to the $N_{part}$ scaling with pre-equilibrium flow, however, the addition of pre-equilibrium flow to the $N_{coll}$ clearly imparts too much transverse momentum to the pions to match the data.
The corresponding $\chi^2_{ndf}$ values are given in Tables~\ref{tab:ch2_spec_npart_etafix} and~\ref{tab:ch2_spec_npart_Tfix}.  Low values of $\chi^2_{ndf}$ are still attainable with $N_{coll}$ scaling.  In this case, the minimum value occurs for a somewhat higher viscosity. 
However, when pre-equilibrium flow is added this is no longer the case.  In both cases, higher initial temperatures are preferred.

\begin{table}[ht]
\begin{tabular}{|c|r r|r r|}
\hline
$T_{\rm cent}$ &
 \multicolumn{2}{c|}{$\chi^2_{ndf}\ N_{coll}$} &
 \multicolumn{2}{c|}{$\chi^2_{ndf}\ N_{coll,preq}$} \\
(GeV)             & PHNX & STAR & PHNX & STAR \\ 
\hline
0.350 & 13.83 & 2.30 & 154.83 & 9.25 \\
\hline
0.345 &   2.77 & 1.75 & 73.04 & 15.56 \\
\hline
0.340 & 15.71 & 8.15 & 33.36 & 30.97 \\
\hline
0.335 & 75.85 & 6.87 & 16.10 & 33.07 \\
\hline
0.330 & 60.26 & 6.94 & 20.87 & 54.94 \\
\hline
\end{tabular}
\caption{$\chi^2_{ndf}$ for evaluation of pion spectra with fixed
  $\eta/s$=0.08 for $N_{coll}$ scaling with and without
  pre-equilibrium flow.}
\label{tab:ch2_spec_ncoll_Tfix}
\end{table}
\begin{table}[ht]
\begin{tabular}{|c|r r|r r|}
\hline
& \multicolumn{2}{c|}{$\chi^2_{ndf}\ N_{coll}$}
& \multicolumn{2}{c|}{$\chi^2_{ndf}\ N_{coll,preq}$} \\
$\eta/s$  & PHNX & STAR & PHNX & STAR \\ 
\hline
0.32 & 8.84 & 2.14 & 308.86 & 10.53 \\
\hline
0.24 & 3.48 & 2.18 & 185.14 & 12.29 \\
\hline
0.16 & 3.50 & 2.34 & 87.03 & 17.52 \\
\hline
0.08 & 15.71 & 8.15 & 33.36 & 30.97 \\
\hline
$10^{-4}$ & 49.31 & 5.54 & 13.15 & 29.61 \\
\hline
\end{tabular}
\caption{$\chi^2_{ndf}$ for evaluation of pion spectra with fixed
$T_{\rm cent}$=0.340 GeV for $N_{coll}$, $N_{coll,preq}$ scaling.}
\label{tab:ch2_spec_ncoll_Tfix}
\end{table}

The $\chi^2_{ndf}$ distributions and paraboloid fits for spectra with $N_{coll}$ scaling are shown in Fig.~\ref{fig:ch2_spec_ncoll}.  In the case without pre-equilibrium flow the minimum is not well constrained for higher values of $\eta/s$.  With the addition of pre-equilibrium flow, the minimum is sharply defined along the diagonal, but with a substantial value for the minimum, indicating the overall poor quality of agreement with the data.
\begin{figure}
\includegraphics[width=0.29\textwidth]{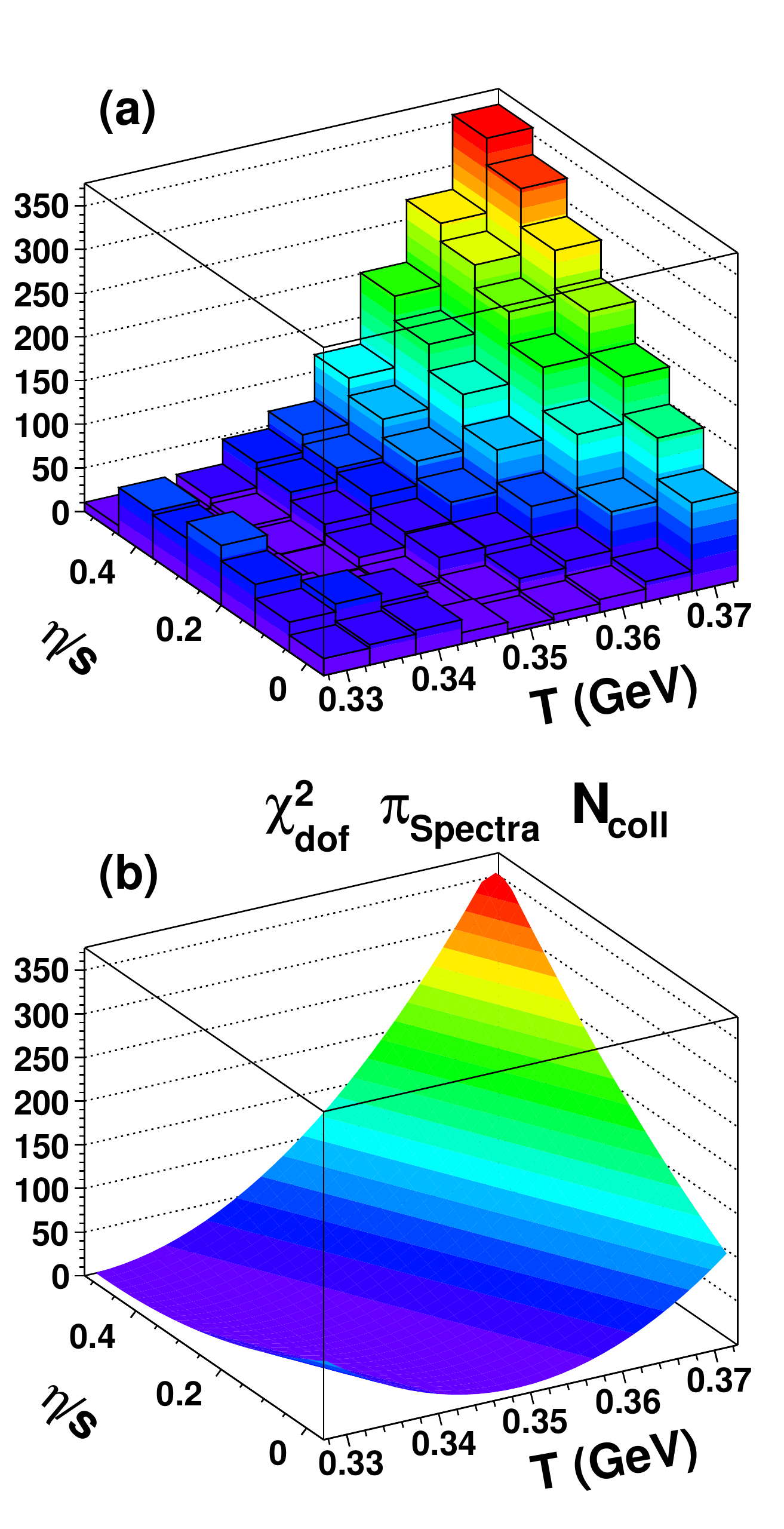}
\caption{(Color online) Model evaluation of pion spectra total $\chi^2_{ndf}$ distribution (a) with paraboloid fit (b) for $N_{coll}$ scaling.}
\label{fig:ch2_spec_ncoll}
\end{figure}
\begin{figure}
\includegraphics[width=0.29\textwidth]{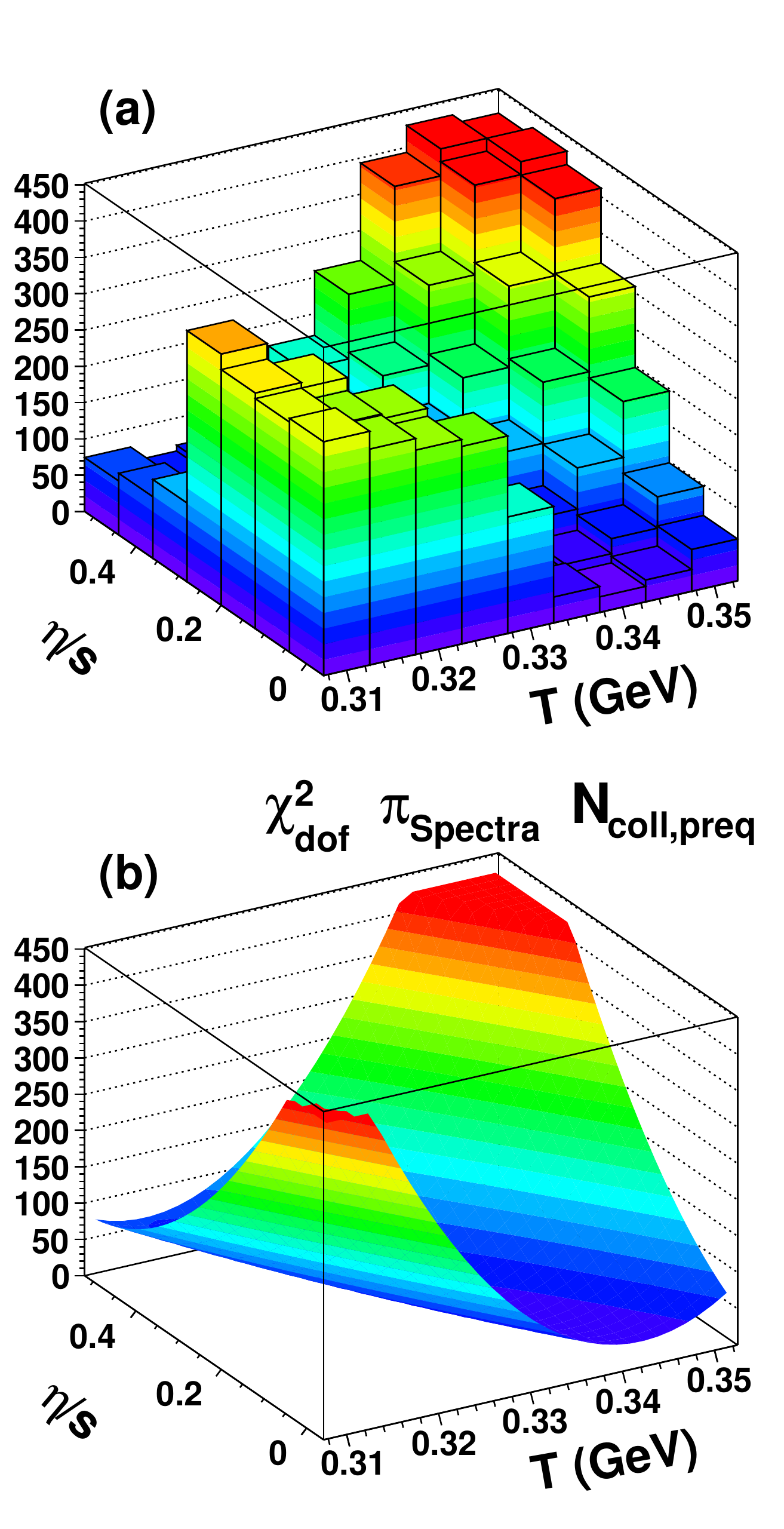}
\caption{(Color online) Model evaluation of pion spectra total $\chi^2_{ndf}$ distribution (a) with paraboloid fit (b) for $N_{coll}$ scaling and pre-equilibrium flow.}
\label{fig:ch2_spec_ncoll_preq}
\end{figure}

Figures~\ref{fig:v2_npart} and~\ref{fig:v2_npart_preq} show the model evaluations for the pion elliptic flow, with the absence and addition of pre-equilibrium flow, respectively.  All comparison are for fixed $T_{cent}$=0.320~GeV --- there is little variation with initial temperature.  For each figure panel (a) shows the pion elliptic flow compared to STAR~\cite{Adams:2005cx} and higher transverse momentum data from PHENIX~\cite{Adare:2012iv} and panel (b) shows a comparison the combined elliptic flow for pions and kaons measured by PHENIX~\cite{Adler:2003gs}.  It is clear from Fig.~\ref{fig:v2_npart_preq} that the additional of pre-equilibrium flow will significantly improve the $\chi^2_{ndf}$ evaluation.  The same is true for the 
$N_{coll}$ scaling comparison, shown in Figures~\ref{fig:v2_ncoll} and~\ref{fig:v2_ncoll_preq} shown without and with pre-equilibrium flow, respectively.

\begin{figure}
\begin{center}
\includegraphics[width=0.40\textwidth]{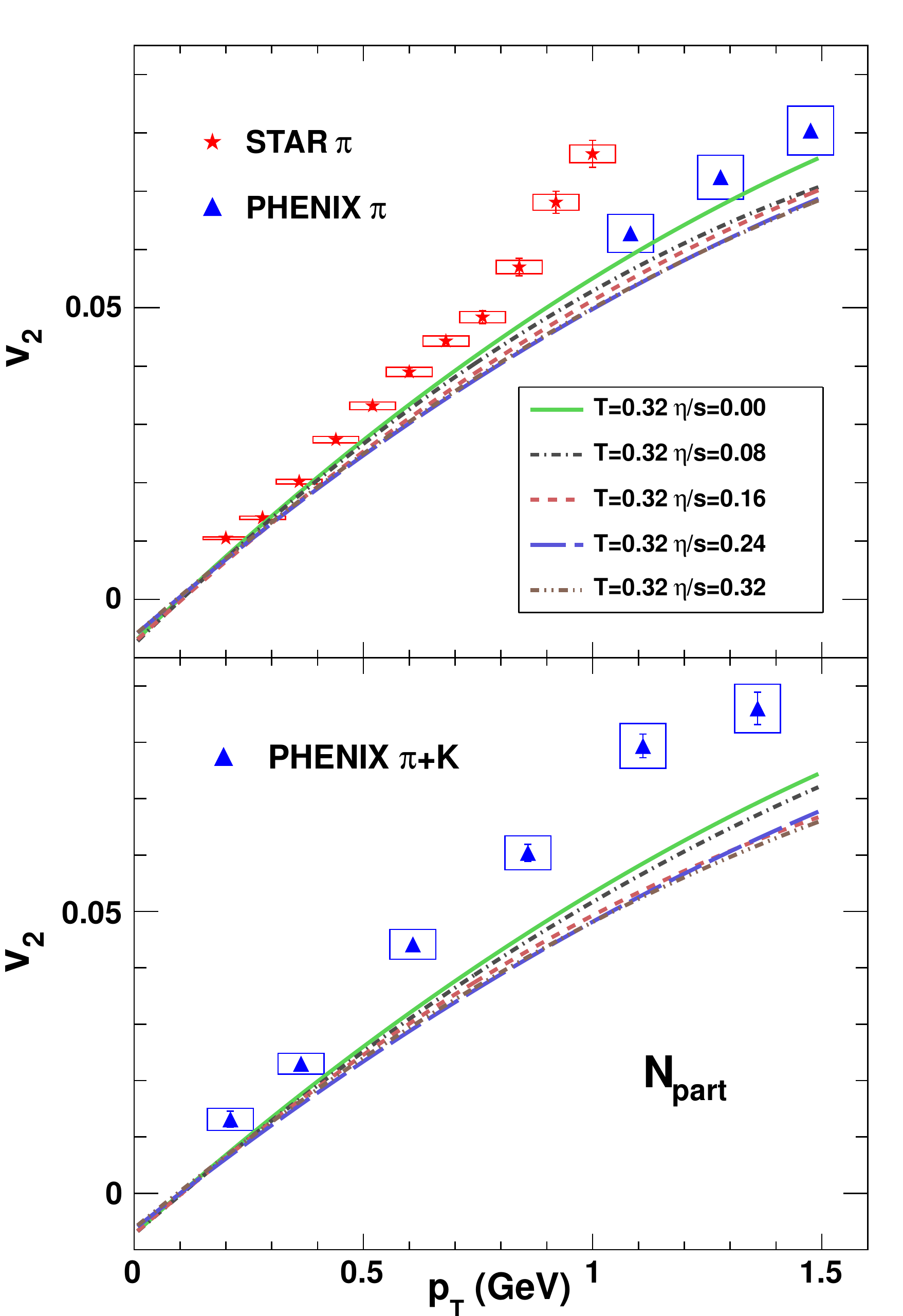}
\end{center}
\caption{(Color online) Model evaluation of pion elliptic flow with $N_{part}$ scaling for fixed $T_{cent}$.}
\label{fig:v2_npart}
\end{figure}
\begin{figure}
\begin{center}
\includegraphics[width=0.40\textwidth]{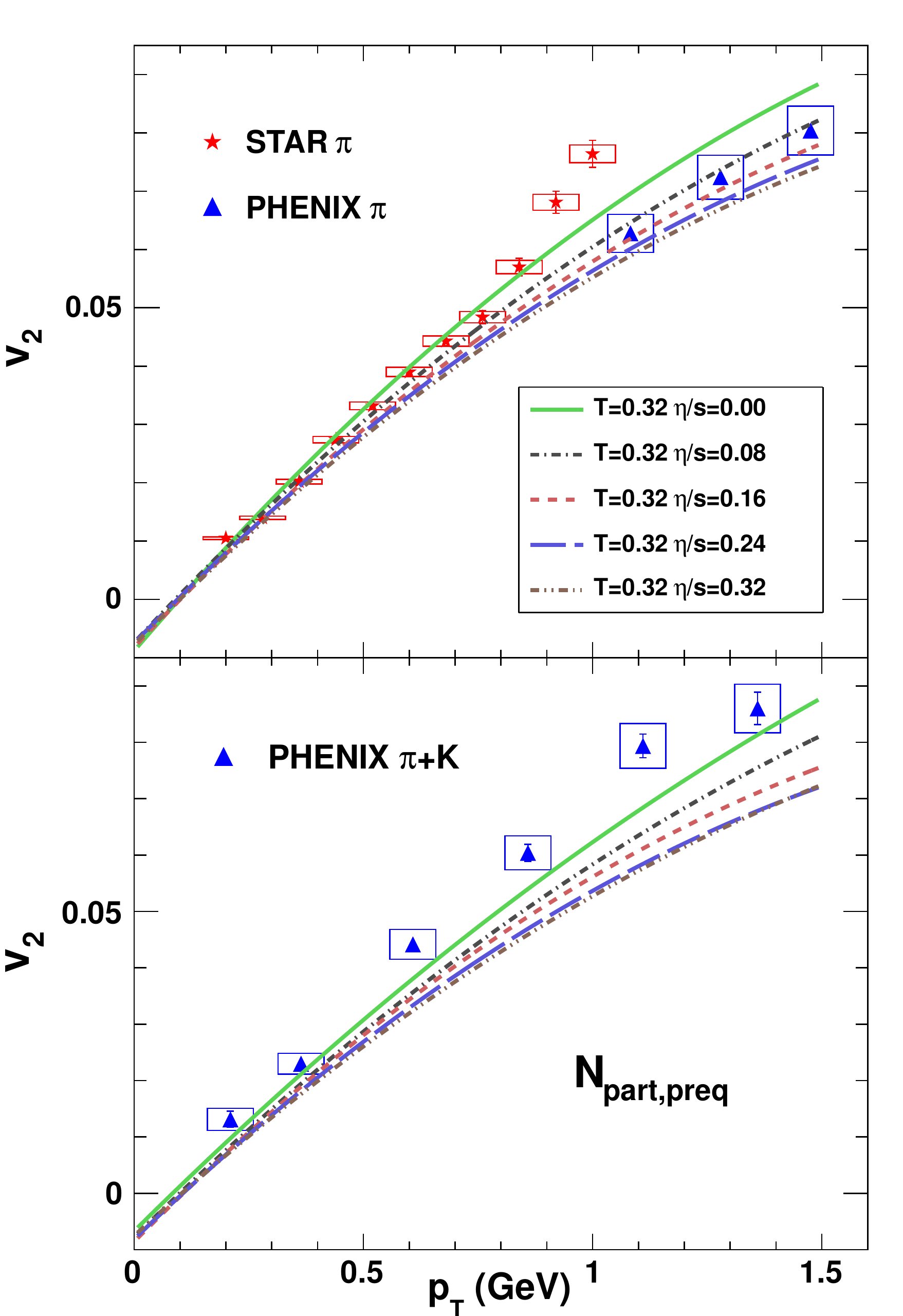}
\end{center}
\caption{(Color online) Model evaluation of pion elliptic flow with $N_{part}$ scaling with pre-equilibrium flow for fixed $T_{cent}$.}
\label{fig:v2_npart_preq}
\end{figure}
\begin{figure}
\begin{center}
\includegraphics[width=0.40\textwidth]{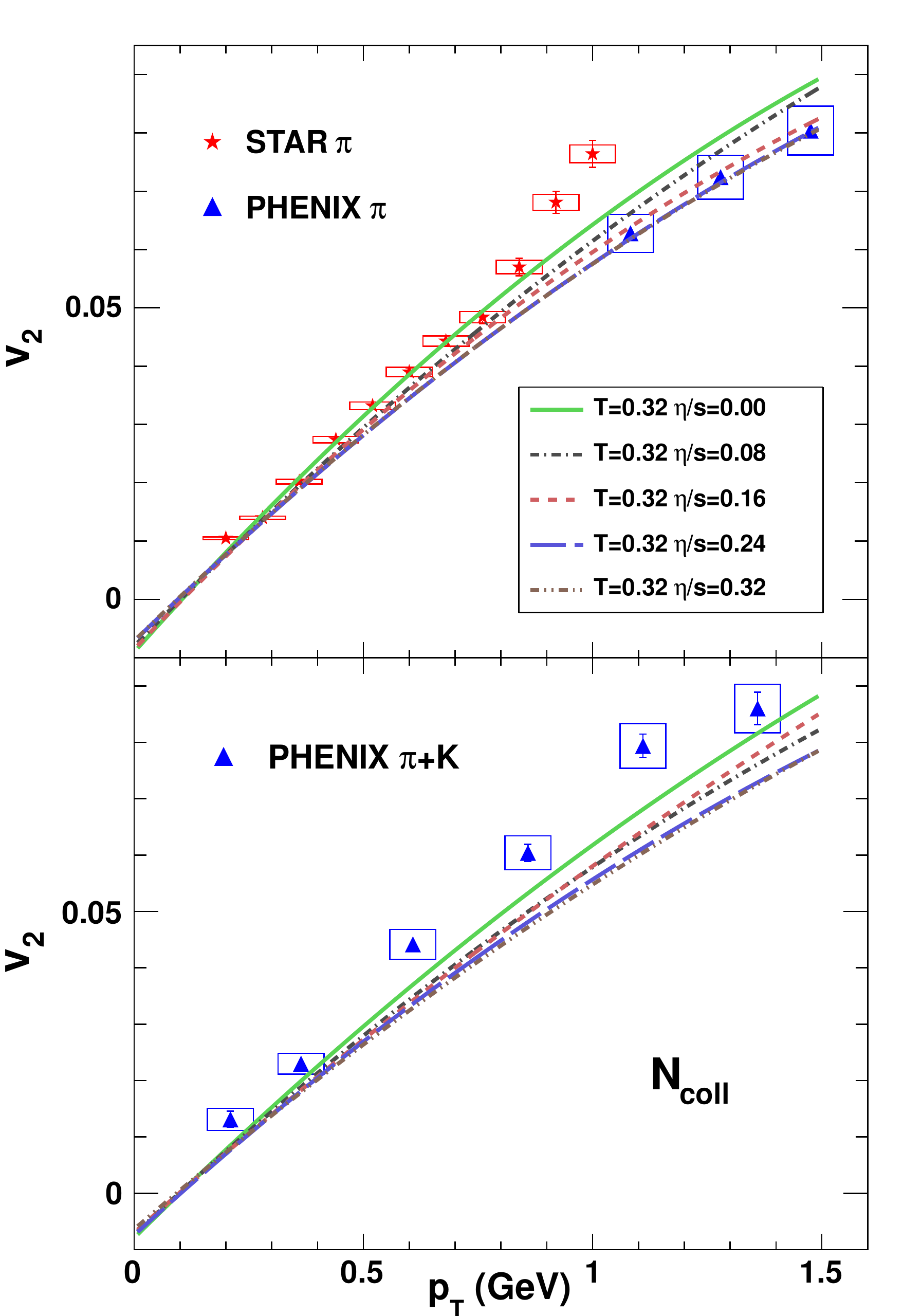}
\end{center}
\caption{(Color online) Model evaluation of pion elliptic flow with $N_{coll}$ scaling for fixed $T_{cent}$.}
\label{fig:v2_ncoll}
\end{figure}
\begin{figure}
\begin{center}
\includegraphics[width=0.40\textwidth]{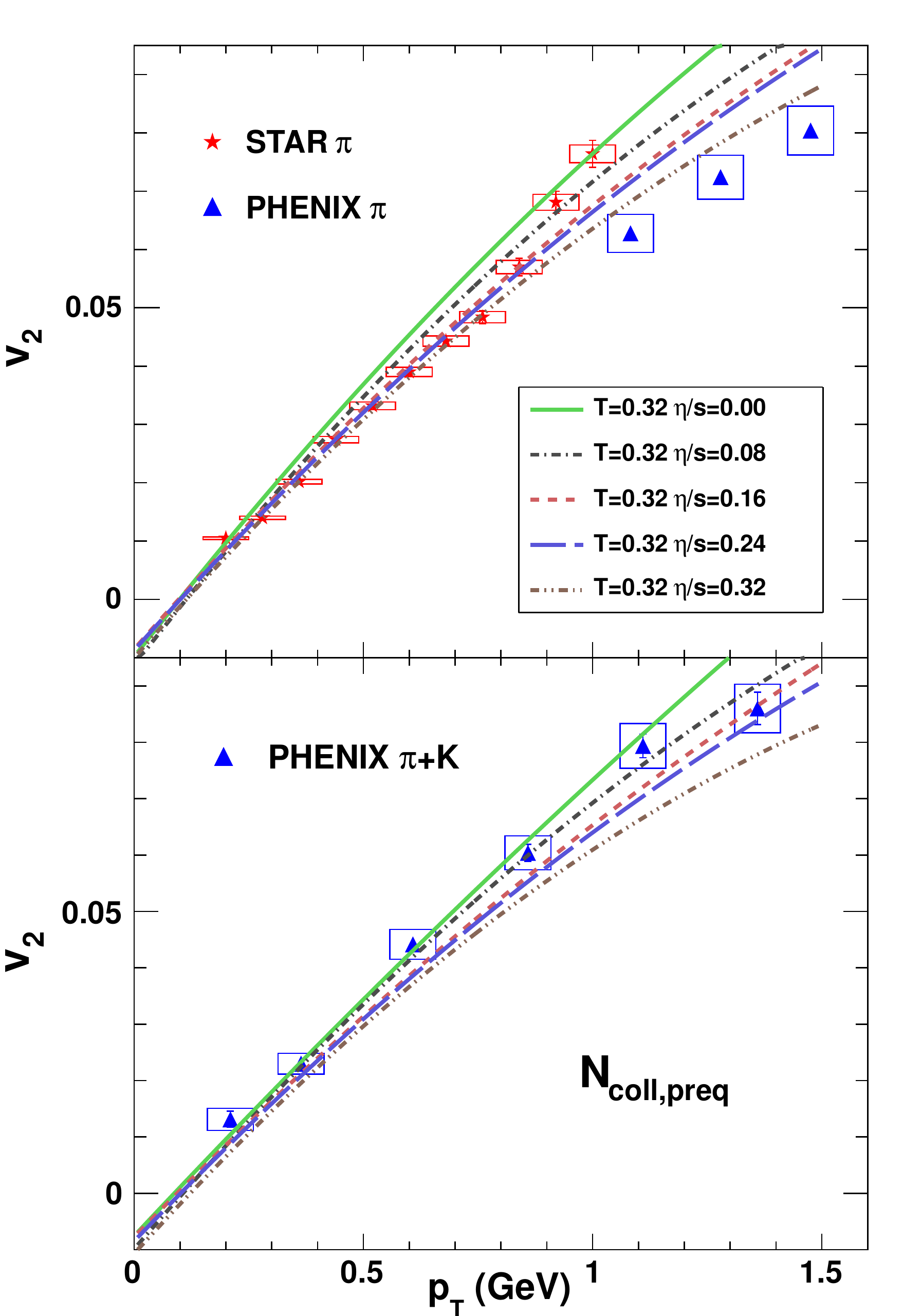}
\end{center}
\caption{(Color online) Model evaluation of pion elliptic flow with $N_{coll}$ scaling with pre-equilibrium flow for fixed $T_{cent}$.}
\label{fig:v2_ncoll_preq}
\end{figure}

The $\chi^2_{ndf}$ values for the elliptic flow are tabulated in Table~\ref{tab:ch2_v2_npart} for $N_{part}$ scaling and in Table~\ref{tab:ch2_v2_ncoll} for $N_{coll}$ scaling.  The initial conditions that include pre-equilibrium flow have lower overall $\chi^2_{ndf}$ values, but the ideal hydrodynamic condition (vanishing viscosity to entropy ratio) is preferred.  The exception is the case of $N_{coll}$ scaling with pre-equilibrium flow, for which all values of $\eta/s$ go through most the data within one standard deviation of the systematic errors.  The STAR data evaluations achieve a minimum $\chi^2_{ndf}$ for $\eta/s$=0.24, but the $\chi^2_{ndf}$ values increase slowly as $\eta/s$ moves away from this value.
\begin{table}[ht]
\begin{tabular}{|c|r r|r r|}
\hline
&
 \multicolumn{2}{c|}{$\chi^2_{ndf}\ N_{part}$} &
 \multicolumn{2}{c|}{$\chi^2_{ndf}\ N_{part,preq}$} \\
$\eta/s$  & PHNX & STAR & PHNX & STAR \\ 
\hline
0.32 & 21.49 & 44.92 & 9.66 & 13.42 \\
\hline        
0.24 & 19.63 & 49.01 & 8.64 & 11.01 \\
\hline        
0.16 & 16.93 & 38.85 & 5.74 & 9.87 \\
\hline        
0.08 & 10.80 & 22.91 & 3.97 & 7.21 \\
\hline
$10^{-4}$ & 8.57 & 15.37 & 2.67 & 7.70 \\
\hline
\end{tabular}
\caption{$\chi^2_{ndf}$ for evaluations of $v_2$ for pions (STAR)
  and pions and kaons (PHENIX) for $N_{part}$, $N_{part,preq}$ scaling
with fixed $T_{\rm cent}$=0.320 GeV.}
\label{tab:ch2_v2_npart}
\end{table}
\begin{table}[ht]
\begin{tabular}{|c|r r|r r|}
\hline
&
 \multicolumn{2}{c|}{$\chi^2_{ndf}\ N_{coll}$} &
 \multicolumn{2}{c|}{$\chi^2_{ndf}\ N_{coll,preq}$} \\
$\eta/s$  & PHNX & STAR & PHNX & STAR \\ 
\hline
0.32 & 7.19 & 9.74 & 2.05 & 8.16 \\
\hline         
0.24 & 5.67 & 9.80 & 2.01 & 5.33 \\
\hline         
0.16 & 4.28 & 8.88 & 2.12 & 6.36 \\
\hline         
0.08 & 4.18 & 7.39 & 2.01 & 9.13 \\
\hline
$10^{-4}$ & 2.53 & 6.64 & 2.32 & 12.9 \\
\hline
\end{tabular}
\caption{$\chi^2_{ndf}$ for evaluations of $v_2$ for pions (STAR)
  and pions and kaons (PHENIX) for $N_{coll}$, $N_{coll,preq}$ scaling
with fixed $T_{\rm cent}$=0.320 GeV.}
\label{tab:ch2_v2_ncoll}
\end{table}

The $\chi^2_{ndf}$ distributions for elliptic flow are shown in Fig.~\ref{fig:ch2_v2_npart} for $N_{part}$ scaling, and Fig.~\ref{fig:ch2_v2_npart_preq} for $N_{part}$ with pre-equilibrium flow.  As evident in Table~\ref{tab:ch2_v2_npart}, the distribution has a shallow minimum and higher values of $\eta/s$ are excluded.  Lower initial temperatures also appear to excluded, at least in the absence of pre-equilibrium flow.   For the case of $N_{coll}$ scaling, as shown in Fig.~\ref{fig:ch2_v2_ncoll} and in Fig.~\ref{fig:ch2_v2_ncoll_preq} the distributions are quite shallow, with the region of higher $T_{cent}$ and $\eta/s$ excluded when there is no pre-equilibrium flow, and the lowest values of $\eta/s$ excluded when there is.
\begin{figure}
\begin{center}
\includegraphics[width=0.34\textwidth]{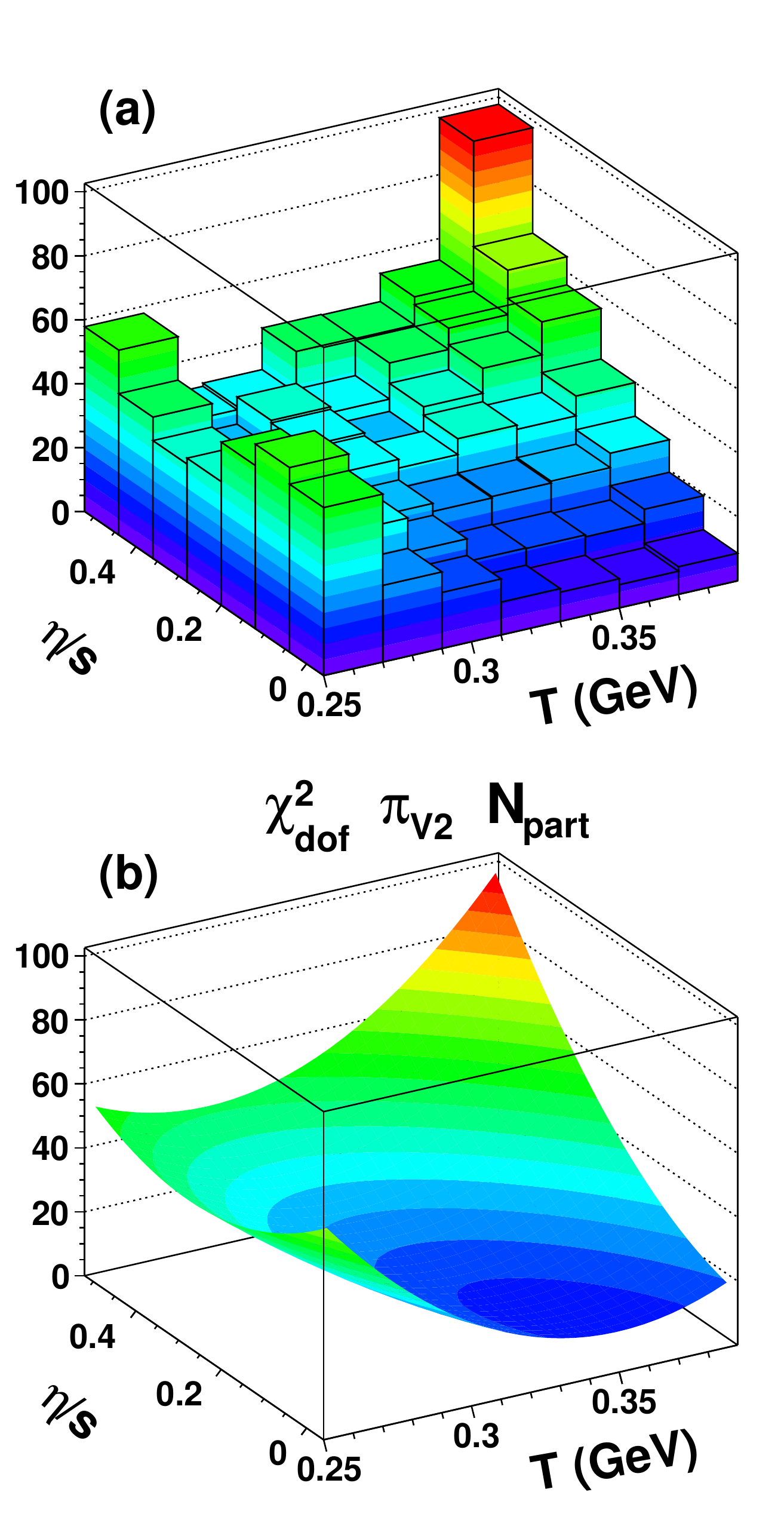}
\end{center}
\caption{(Color online) Model evaluation of pion elliptic flow $\chi^2_{ndf}$ distribution (a) with paraboloid fit (b) for $N_{part}$ scaling without pre-equilibrium flow.}
\label{fig:ch2_v2_npart}
\end{figure}
\begin{figure}
\begin{center}
\includegraphics[width=0.34\textwidth]{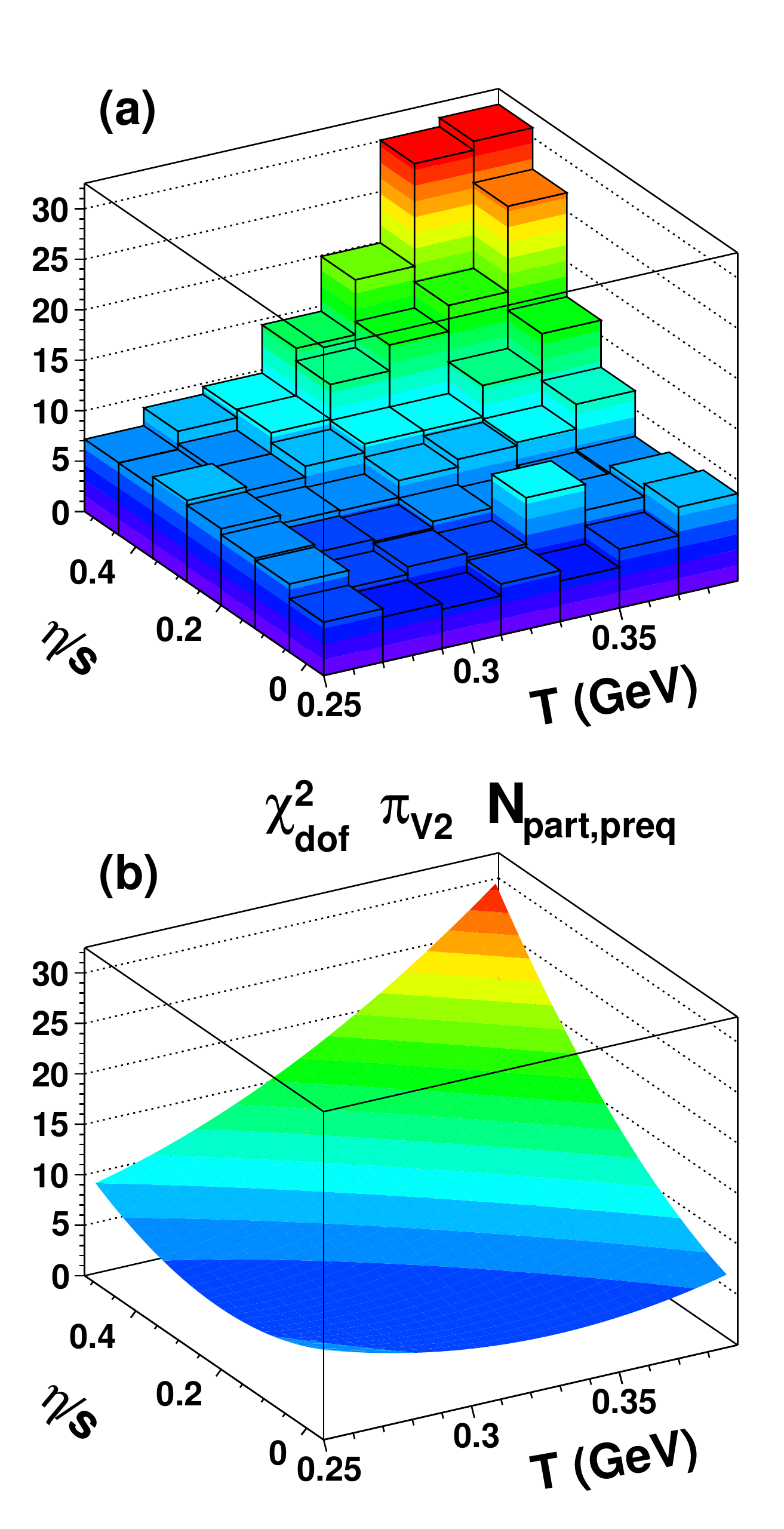}
\end{center}
\caption{(Color online) Model evaluation of pion elliptic flow $\chi^2_{ndf}$ distribution (a) with paraboloid fit (b) for $N_{part}$ scaling with pre-equilibrium flow.}
\label{fig:ch2_v2_npart_preq}
\end{figure}
\begin{figure}
\begin{center}
\includegraphics[width=0.34\textwidth]{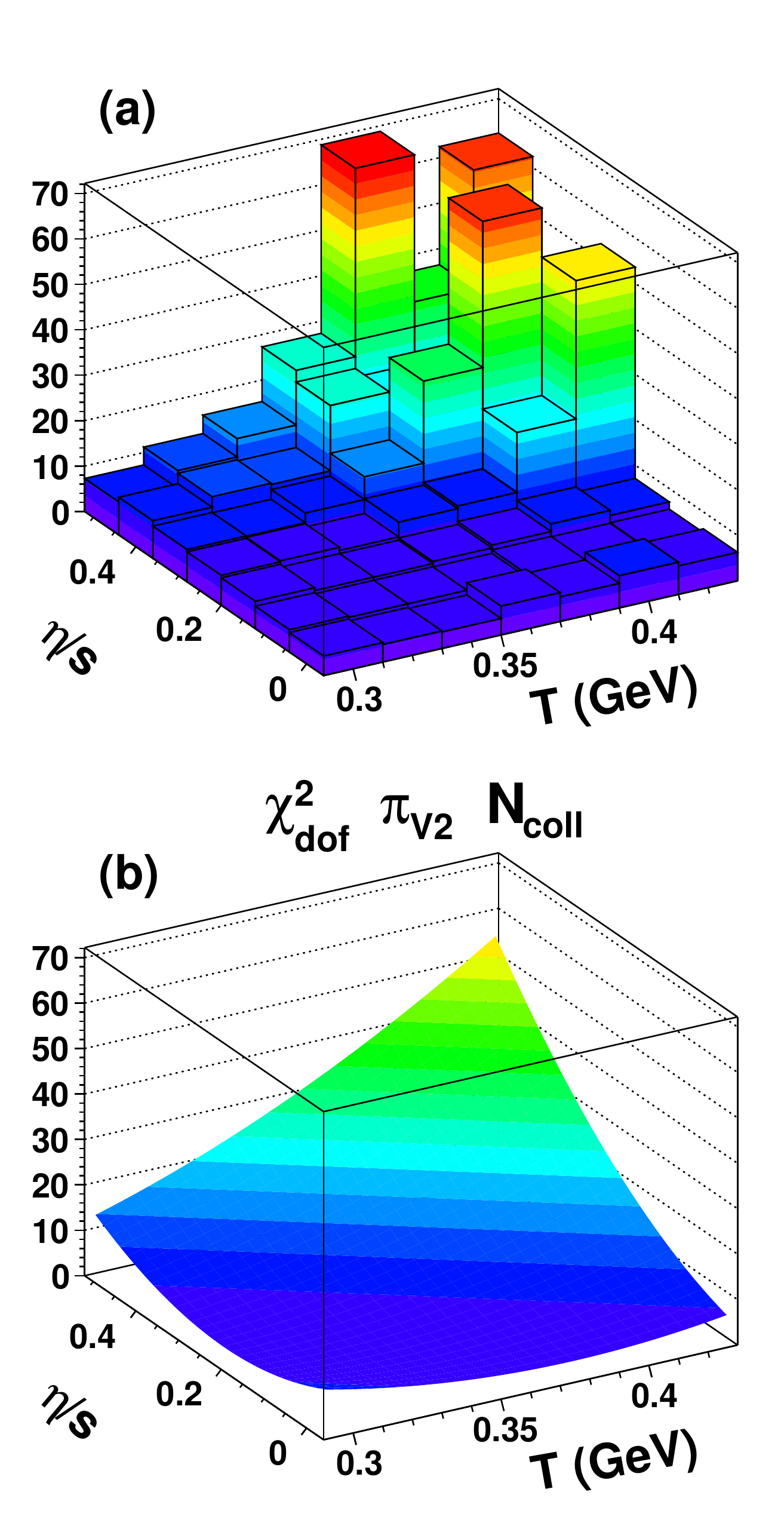}
\end{center}
\caption{(Color online) Model evaluation of pion elliptic flow $\chi^2_{ndf}$ distribution (a) with paraboloid fit (b) for $N_{coll}$ scaling without pre-equilibrium flow.}
\label{fig:ch2_v2_ncoll}
\end{figure}
\begin{figure}
\begin{center}
\includegraphics[width=0.34\textwidth]{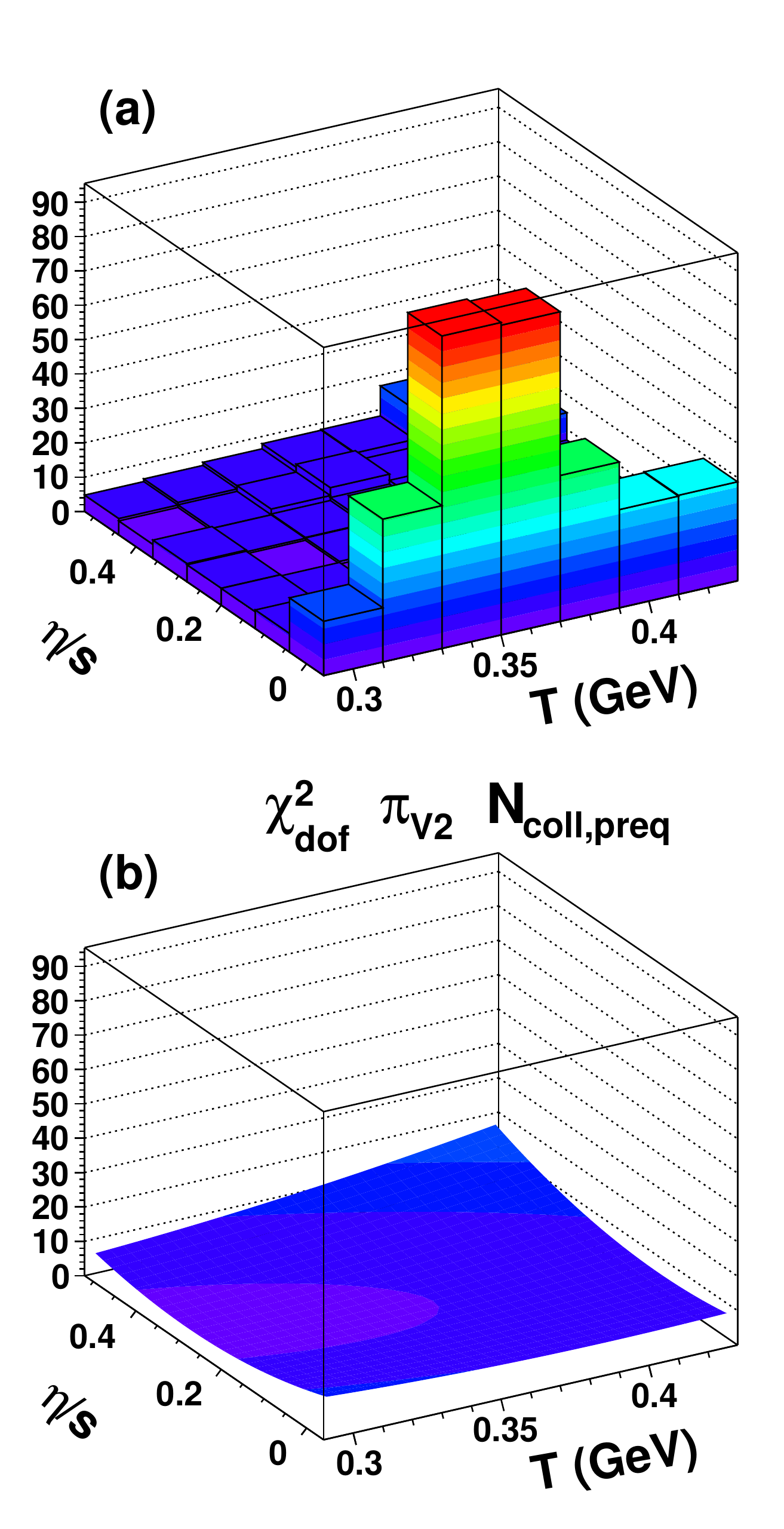}
\end{center}
\caption{(Color online) Model evaluation of pion elliptic flow $\chi^2_{ndf}$ distribution (a) with paraboloid fit (b) for $N_{coll}$ scaling with pre-equilibrium flow.}
\label{fig:ch2_v2_ncoll_preq}
\end{figure}

In Fig.~\ref{fig:hbt_npart} we show for the first time a direct comparison of VH2+UrQMD values of $R_{long}$, $R_{side}$, and $R_{out}$ compared to experimental data for 200 GeV Au+Au collisions.  The comparison displays the characteristic discrepancy (referred to as the HBT puzzle~\cite{Lisa:2005cg}) in which $R_{out}$ trends above the data and $R_{side}$ trends below.  The addition of pre-equilibrium flow shown in Fig.~\ref{fig:hbt_npart_preq} leads to a significant improvement in model comparison for $R_{out}$.  This was first observed for a one-dimensional model in~\cite{Pratt:2009bk}.  Similar results were achieved for a two-dimensional model without pre-equilibrium flow using a Gaussian initial density profile and a free-streaming final state~\cite{Broniowski:2008iq}.  This is the first time that agreement with HBT results has been achieved for a two-dimensional hydrodynamic model with a hadronic cascade.  A similar improvement in $R_{out}$ occurs for $N_{coll}$ scaling; see Figure~\ref{fig:hbt_ncoll} and~\ref{fig:hbt_ncoll_preq}.  There is little dependence of the radii on $\eta/s$; therefore only the dependence on $T_{cent}$ is shown in these figures.
\begin{figure}
\includegraphics[width=0.40\textwidth]{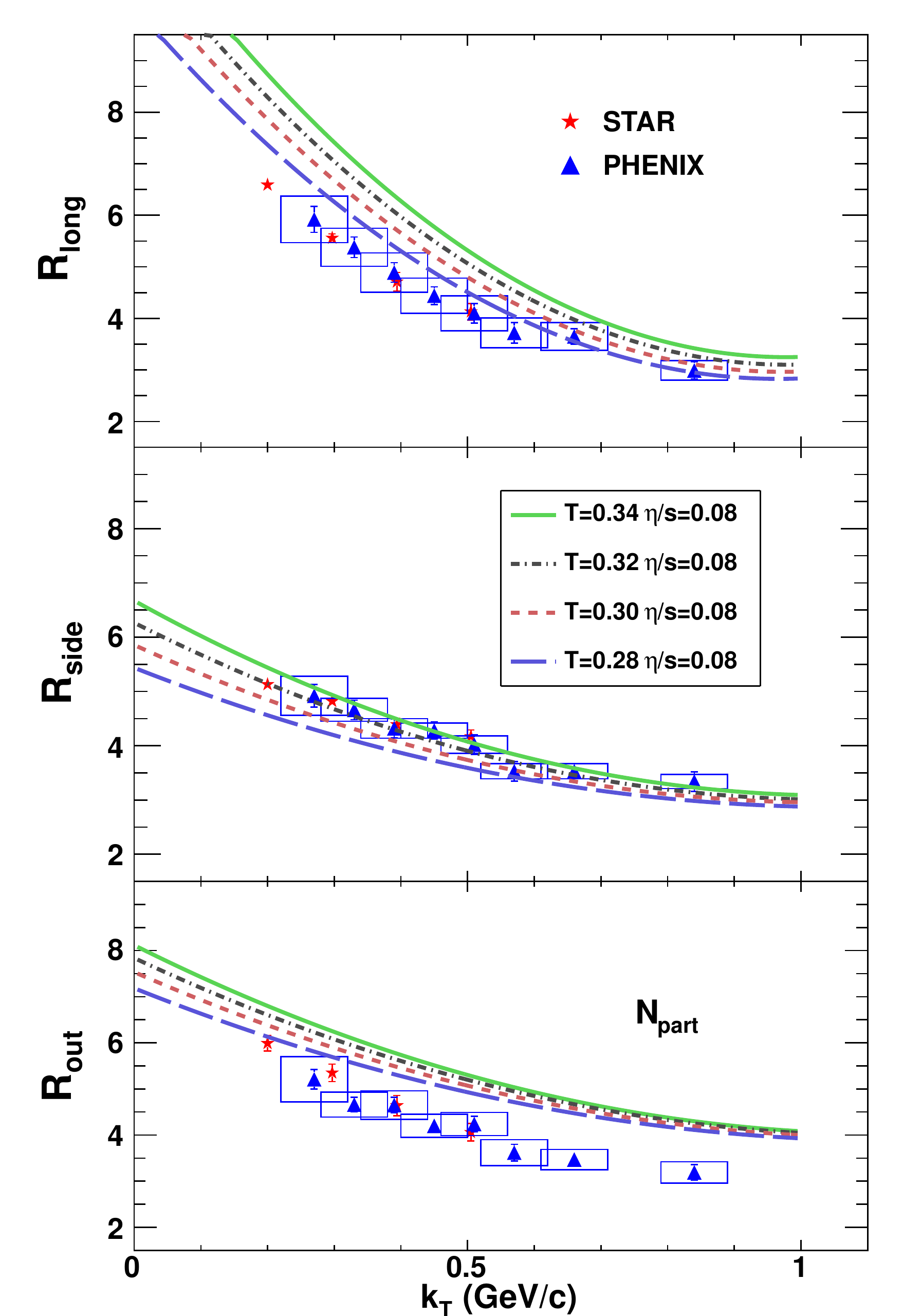}
\caption{(Color online) Model evaluation of pion radii with $N_{part}$ scaling for fixed $\eta/s$.}
\label{fig:hbt_npart}
\end{figure}
\begin{figure}
\includegraphics[width=0.40\textwidth]{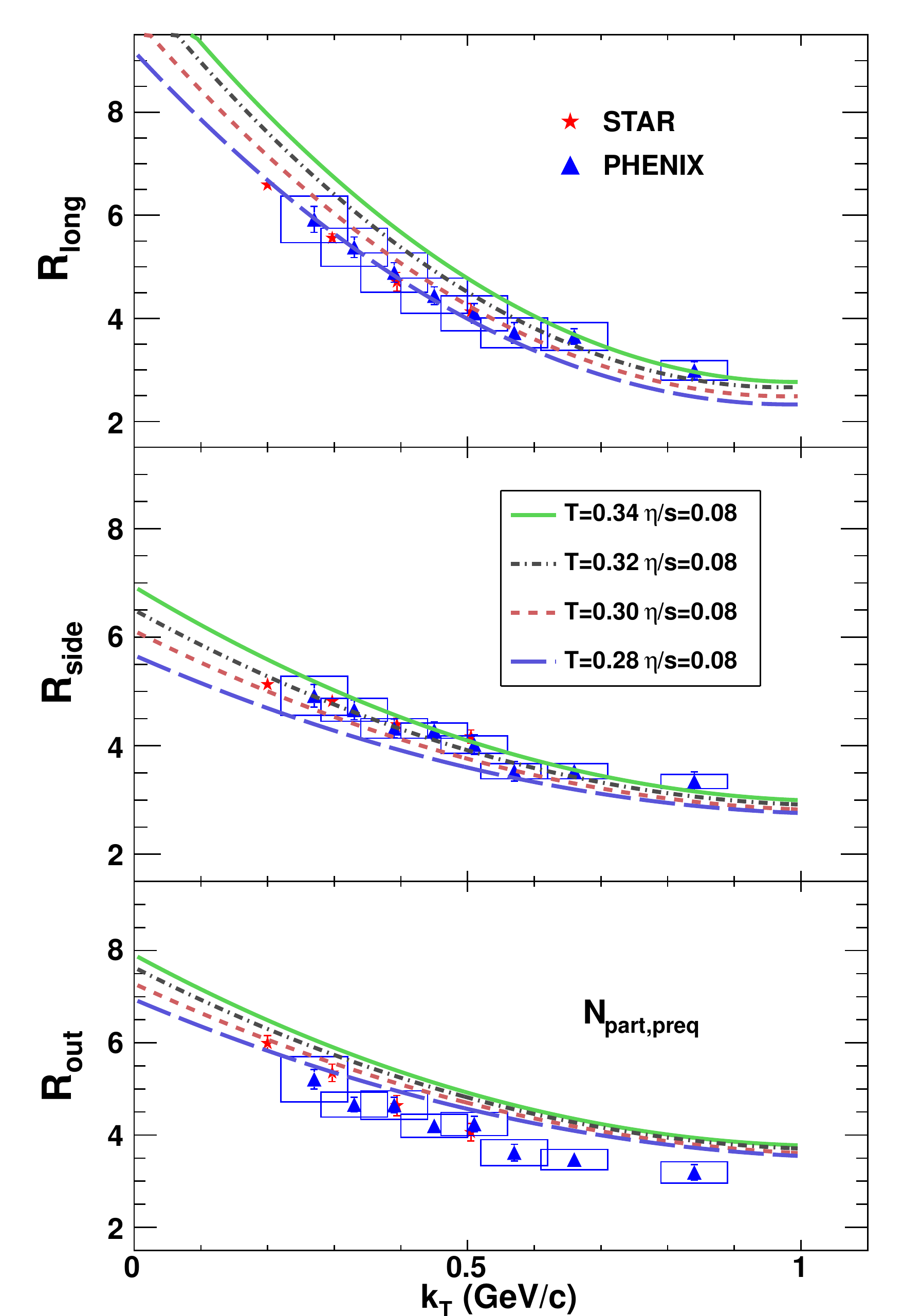}
\caption{(Color online) Model evaluation of pion radii with $N_{part}$ scaling with pre-equilibrium flow for fixed $\eta/s$.}
\label{fig:hbt_npart_preq}
\end{figure}
\begin{figure}
\includegraphics[width=0.40\textwidth]{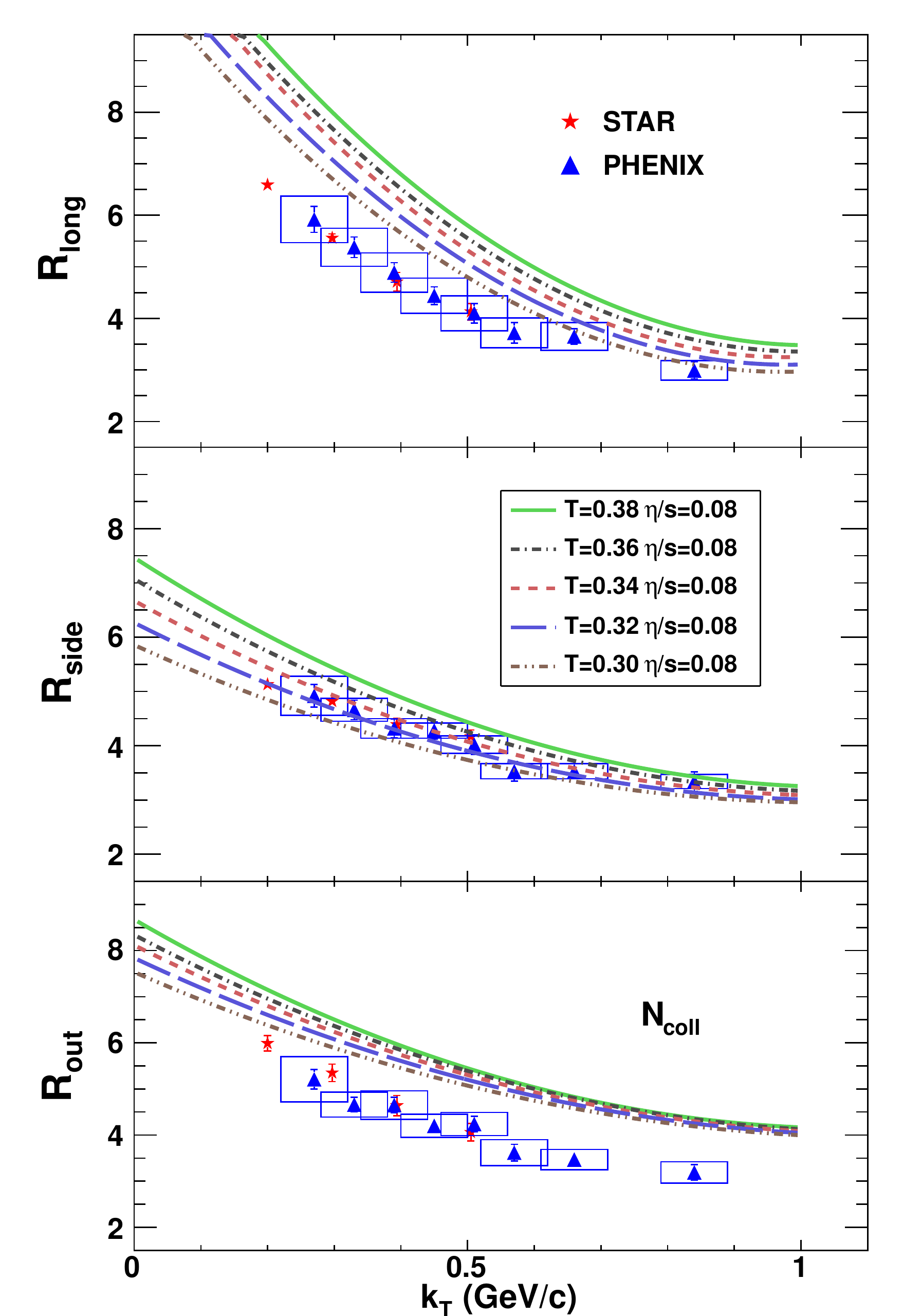}
\caption{(Color online) Model evaluation of pion radii with $N_{coll}$ scaling for fixed fixed $\eta/s$.}
\label{fig:hbt_ncoll}
\end{figure}
\begin{figure}
\includegraphics[width=0.40\textwidth]{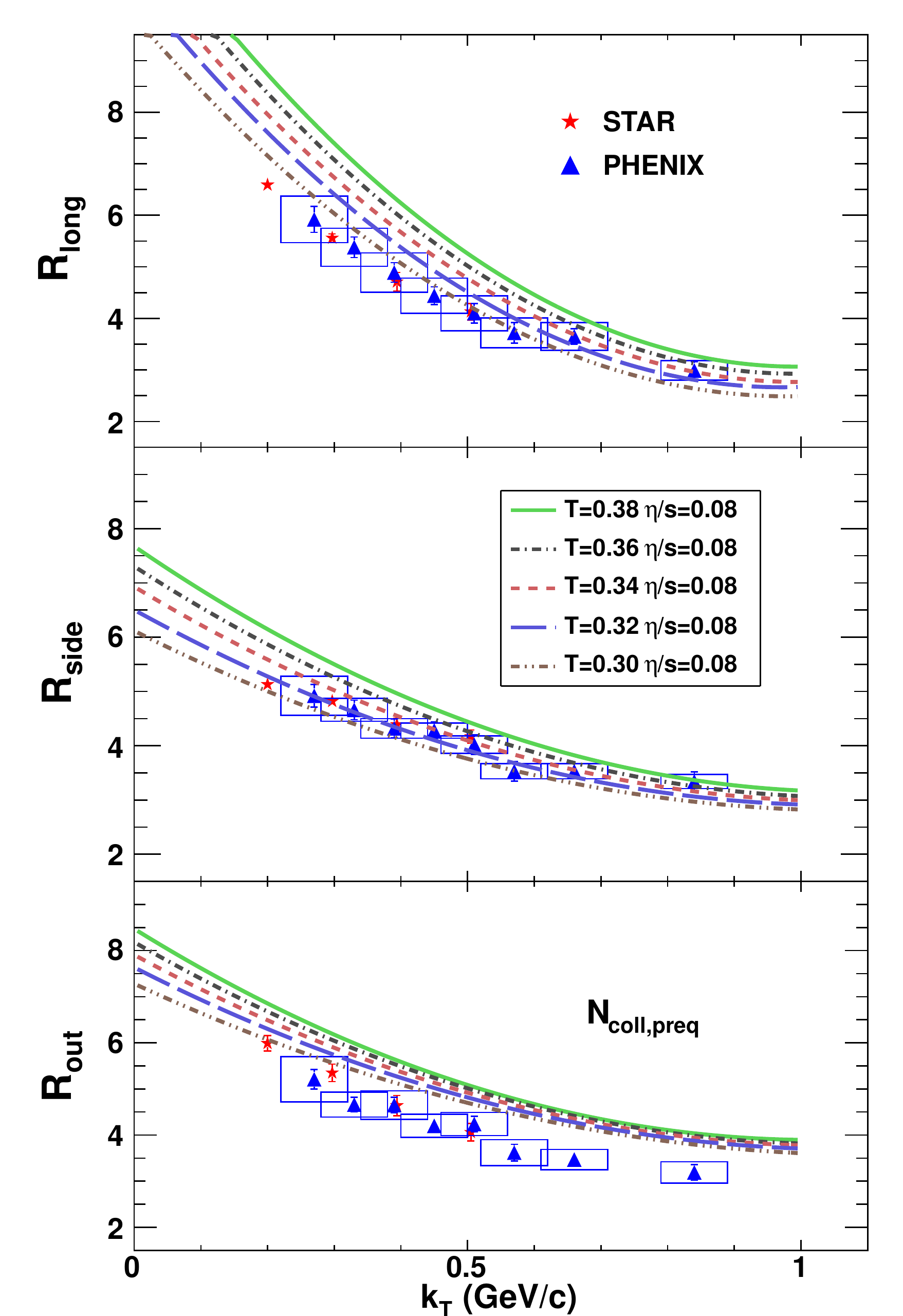}
\caption{(Color online) Model evaluation of pion radii with $N_{coll}$ scaling with pre-equilibrium flow for fixed fixed $\eta/s$.}
\label{fig:hbt_ncoll_preq}
\end{figure}

The $\chi^2_{ndf}$ values are given in Tables~\ref{tab:hbt_npart} and~\ref{tab:hbt_npart_preq} for $N_{part}$ scaling without and with pre-equilibrium flow, respectively, and Tables~\ref{tab:hbt_ncoll} and~\ref{tab:hbt_ncoll_preq} for $N_{coll}$ scaling without and with pre-equilibrium flow, respectively.  In nearly all cases, the lower temperatures lead to better agreement.  Without pre-equilibrium flow there are larger $\chi^2_{ndf}$ values for the evaluation of $R_{long}$ with the STAR data, as well as higher temperatures for $R_{side}$.  In nearly all cases, the PHENIX data, with the more generous systematic errors lead to lower $\chi^2_{ndf}$ values.
\begin{table}[ht]
\begin{tabular}{|c|r r|r r|r r|}
\hline
$T_{\rm cent}$ &
 \multicolumn{2}{c|}{$R_{long}$} &
 \multicolumn{2}{c|}{$R_{side}$} &
 \multicolumn{2}{c|}{$R_{out}$} \\
(GeV)             & PHNX & STAR & PHNX & STAR & PHNX & STAR \\ 
\hline
0.340 & 2.84 & 493 & 0.45 & 10.9 & 4.60 & 28.2 \\
\hline                
0.320 & 2.16 & 312 & 0.46 & 2.85 & 4.28 & 20.9 \\
\hline                
0.300 & 1.69 & 176 & 0.77 & 23.4 & 3.95 & 13.9 \\
\hline                
0.280 & 1.26 & 67.3 & 1.35 & 69.2 & 3.66 & 8.28 \\
\hline
\end{tabular}
\caption{$\chi^2_{ndf}$ for evaluation of pion radii for
  $N_{part}$ scaling without pre-equilibrium flow for fixed
$\eta/s$=0.08.}
\label{tab:chi2_hbt_npart}
\end{table}
\begin{table}[ht]
\begin{tabular}{|c|r r|r r|r r|}
\hline
$T_{\rm cent}$ &
 \multicolumn{2}{c|}{$R_{long}$} &
 \multicolumn{2}{c|}{$R_{side}$} &
 \multicolumn{2}{c|}{$R_{out}$} \\
(GeV)             & PHNX & STAR & PHNX & STAR & PHNX & STAR \\ 
\hline
0.340 & 2.44 & 199 & 0.64 & 24.9 & 2.61 & 12.2 \\
\hline                
0.320 & 2.36 & 107 & 0.53 & 3.66 & 2.33 & 7.77 \\
\hline                
0.300 & 2.28 & 33.5 & 0.86 & 11.1 & 2.09 & 4.24 \\
\hline                
0.280 & 1.10 & 1.40 & 1.39 & 48.4 & 1.94 & 2.41 \\
\hline
\end{tabular}
\caption{$\chi^2_{ndf}$ for evaluation of pion radii for
  $N_{part}$ scaling with pre-equilibrium flow for fixed
$\eta/s$=0.08.}
\label{tab:chi2_hbt_npartpreq}
\end{table}
\begin{table}[ht]
\begin{tabular}{|c|r r|r r|r r|}
\hline
$T_{\rm cent}$ &
 \multicolumn{2}{c|}{$R_{long}$} &
 \multicolumn{2}{c|}{$R_{side}$} &
 \multicolumn{2}{c|}{$R_{out}$} \\
(GeV)             & PHNX & STAR & PHNX & STAR & PHNX & STAR \\ 
\hline
0.380 & 3.93 & 808 & 1.36 & 114 & 5.09  & 43.2 \\
\hline                
0.360 & 3.17 & 612 & 0.76 & 48.7 & 4.94 & 35.2 \\
\hline                
0.340 & 2.84 & 493 & 0.45 & 10.9 & 4.60 & 28.2 \\
\hline                
0.320 & 2.16 & 312 & 0.46 & 2.85 & 4.28 & 20.9 \\
\hline
0.300 & 1.69 & 176 & 0.77 & 23.4 & 3.95 & 13.9 \\
\hline
\end{tabular}
\caption{$\chi^2_{ndf}$ for evaluation of pion radii for
  $N_{coll}$ scaling without pre-equilibrium flow for fixed
$\eta/s$=0.08.}
\label{tab:chi2_hbt_ncoll}
\end{table}

\begin{table}[ht]
\begin{tabular}{|c|r r|r r|r r|}
\hline
$T_{\rm cent}$ &
 \multicolumn{2}{c|}{$R_{long}$} &
 \multicolumn{2}{c|}{$R_{side}$} &
 \multicolumn{2}{c|}{$R_{out}$} \\
(GeV)             & PHNX & STAR & PHNX & STAR & PHNX & STAR \\ 
\hline
0.380 & 3.23 & 488  & 1.55 & 144  & 3.15 & 23.4 \\
\hline                
0.360 & 2.80 & 334  & 0.97 & 71.4 & 2.89 & 17.6 \\
\hline                
0.340 & 2.44 & 199  & 0.64 & 24.8 & 2.61 & 12.2 \\
\hline                
0.320 & 2.36 & 107  & 0.53 & 3.66 & 2.33 & 7.77 \\
\hline
0.300 & 2.28 & 33.5 & 0.86 & 11.1 & 2.09 & 4.24 \\
\hline
\end{tabular}
\caption{$\chi^2_{ndf}$ for evaluation of pion radii for
  $N_{coll}$ scaling with pre-equilibrium flow for fixed
$\eta/s$=0.08.}
\label{tab:chi2_hbt_ncollpreq}
\end{table}

The $\chi^2_{ndf}$ distributions are shown in Figures~\ref{fig:ch2_hbt_npart} and~\ref{fig:ch2_hbt_npart_preq} for $N_{part}$ scaling without and with pre-equilibrium flow, respectively, and Figures~\ref{fig:ch2_hbt_ncoll} and~\ref{fig:ch2_hbt_ncoll_preq} for $N_{coll}$ scaling without and with pre-equilibrium flow.  The minimum values for $\chi^2_{ndf}$ vary for each set of of initial conditions, moving to higher initial temperatures for $N_{coll}$ scaling as well as with the addition of pre-equilibrium flow.  There is little constraining power for $\eta/s$, however in each case there are regions of $T_{cent}$ that are clearly excluded.  The ability to constrain the temperature and exclude regions that may be allowed by the comparisons to the spectra alone provide the main reason for continuing to compare to femtoscopic radii at this time.

\begin{figure}
\includegraphics[width=0.29\textwidth]{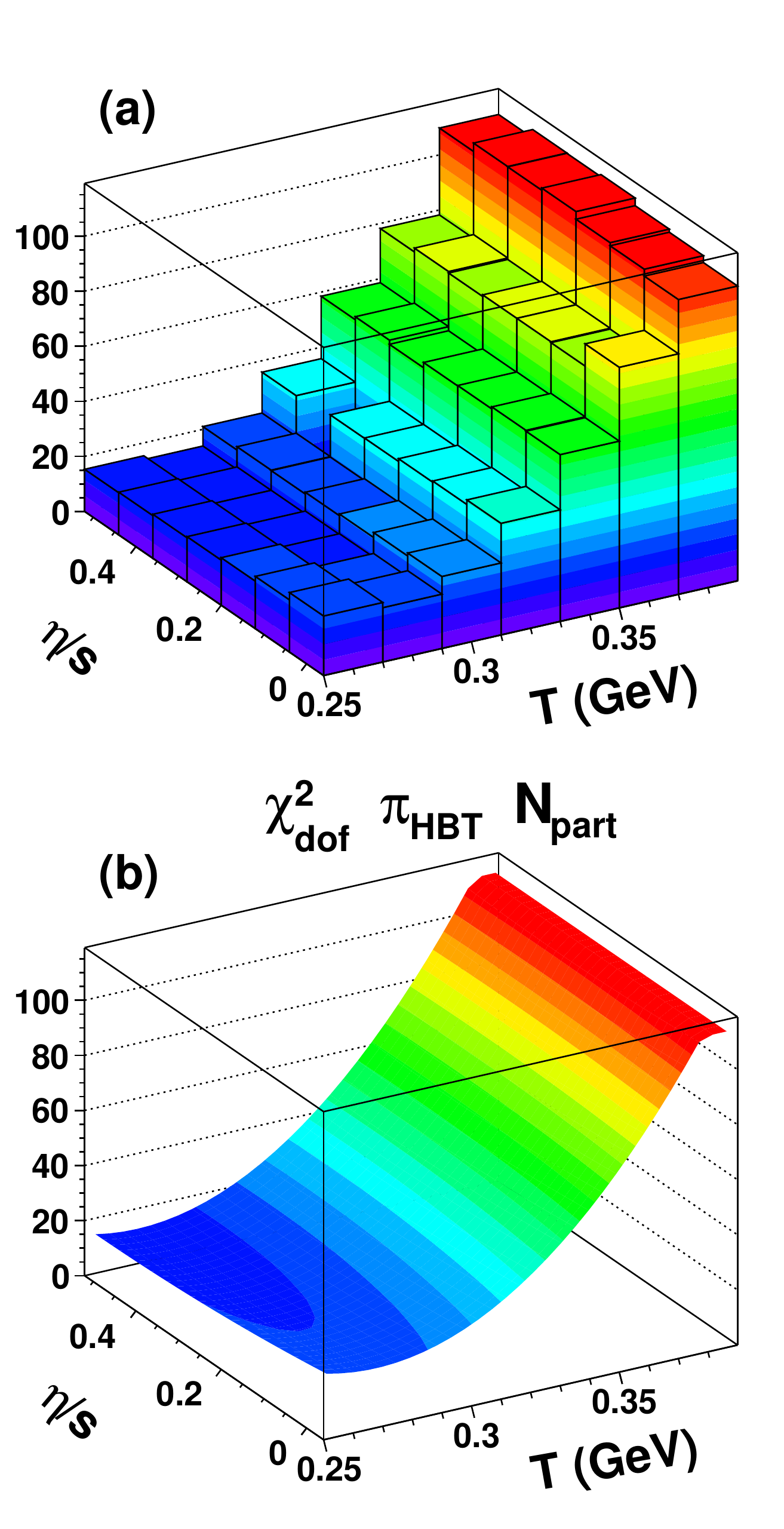}
\caption{(Color online) Model evaluation of pion radii $\chi^2_{ndf}$ distribution (a) with paraboloid fit (b) for $N_{part}$ scaling.}
\label{fig:ch2_hbt_npart}
\end{figure}
\begin{figure}
\includegraphics[width=0.29\textwidth]{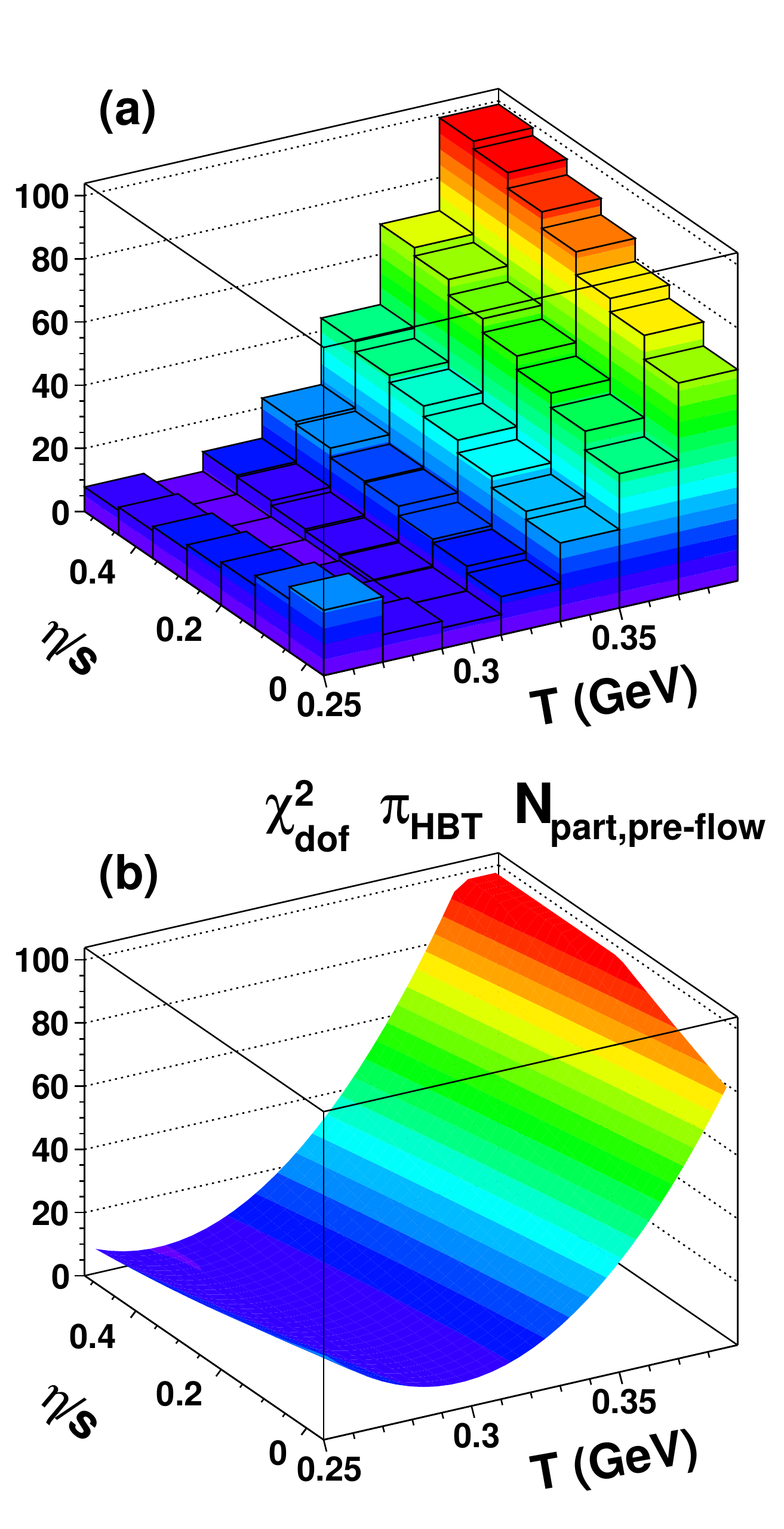}
\caption{(Color online) Model evaluation of pion radii $\chi^2_{ndf}$ distribution (a) with paraboloid fit (b) for $N_{part}$ scaling with pre-equilibrium flow.}
\label{fig:ch2_hbt_npart_preq}
\end{figure}
\begin{figure}
\includegraphics[width=0.29\textwidth]{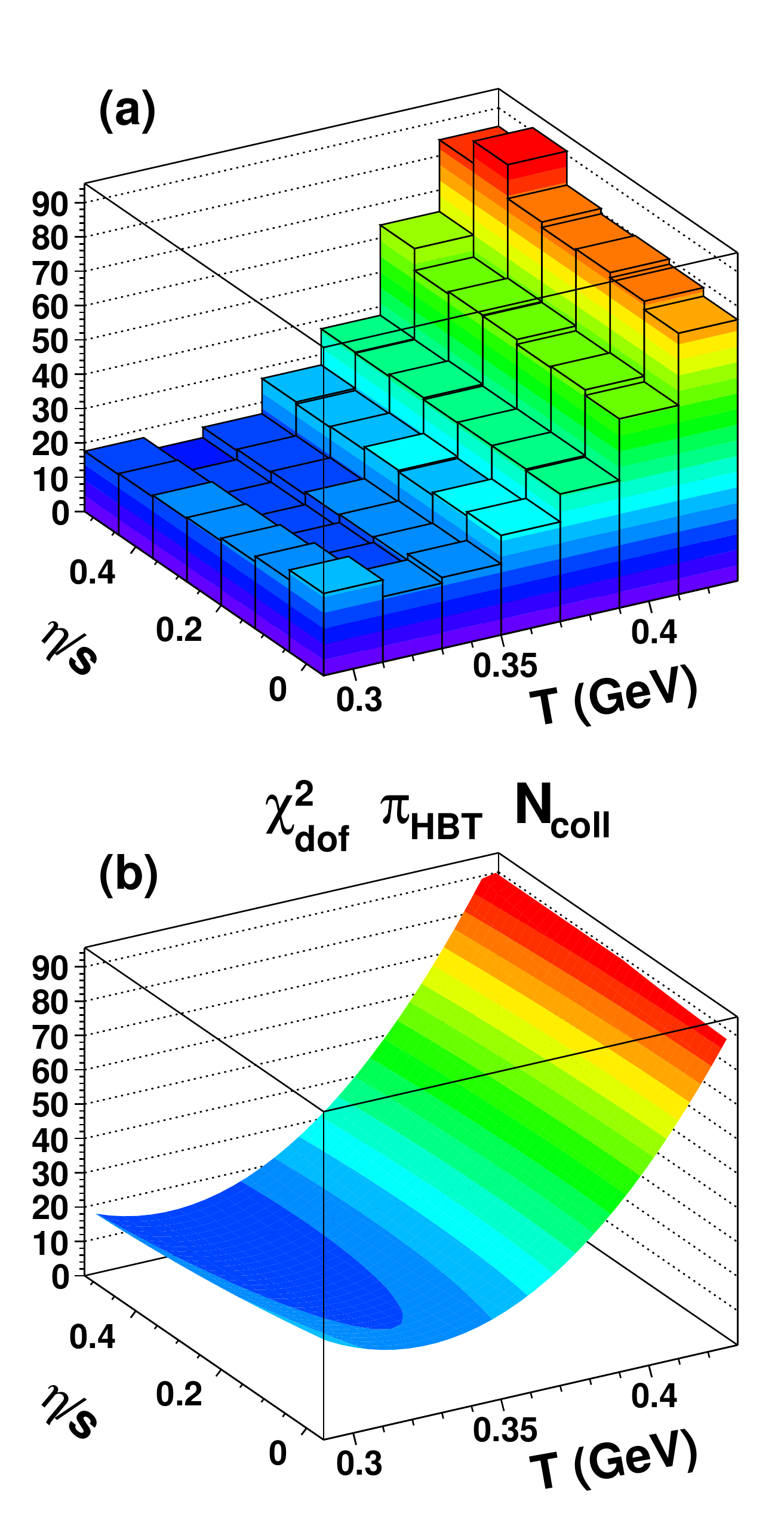}
\caption{(Color online) Model evaluation of pion radii $\chi^2_{ndf}$ distribution (a) with paraboloid fit (b) for $N_{coll}$ scaling.}
\label{fig:ch2_hbt_ncoll}
\end{figure}
\begin{figure}
\includegraphics[width=0.29\textwidth]{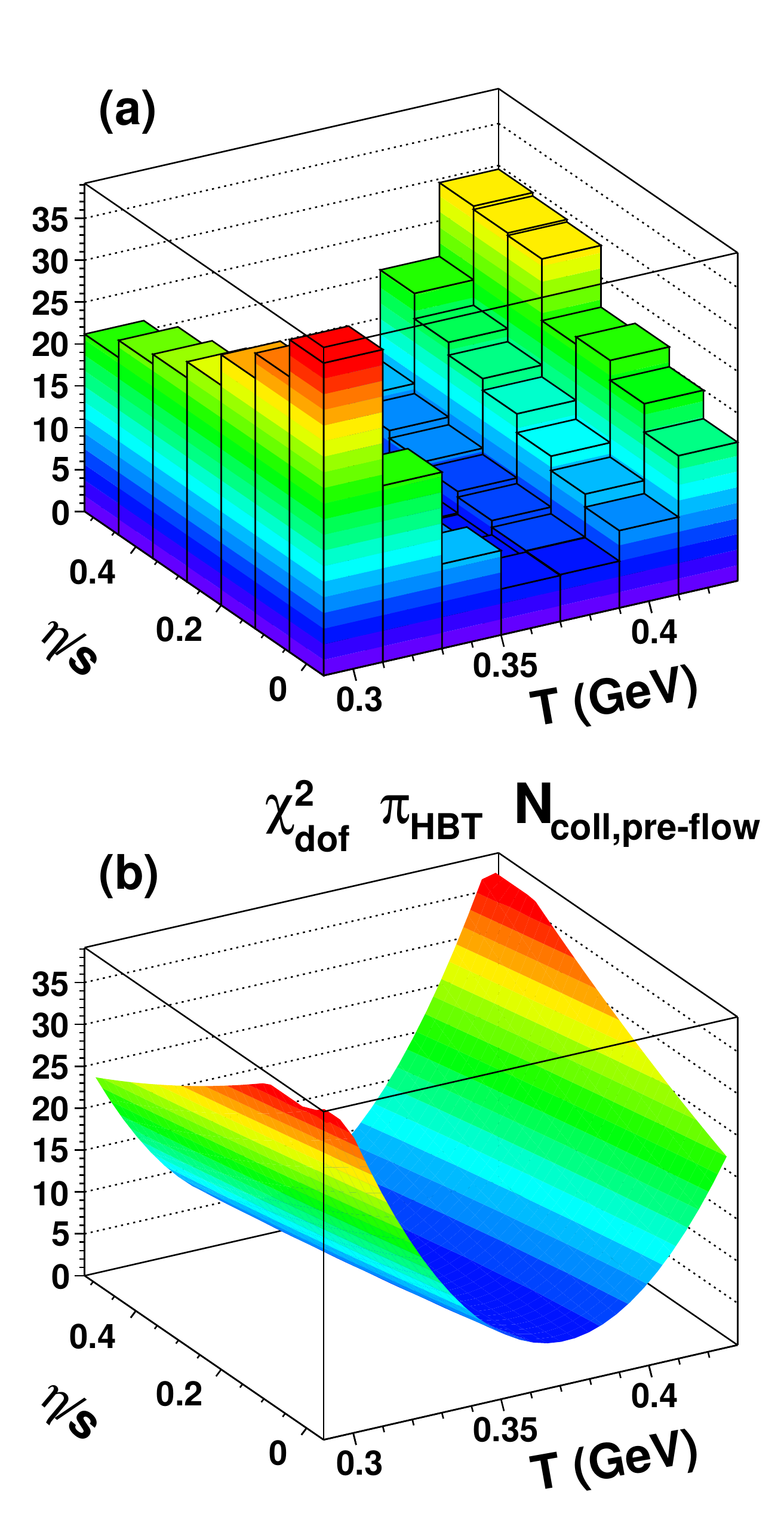}
\caption{(Color online) Model evaluation of pion radii $\chi^2_{ndf}$ distribution (a) with paraboloid fit (b) for $N_{coll}$ scaling with pre-equilibrium flow.}
\label{fig:ch2_hbt_ncoll_preq}
\end{figure}

The paraboloid fits to the $\chi^2_{ndf}$ distributions for each observable are used to estimate the location of the optimum values and ranges for the $T_{cent}$ and $\eta/s$ parameters.  The paraboloid fit parameters are summarized in Table~\ref{tab:fitparams}, which gives the major and minor axes, angle of rotation as well as the location of the minimum in $T_{cent}$ and $\eta/s$ and minimum value of the $\chi^2_{ndf}$.  Results are shown for each observalbe, as well as for the sum of $\chi^2$ divided by total number of degrees of freedom for all observables and sum in which the spectra are de-weighted by 10\%.  The latter two sums are discussed further below.  

\begin{table*}[th]
\begin{tabular}{|l|r||c|c|c||c|c|c|}
\hline
Initial profile & Evaluation & Major axis & Minor axis & angle (deg) &\ $T_{\rm min}$ \ & $\ \eta/s_{\rm min}$ \ & $\chi^2/{ndf}_{\rm min}$ \\
\hline
\multirow{4}{*}{$N_{part}$} & HBT & 0.211 & 0.0114 & 0.28 & 0.266 & 0.5 & 13.9 \\
\cline{2-8}
 & V$_{2}$ & 0.088 & 0.0134 & 6.52 & 0.347 & 0.0 & 11.5 \\
\cline{2-8}
 & Spectra & 0.153 & 0.0080 & 2.3 & 0.297 & 0.29 & 2.7 \\
\cline{2-8}
 & Sum & 0.126 & 0.0049 & 2.18 & 0.297 & 0.22 & 12.7 \\
\cline{2-8}
 & Sum$_{10}$ & 0.127 & 0.0074 & 1.24 & 0.29 & 0.21 & 19.2 \\
\hline \hline
\multirow{4}{*}{$N_{coll}$} & HBT & 0.237 & 0.0121 & 1.01 & 0.317 & 0.52 & 14.1 \\
\cline{2-8}
 & V$_{2}$ & 279 & 0.0318 & 19.2 & 0.324 & 0.12 & 3.9 \\
\cline{2-8}
 & Spectra & 210 & 0.0023 & 2.27 & 0.348 & 0.0 & 4.4 \\
\cline{2-8}
 & Sum & 620 & 0.0032 & 2.15 & 0.341 & 0.18 & 10.6 \\
\cline{2-8}
 & Sum$_{10}$ & 13.8 & 0.0068 & 2.01 & 0.334 & 0.28 & 14.4 \\
\hline \hline
\multirow{4}{*}{$N_{part,preq}$} & HBT & 0.374 & 0.0105 & 2.51 & 0.273 & 0.52 & 4.8 \\
\cline{2-8}
 & V$_{2}$ & 0.335 & 0.039 & 14.2 & 0.322 & 0.0 & 4.8 \\
\cline{2-8}
 & Spectra & 0.094 & 0.0019 & 3.11 & 0.303 & 0.0 & 4.9 \\
\cline{2-8}
 & Sum & 0.128 & 0.0027 & 2.96 & 0.302 & 0.0 & 6.3 \\
\cline{2-8}
 & Sum$_{10}$ & 0.246 & 0.0059 & 2.69 & 0.300 & 0.0 & 6.4 \\
\hline \hline
\multirow{4}{*}{$N_{coll,preq}$} & HBT & 0.434 & 0.013 & 2.52 & 0.361 & 0.24 & 4.3 \\
\cline{2-8}
 & V$_{2}$ & 0.156 & 0.0552 & 19.3 & 0.29 & 0.30 & 3.5 \\
\cline{2-8}
 & Spectra & 0.086 & 0.0015 & 3.05 & 0.341 & 0.0 & 29.8 \\
\cline{2-8}
 & Sum & 0.138 & 0.0022 & 3.05 & 0.341 & 0.0 & 18.0 \\
\cline{2-8}
 & Sum$_{10}$ & 0.182 & 0.0050 & 2.91 & 0.328 & 0.26 & 8.4 \\
\hline \hline
\end{tabular}
\caption{(Color online) Ellipse parameters determined from paraboloid fits to $\chi^2_{ndf}$ distributions in $T$ and $\eta/s$ for each set of initial condition profile and for each measurement.}
\label{tab:fitparams}
\end{table*}

To examine the relationship between the $\chi^2_{ndf}$ distributions for each measurement more closely we show in Fig.~\ref{fig:ch2_contours} the one- and two-sigma contours obtained from the paraboloid fits.  Contours are drawn for the femtoscopic radii, elliptic flow, spectra, with each set of concentric ellipses labeled corresponding minimum value of $\chi^2_{ndf}$ achieved for each evaluation.  Fig.~\ref{fig:ch2_contours} shows the evaluations for $N_{part}$ scaling (left) and $N_{coll}$ scaling (right) with pre-equilibrium flow (top) and without (bottom).  The regions that fall within the two-sigma contours for all measurements are shaded.  This occurs for $N_{part}$ scaling with pre-equilibrium flow as well as for the $N_{coll}$ scaling without.  These are also initial conditions that lead the best overall agreement with the three data sets.  The $N_{part}$ evaluation in the lower left quadrant of Fig.~\ref{fig:ch2_contours} provides a different perspective on what has been referred to as the "HBT puzzle"~\cite{Lisa:2005cg}.  The optimal initial temperatures for radii, spectra, and flow are inconsistent.  The addition of pre-equilibrium flow shifts the HBT minimum to higher initial temperature, and broadens the v2 distribution to achieve a consistent, albeit still imperfect agreement between the different measurements.  The use of $N_{coll}$ scaling has a similar effect, but adding pre-equilibrium flow to this scaling overcompensates.

\begin{figure*}
\includegraphics[width=0.45\textwidth]{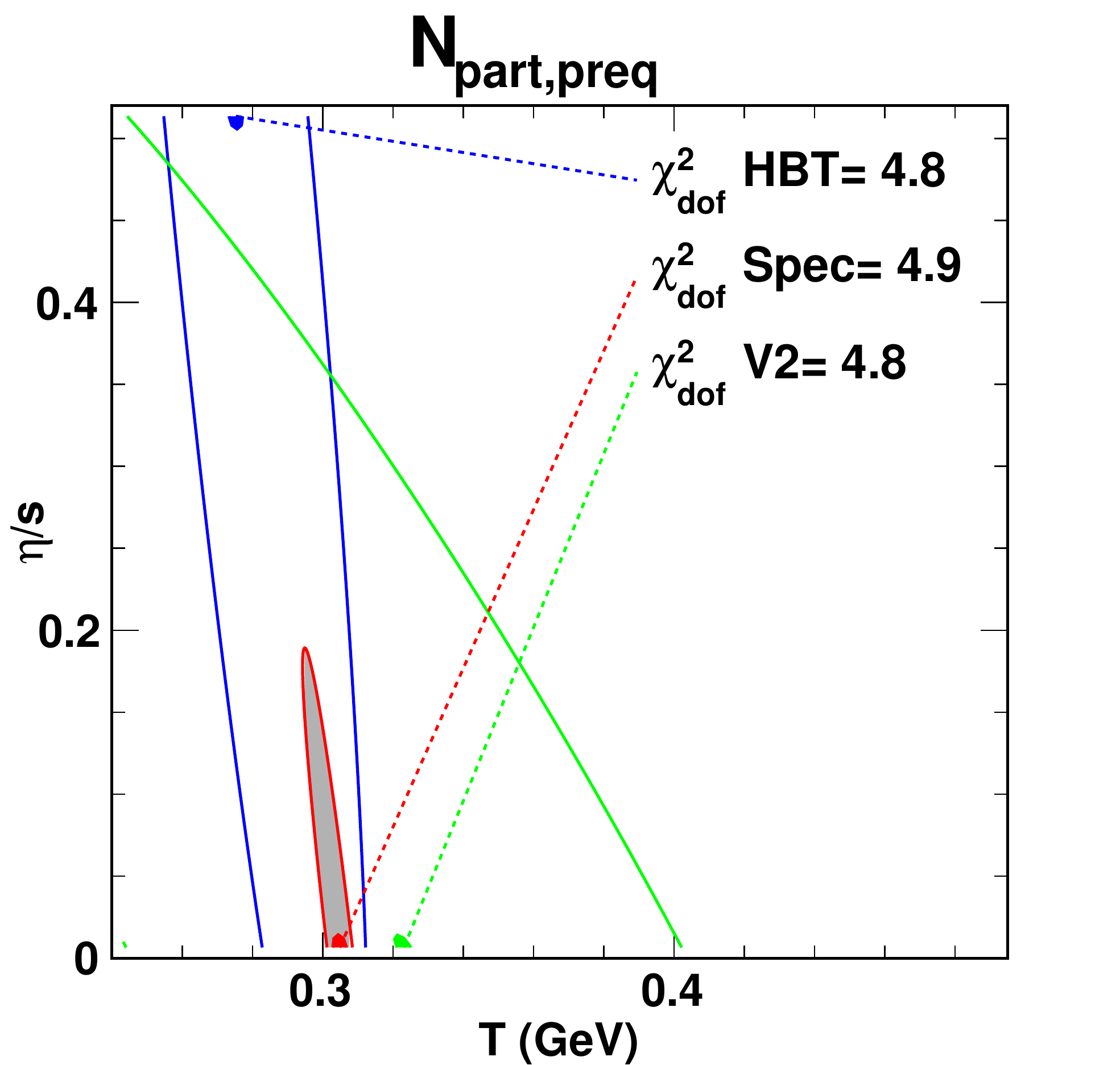}
\includegraphics[width=0.45\textwidth]{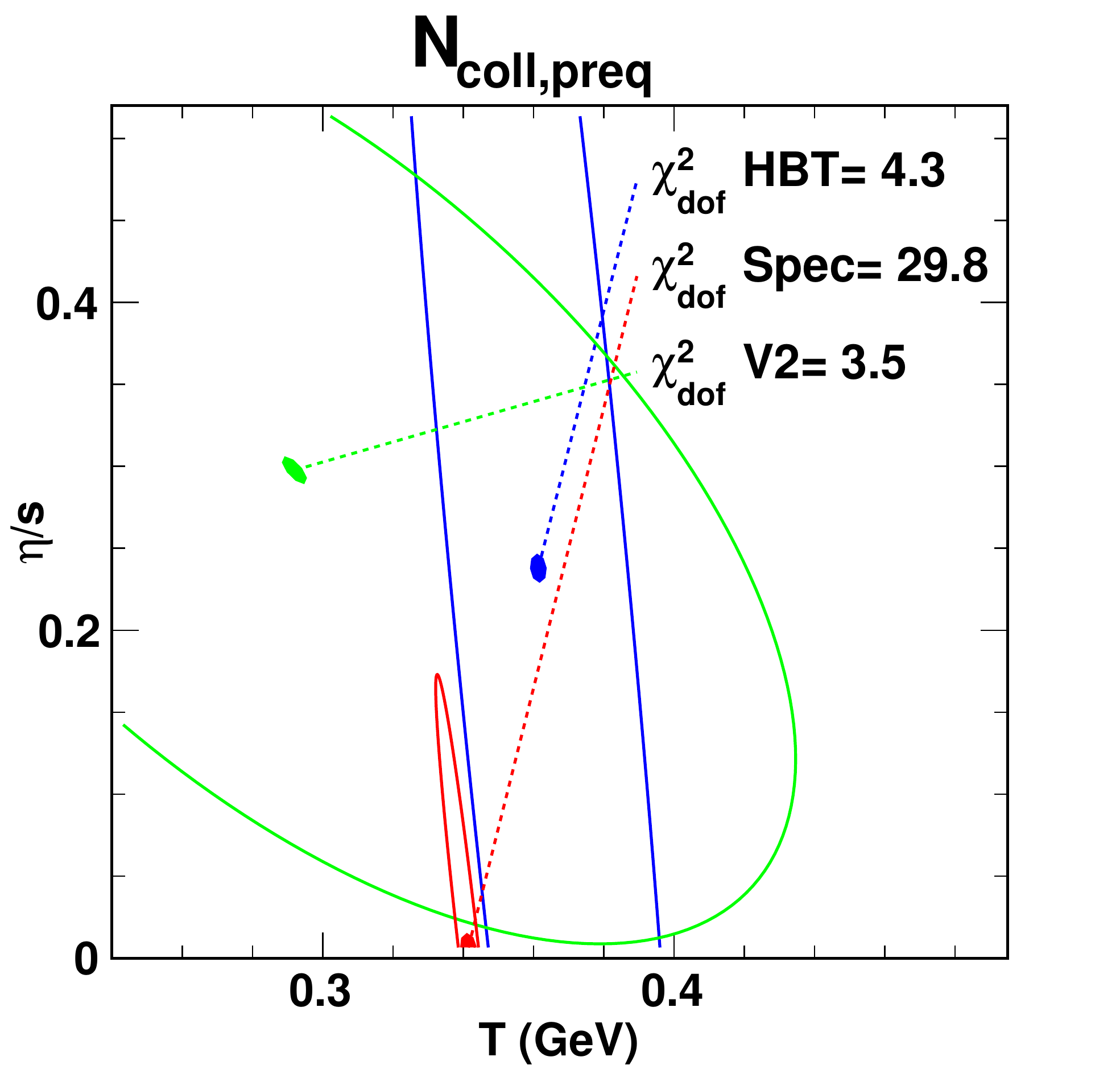}
\includegraphics[width=0.45\textwidth]{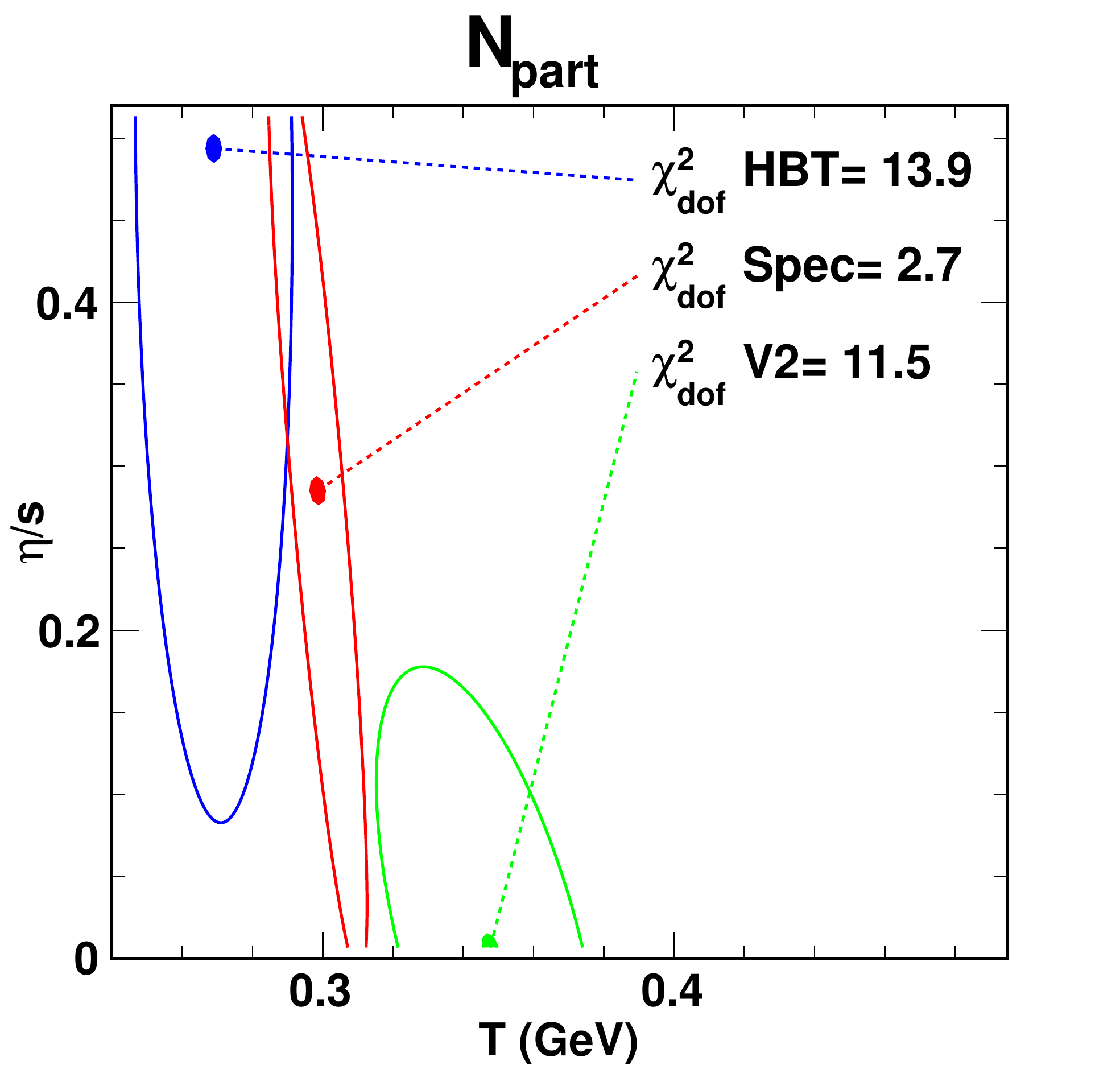}
\includegraphics[width=0.45\textwidth]{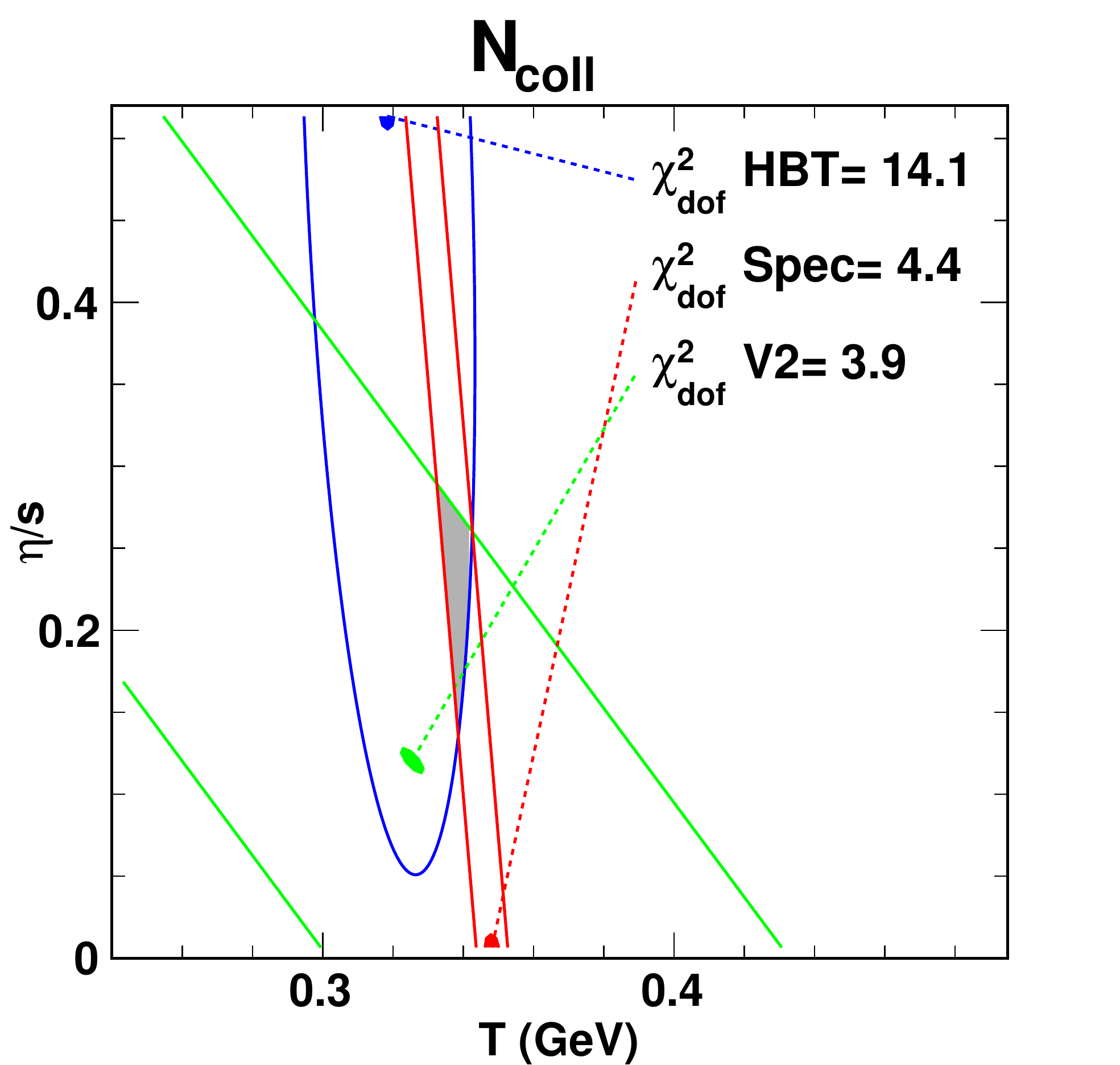}
\caption{(Color online) $\chi^2_{ndf}$ contours for pions spectra, elliptic flow, radii, and a weighted sum in which the spectra $\chi^2_{ndf}$ are weighted by 10\% relative to the other measurements.  The four panels show contours for $N_{part}$ scaling (upper left), $N_{coll}$ scaling (lower left), $N_{part}$ with pre-equilibrium flow (upper right), and $N_{coll}$ with pre-equilibrium flow (lower right).  The minimum $\chi^2_{ndf}$ values are labeled for each fit, and regions which overlap at the two-sigma level are shaded.}
\label{fig:ch2_contours}
\end{figure*}

Although the existence of a two-sigma overlap region is encouraging, it is obvious that we have not yet achieved a truly acceptable evaluation of $\chi^2_{ndf} \sim 1$ for a single measurement, and even if this were achieved, it would be necessary to evaluate a greater breadth of measurements and centrality bins make a compelling case for the validity of the model and initial conditions.  However, when this is achieved, the maximum constraining power will be realized by combining the $\chi^2$ evaluations from all measurements.  Fig.~\ref{fig:sum_ch2} shows the distributions for the sum over $\chi^2$ for each of the three evaluations: spectra, flow, and radii, divided by the total degrees of freedom.  A close inspection reveals that these distributions are nearly identical to the $\chi^2_{ndf}$ distributions for the spectra show in Figures~\ref{fig:ch2_spec_npart}, \ref{fig:ch2_spec_npart_preq}, \ref{fig:ch2_spec_ncoll}, and~\ref{fig:ch2_spec_ncoll_preq}.  In other words, the $\chi^2$ sums are completely dominated by the spectra.

\begin{figure*}
\includegraphics[width=0.24\textwidth]{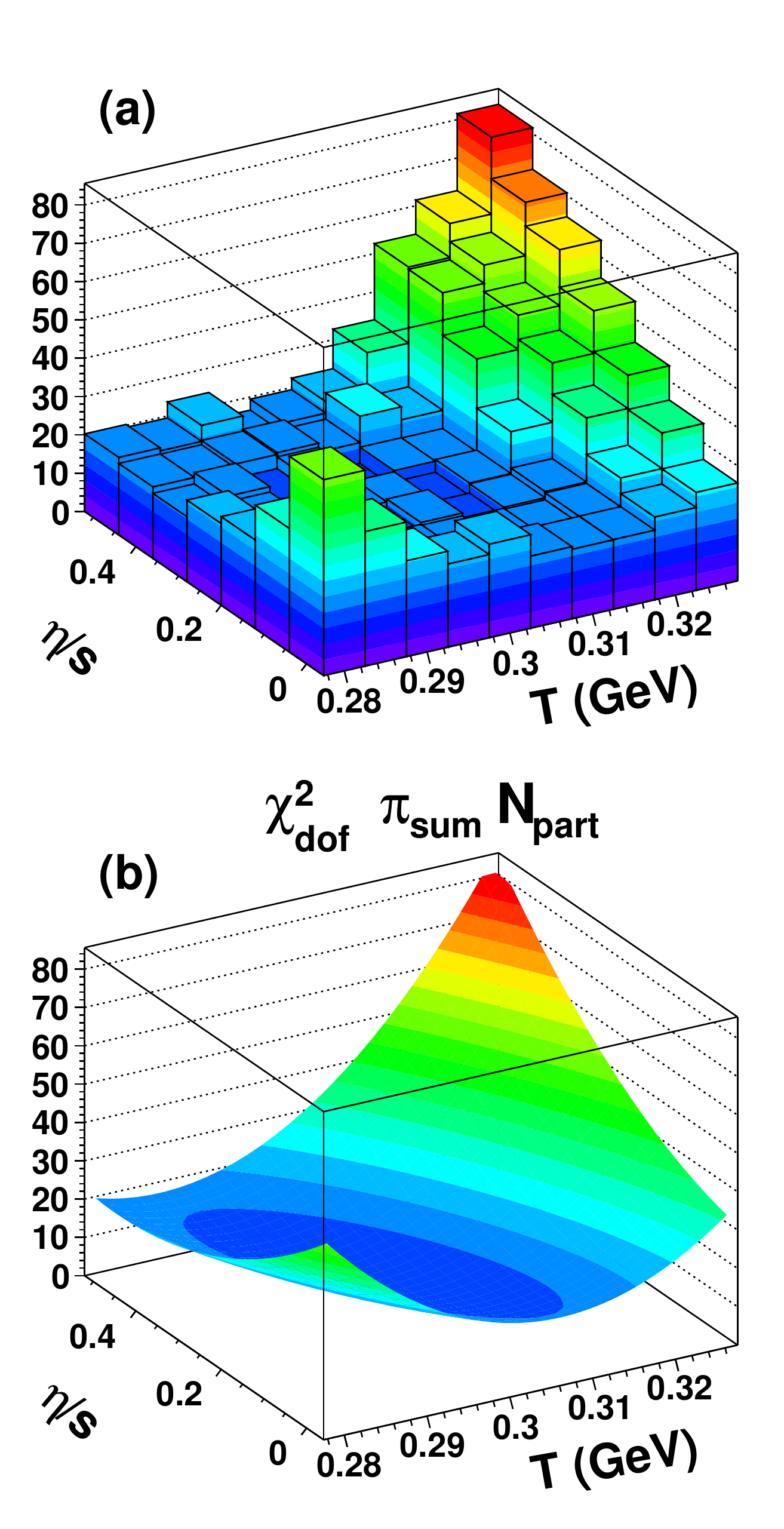}
\includegraphics[width=0.24\textwidth]{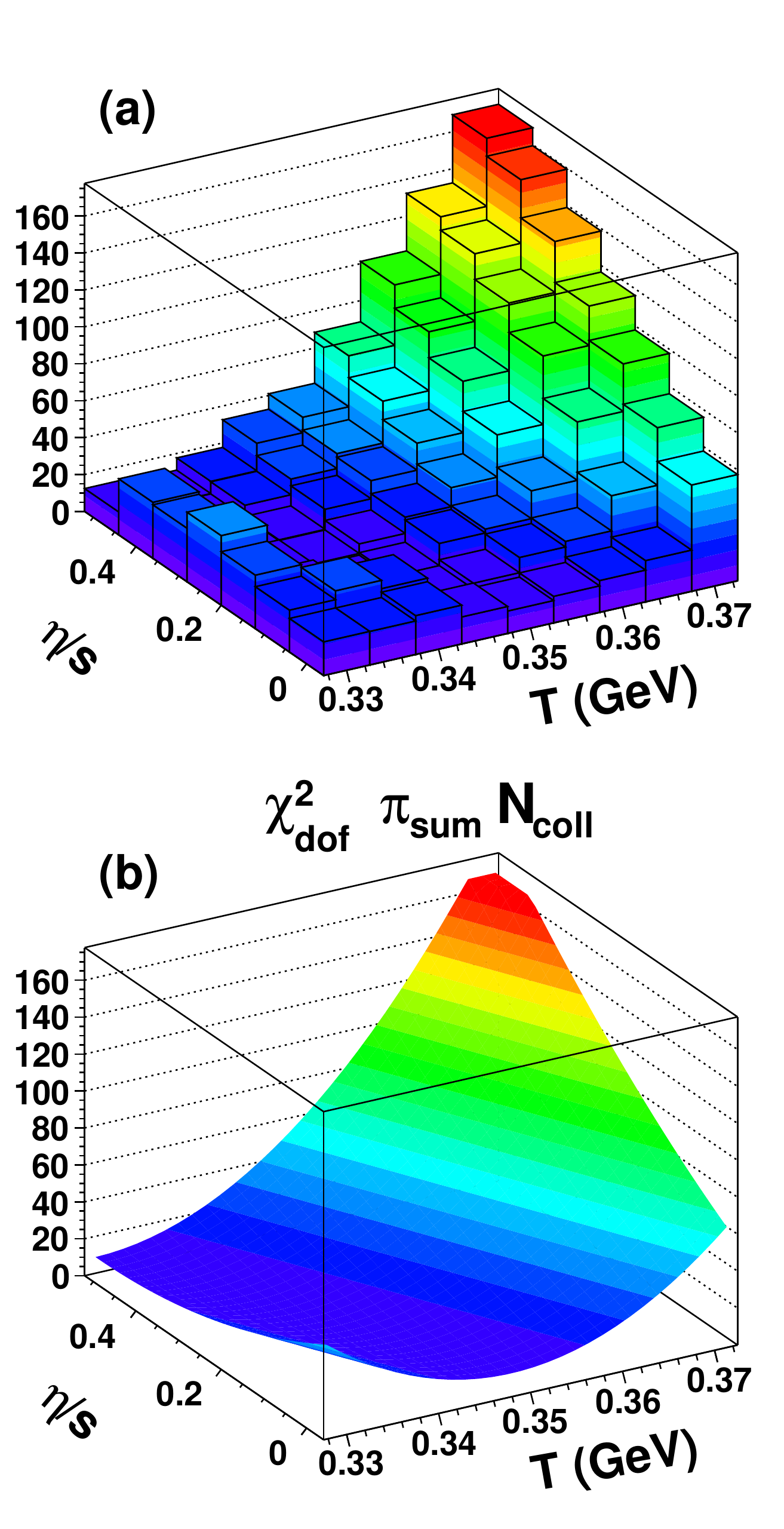}
\includegraphics[width=0.24\textwidth]{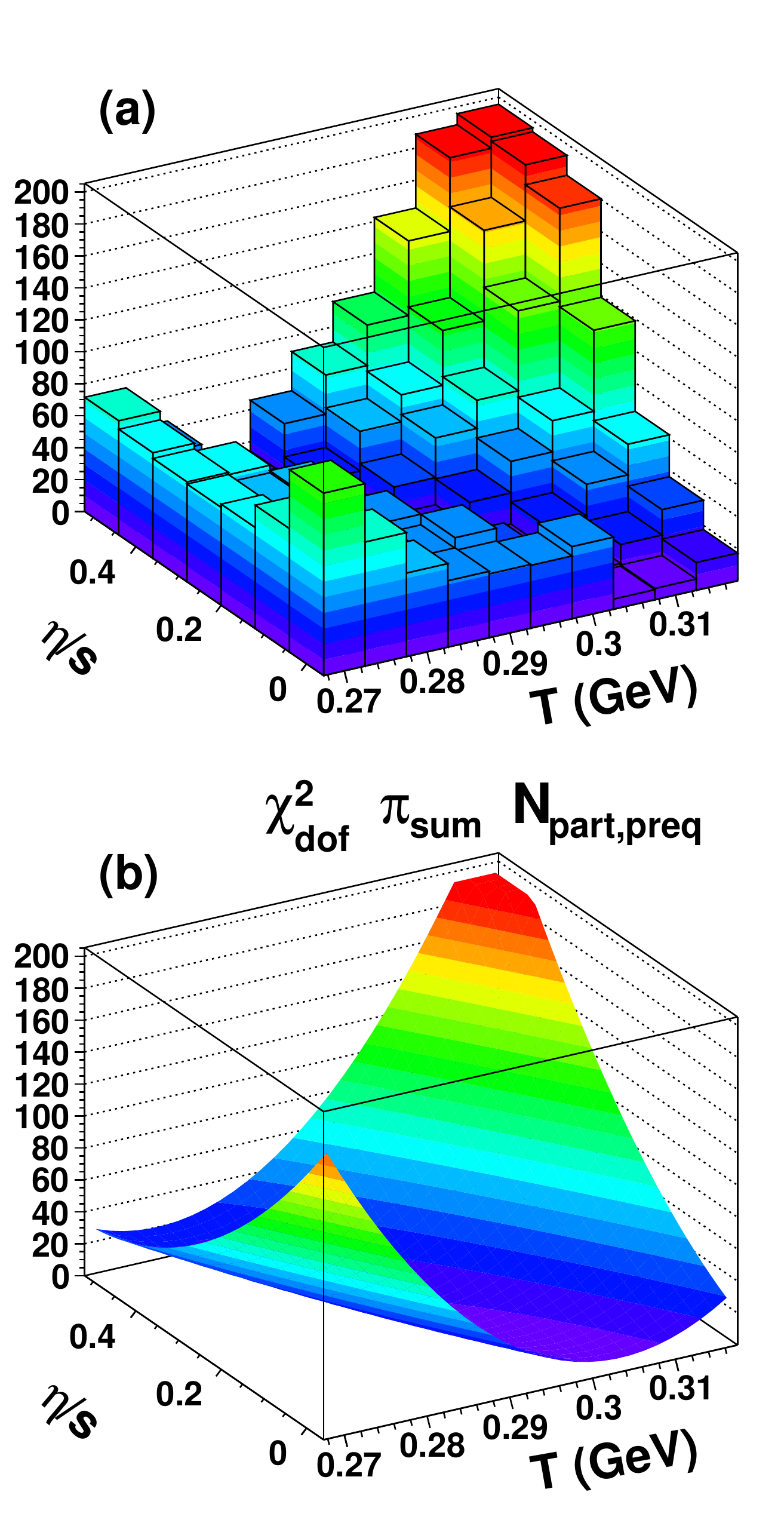}
\includegraphics[width=0.24\textwidth]{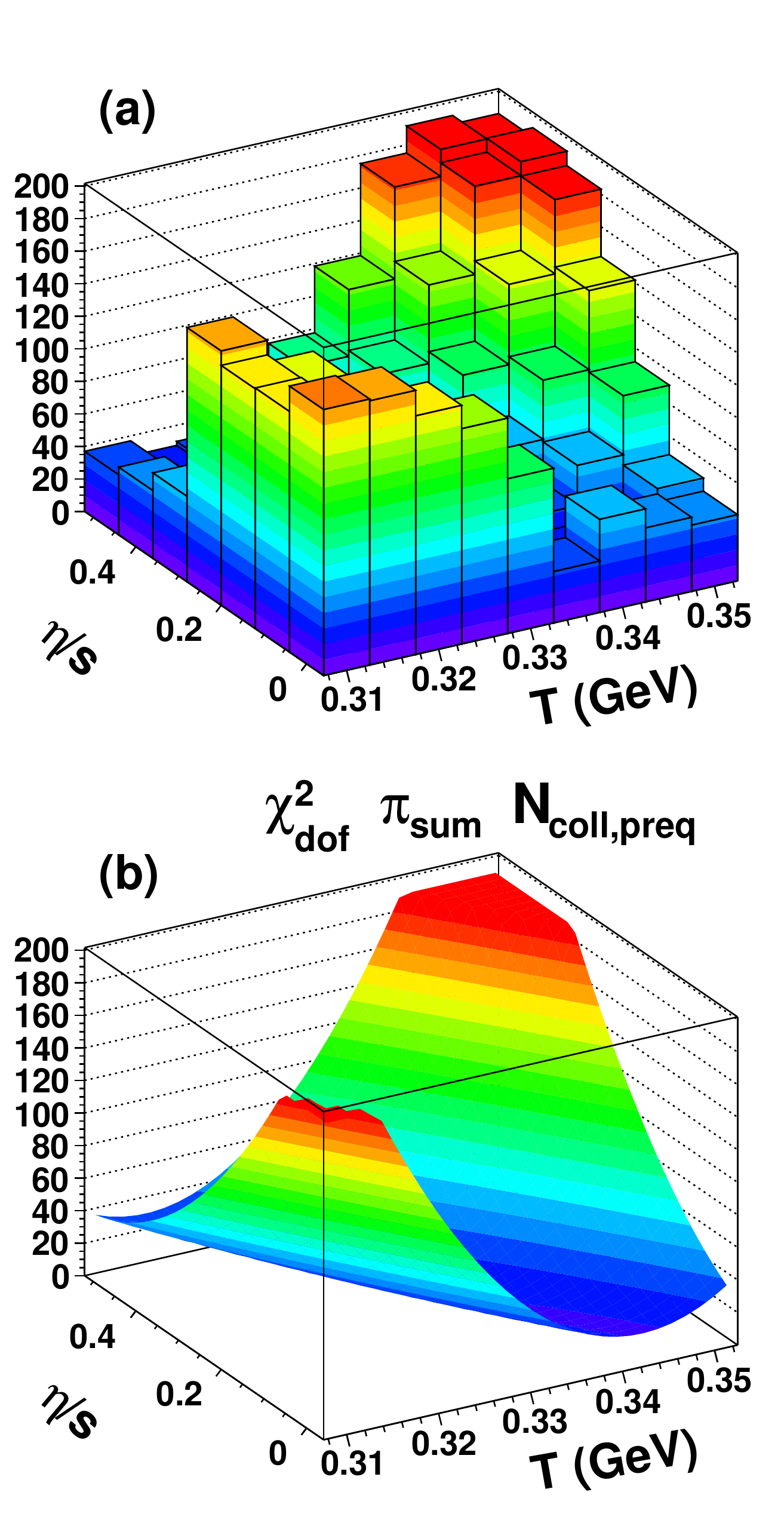}
\caption{(Color online) Equal weight sum of $\chi^2_{ndf}$ for pion spectra, elliptic flow, and femtoscopic radii.  From left to right the distributions are for $N_{part}$ scaling, $N_{coll}$ scaling, $N_{part}$ scaling with pre-equilibrium flow, and $N_{coll}$ scaling and pre-equilibrium flow.  The summed $\chi^2_{ndf}$ distributions are on top, and the paraboloid fits are below.}
\label{fig:sum_ch2}
\end{figure*}

To achieve a better balance between the three measurements, we repeat the $\chi^2$ sum with the spectra given a 10\% weight relative to the other measurements.  These distributions are shown in Fig.~\ref{fig:sum10_ch2}.  These distributions are still quite close to the spectra, but the parameters listed in Table~\ref{tab:fitparams} are slightly different, indicating that information from the elliptic flow and femtoscopic radii evaluation is being given greater weight.  The overall $\chi^2_{ndf}$ for the weighted sum, denoted as ${\rm Sum}_{10}$ in the table, is below 20 for the conditions without pre-equilibrium flow, and less than 10 for the initial conditions that include it.  A comparison of the one- and two-sigma contours for the weighted sum $\chi^2_{ndf}$ distributions is shown in Fig.~\ref{fig:sum10_contour}.

\begin{figure*}
\includegraphics[width=0.24\textwidth]{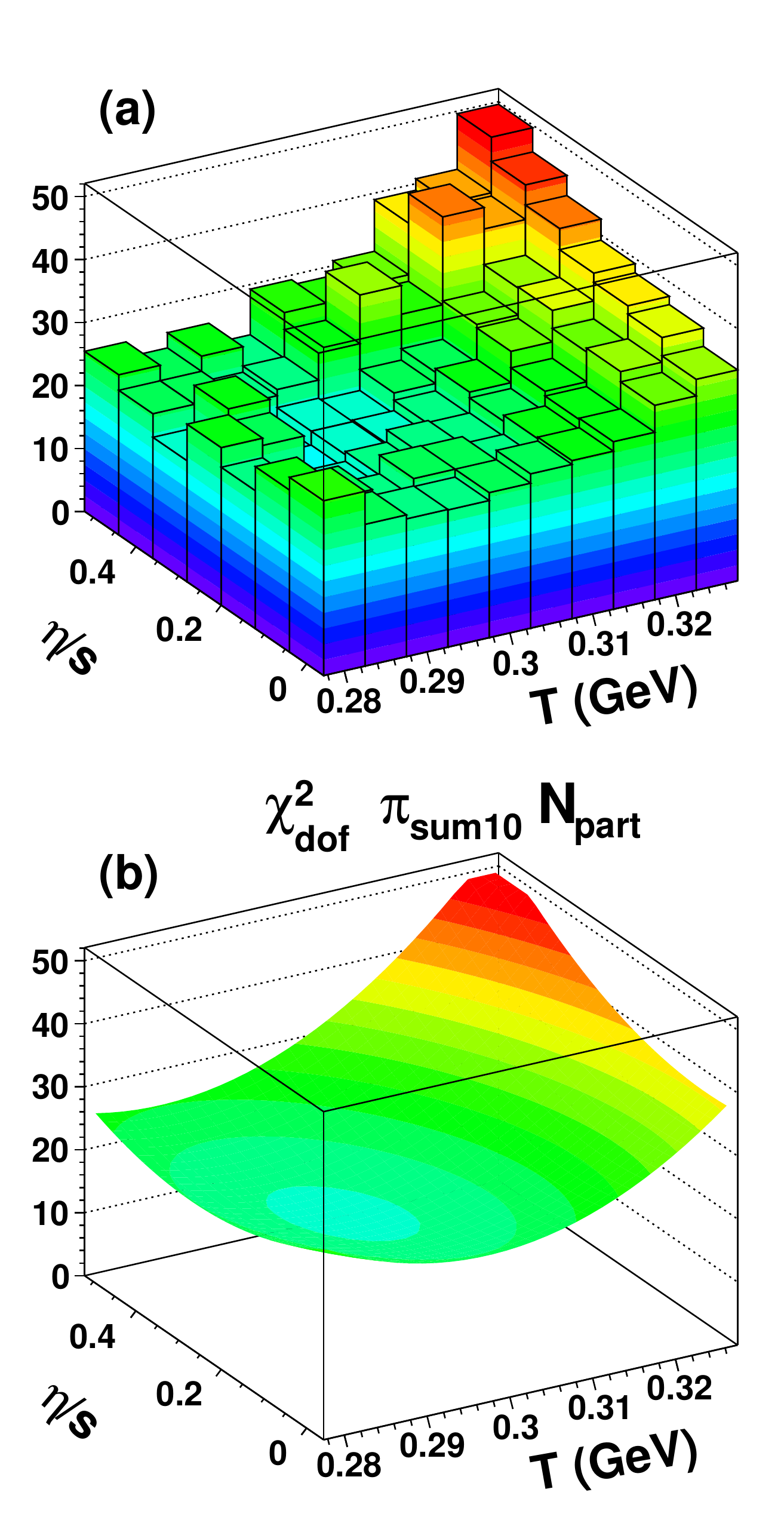}
\includegraphics[width=0.24\textwidth]{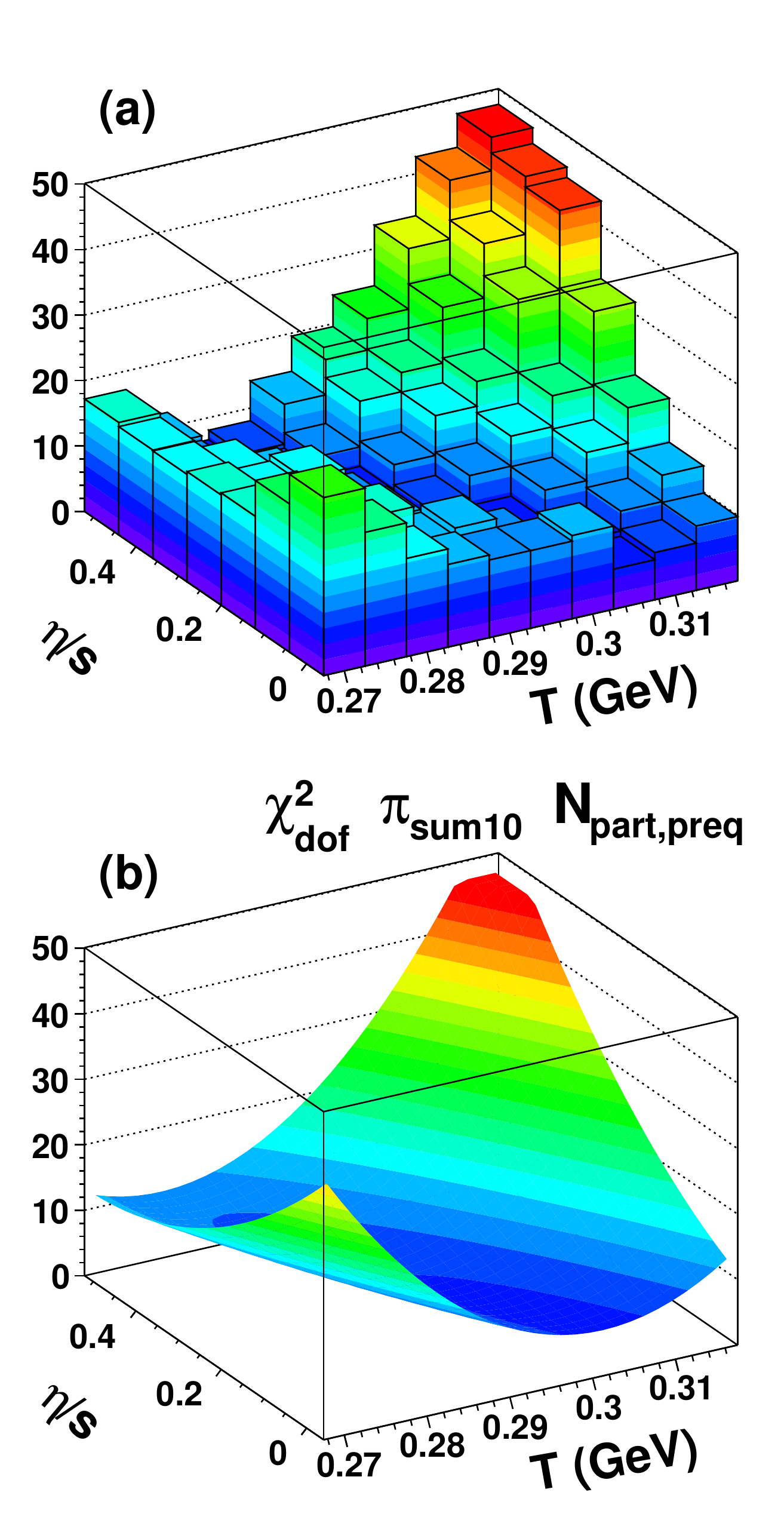}
\includegraphics[width=0.24\textwidth]{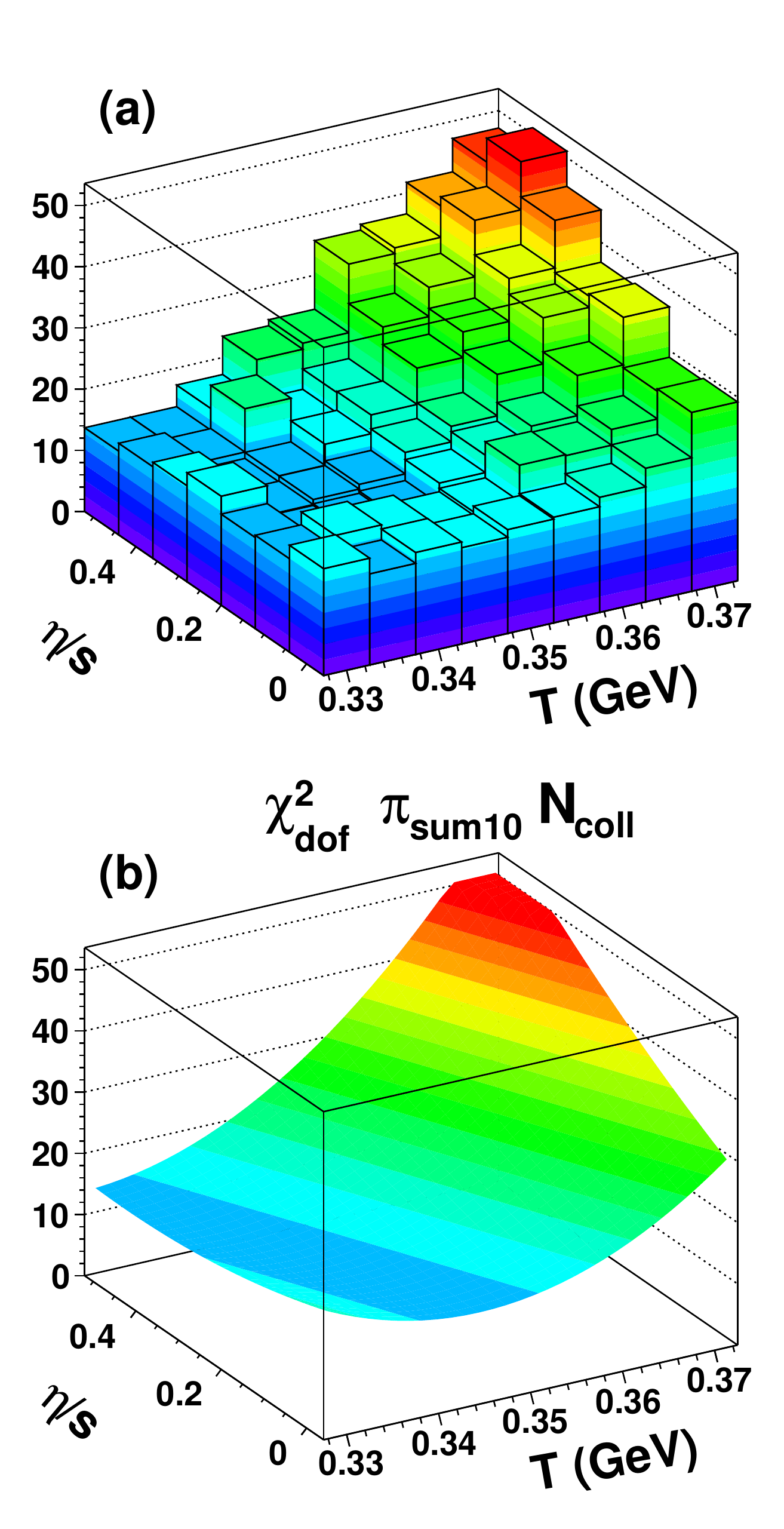}
\includegraphics[width=0.24\textwidth]{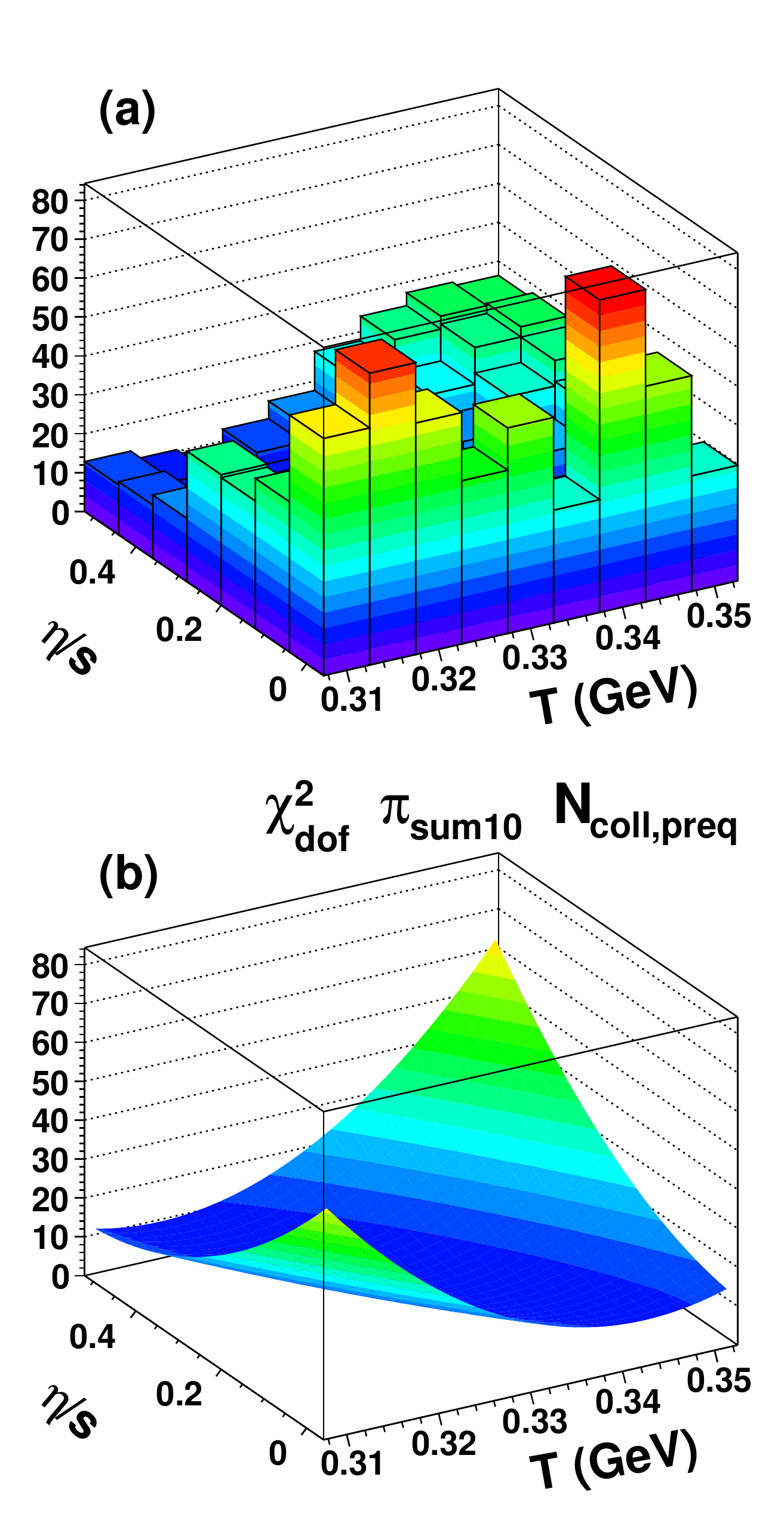}
\caption{(Color online) Sum of $\chi^2_{ndf}$ for $N_{part}$ scaling with spectra given 10\% weight.  From left to right the distributions are for $N_{part}$ scaling, $N_{coll}$ scaling, $N_{part}$ scaling with pre-equilibrium flow, and $N_{coll}$ scaling and pre-equilibrium flow.  The weighted sum $\chi^2_{ndf}$ distributions are on top, and the paraboloid fits are below.}
\label{fig:sum10_ch2}
\end{figure*}

\begin{figure}
\includegraphics[width=0.45\textwidth]{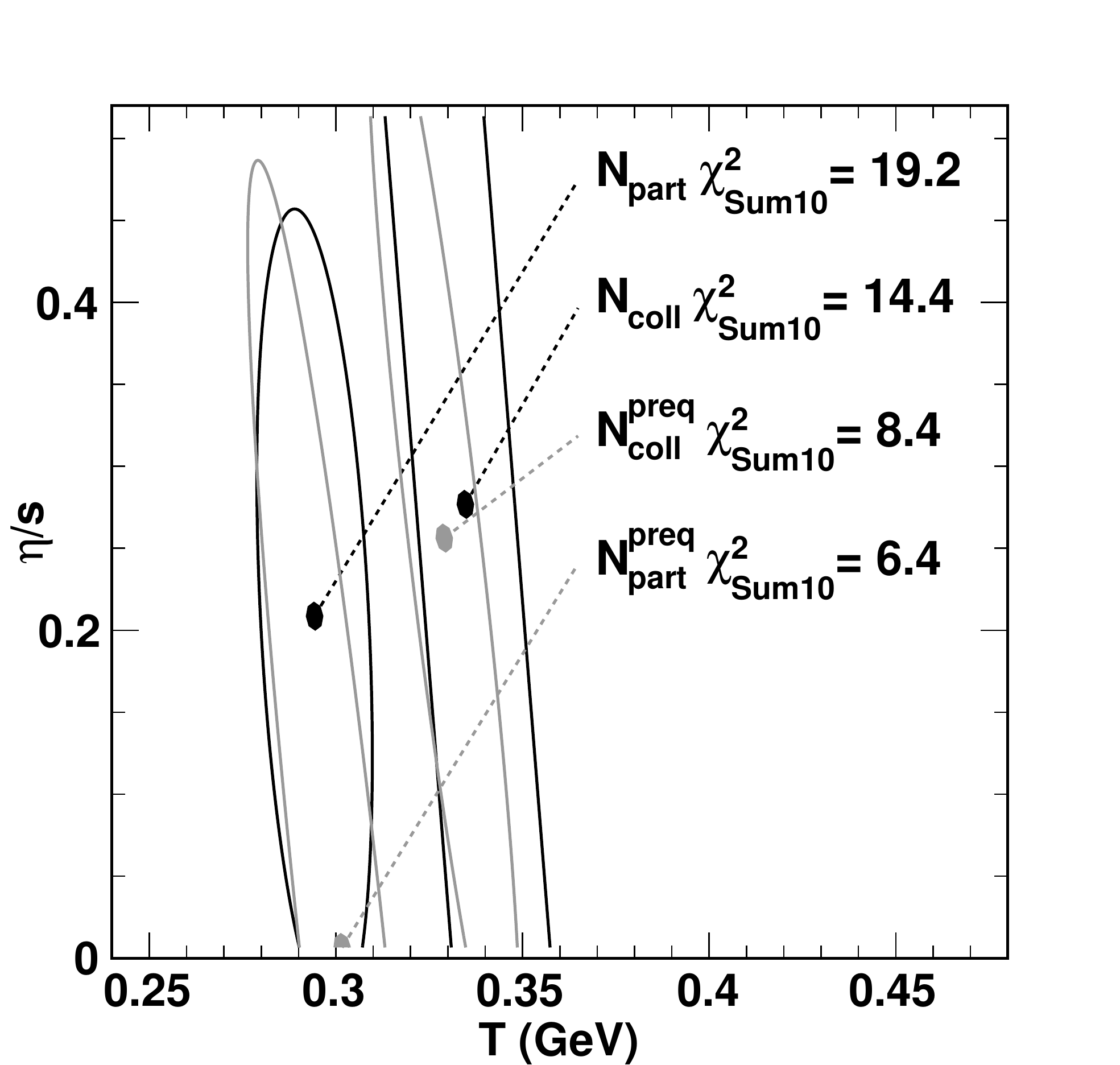}
\caption{(Color online) $\chi^2_{ndf}$ contours for the weighted sum in which the spectra are given a 10\% weight relative the $\chi^2$ distributions from elliptic flow and femtoscopic radii.  One- and two-sigma contours are drawn with solid lines for the $N_{part}$ (left) and $N_{coll}$ scaling, and with grey dashed lines for the addition of pre-equilibrium flow.}
\label{fig:sum10_contour}
\end{figure}

\section{Systematic Errors}\label{sec:sys}

The incorporation of systematic errors from the measurements is an important component of the evaluation that has been presented.  A separate source of systematic errors comes from the models employed.  These include the systematic errors of Glauber model used to generate the initial conditions, the parameterization of the initial state flow, the choice of hydrodynamic solvers, the freeze-out conditions, and cascade model assumptions.  A full investigation of these effects is well beyond the scope of the current evaluation and is left for future study.  However, there are additional set of systematic errors that are specific to the CHIMERA framework:
\begin{itemize}
\item the criteria for centrality matching
\item the use of Chebyshev polynomials to fit momentum dependence of model results,
\item the momentum range over which evaluations were performed,
\item the paraboloid functions fit to the $\chi^2_{ndf}$ distributions in $T$ and $\eta/s$,
\item the ranges in $T$ and $\eta/s$ used for the paraboloid fits.
\end{itemize}

In this work we address the first of these items, the centrality matching criteria.  As mentioned earlier, the difference between the model centrality of  $\langle N_{part} \rangle$=276 and $\langle N_{part} \rangle$=298 for the femtoscopic radii and elliptic flow measurements by STAR is large enough to influence the evaluation procedure.  To gauge the impact of this effect on the $\chi^2_{ndf}$ evaluations of femtoscopic radii we repeat the analysis after performing a linear interpolation of the radii and $v_2$ measurements to $\langle N_{part} \rangle$=276.  For the radii, the interpolation is a function of $\langle N_{part} \rangle^{1/3}$ using parameters derived from fits to the the centrality dependence~\cite{Adler:2004ii}.  For the $v_2$ measurements, the interpolation was performed using the integrated $v_2$ centrality dependence measured in~\cite{Adams:2005cx}.  As stated previously, these adjustment amount to a 2\% decrease for the STAR radii and an 8\% increase in $v_2$.  The CHIMERA evaluation is repeated for only the $N_{part}$ scaling with pre-equilibirum flow, which was shown to have the best overall agreement with the data.  The net effect of applying this centrality adjustment is shown in Fig.~\ref{fig:npadj_contour} and the parameters for the paraboloid fits are listed in Table~\ref{tab:npadj_params}.  

\begin{table}[h]
\begin{tabular}{|r||c|c|c||c|c|c|}
\hline
Evaluation & Major & Minor & angle$^\circ$ &\ $T_{\rm min}$ \ & $\ \eta/s_{\rm min}$ \ & $\chi^2_{ndf}$ \\
\hline
HBT$^{adj}$ & 0.366 & 0.0103 & 2.52 & 0.265 & 0.52 & 4.8 \\
\hline
HBT & 0.374 & 0.0105 & 2.51 & 0.273 & 0.52 & 4.8 \\
\hline \hline
V$_{2}^{adj}$ & 0.180 & 0.0315 & 15.5 & 0.322 & 0.0 & 6.3 \\
\hline
V$_{2}$ & 0.335 & 0.0393 & 14.2 & 0.322 & 0.0 & 4.8 \\
\hline \hline
Sum$_{10}^{adj}$ & 0.197 & 0.0059 & 2.69 & 0.298 & 0.0 & 7.4 \\
\hline
Sum$_{10}$ & 0.246 & 0.0059 & 2.69 & 0.300 & 0.0 & 6.4 \\
\hline
\end{tabular}
\caption{(Color online) Ellipse parameters determined from paraboloid fits to $\chi^2_{ndf}$ distributions in $T$ and $\eta/s$ for $N_{part}$ scaling with pre-equilibrium flow and with the centrality adjustment described in the text.}
\label{tab:npadj_params}
\end{table}

The differences between the unadjusted and adjusted centrality evaluations are minor.  The best fit initial temperature for the radii shifts down by 8~MeV, and the contour for the $V_2$ becomes narrower by 40\% along the major axis and 20\% along the minor axis.  The overall $\chi^2_{ndf}$ minimum rises slightly.  We conclude that the overall centrality match between the model and data is sufficient for this limited evaluation, however, this issue will require further scrutiny as the CHIMERA evaluations become more precise.

\begin{figure}
\includegraphics[width=0.45\textwidth]{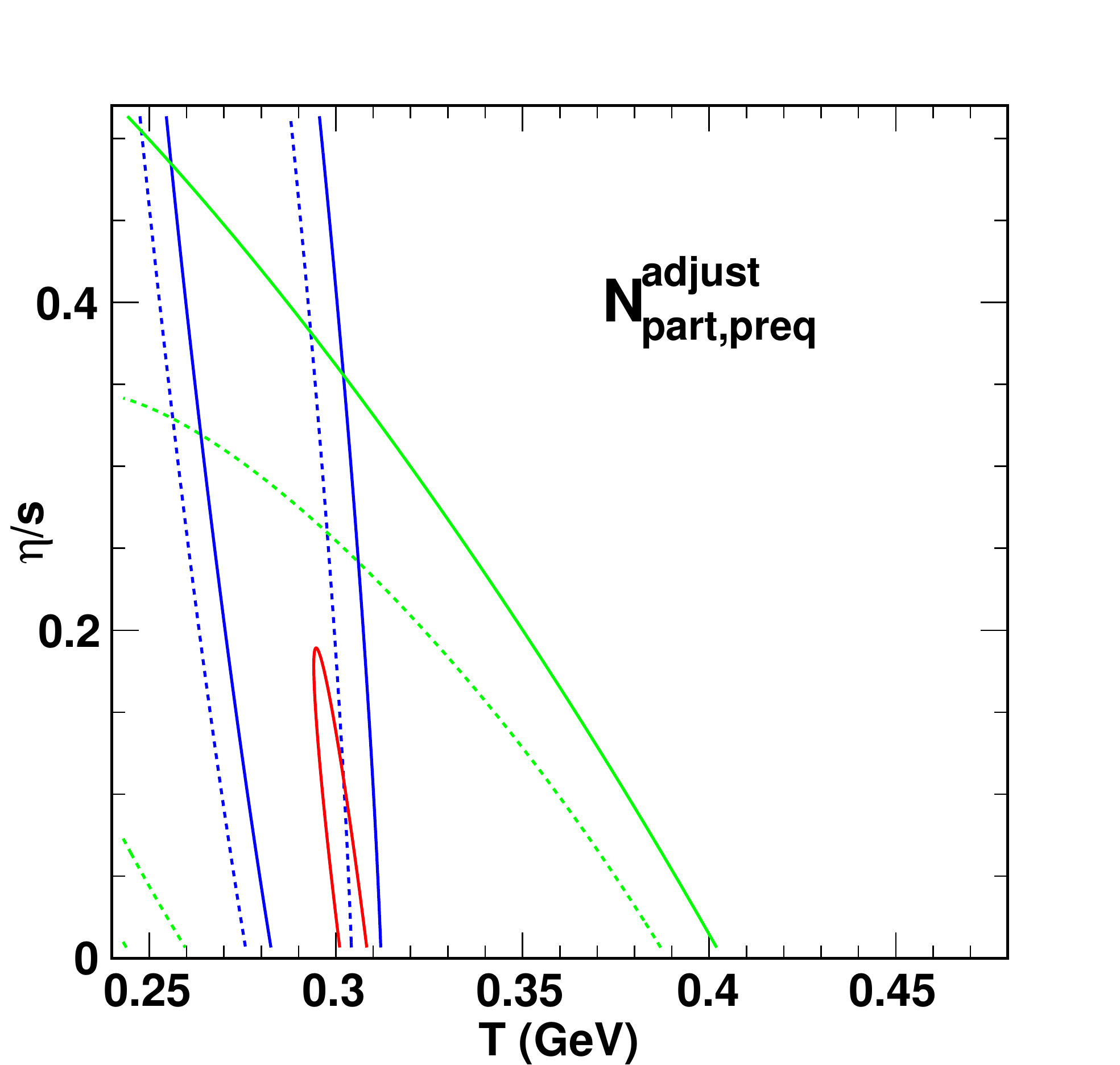}
\caption{(Color online) $\chi^2_{ndf}$ contours for pions spectra, elliptic flow, radii, and a weighted sum for $N_{part}$ scaling with pre-equilbrium flow with an additional adjustment to improve the centrality match with the STAR femtoscopic radii and elliptic flow.}
\label{fig:npadj_contour}
\end{figure}

\section{Conclusions}\label{sec:conc}

We have performed simultaneous $\chi^2$ evaluations of three measurements: spectra, elliptic flow, and femtoscopic radii from 200~GeV Au+Au collisions for a the 0-20\% centrality bin, using using an augmented version of the VH2 viscous 2D+1 hydrodynamic model that incorporates pre-equilibrium flow and initial state eccentricity coupled to the UrQMD hadronic cascade.The evaluations were performed for four sets of initial conditions: $N_{part}$ and $N_{coll}$ initial density profiles with and without the addition of pre-equilibrium flow.  The $\chi^2$ evaluations were performed for measurements by the PHENIX and STAR collaborations, and include both statistical and systematic errors.  For two of the initial conditions, $N_{part}$ without pre-equilibrium flow, and $N_{coll}$ with pre-equilibrium flow, the constrained regions of $T$ and $\eta/s$ were mutually exclusive.  However, for the $N_{part}$ scaling with pre-equilibrium flow and $N_{coll}$ without, the constraints regions overlap at the level of $2\sigma$.  This is also the first time that pion radius measurements have been successfully reproduced by a 2D+1 viscous hydrodynamic model coupled to a hadronic cascade.

We regard the main conclusion of this work to be the successful demonstration of an evaluation technique for constraining the multi-parameter space of heavy ion models by comparing to multiple data sets.  Working with a relatively small data sample, we were able to achieve a set of constraints for two sets of initial conditions, and to exclude two others.  However, the current implementation of CHIMERA should be regarded as incomplete.  The centrality matching has room for improvement, the equation of state is not yet specified according to the most recent lattice QCD results, and the full range of initial state profiles has not been explored.   There are also additional model parameters to be explored, such as the initial start time, $\tau_{start}$, and the switching temperature, $T_{sw}$.  In addition, the comparisons to experimental data were limited, covering only a single particle species, and a single centrality range.  We expect to address these issues in the near future.

Perhaps a more severe limitation on the current implementation of CHIMERA is the inability of VH2 to handle non-smooth initial conditions.  Qiu and Heinz have shown that~\cite{Qiu:2011js} a proper accounting of event by event fluctuations is required to compare to non-central ($>$60\%) collisions and is also needed to generate the higher order harmonics that promise to improve our ability to constrain $\eta/s$~\cite{Shen:2011zc,Schenke:2012jr}.  In order for CHIMERA or a similar framework to ultimately succeed, a more sophisticated hydrodynamic code will need to be employed, one that can accommodate initial state fluctuations, and its performance will need to be evaluated against the fullest possible set of physics observables.

\section{Acknowledgements}
\label{sec:ack}
This work was supported by the U.S. Department of Energy, Grant No. DE-FG02-03ER41259.  The authors wish to acknowledge P. Huovinen and D. Molnar for help received when working with the AZHYDRO code, P. Ramatschke for generously sharing his VH2 code, and also S. Bass and the UrQMD working group for the use of UrQMD and input conversion routines.  We also wish to thank J. Vredevoogd, B. M\"{u}ller, U. Heinz for insightful discussions.
\clearpage
\section{References}

\bibliography{chm_prc0}

\begin{thebibliography}{48}%
\makeatletter
\providecommand \@ifxundefined [1]{%
 \@ifx{#1\undefined}
}%
\providecommand \@ifnum [1]{%
 \ifnum #1\expandafter \@firstoftwo
 \else \expandafter \@secondoftwo
 \fi
}%
\providecommand \@ifx [1]{%
 \ifx #1\expandafter \@firstoftwo
 \else \expandafter \@secondoftwo
 \fi
}%
\providecommand \natexlab [1]{#1}%
\providecommand \enquote  [1]{``#1''}%
\providecommand \bibnamefont  [1]{#1}%
\providecommand \bibfnamefont [1]{#1}%
\providecommand \citenamefont [1]{#1}%
\providecommand \href@noop [0]{\@secondoftwo}%
\providecommand \href [0]{\begingroup \@sanitize@url \@href}%
\providecommand \@href[1]{\@@startlink{#1}\@@href}%
\providecommand \@@href[1]{\endgroup#1\@@endlink}%
\providecommand \@sanitize@url [0]{\catcode `\\12\catcode `\$12\catcode
  `\&12\catcode `\#12\catcode `\^12\catcode `\_12\catcode `\%12\relax}%
\providecommand \@@startlink[1]{}%
\providecommand \@@endlink[0]{}%
\providecommand \url  [0]{\begingroup\@sanitize@url \@url }%
\providecommand \@url [1]{\endgroup\@href {#1}{\urlprefix }}%
\providecommand \urlprefix  [0]{URL }%
\providecommand \Eprint [0]{\href }%
\providecommand \doibase [0]{http://dx.doi.org/}%
\providecommand \selectlanguage [0]{\@gobble}%
\providecommand \bibinfo  [0]{\@secondoftwo}%
\providecommand \bibfield  [0]{\@secondoftwo}%
\providecommand \translation [1]{[#1]}%
\providecommand \BibitemOpen [0]{}%
\providecommand \bibitemStop [0]{}%
\providecommand \bibitemNoStop [0]{.\EOS\space}%
\providecommand \EOS [0]{\spacefactor3000\relax}%
\providecommand \BibitemShut  [1]{\csname bibitem#1\endcsname}%
\let\auto@bib@innerbib\@empty
\bibitem [{\citenamefont {Arsene}\ \emph {et~al.}(2005)\citenamefont {Arsene}
  \emph {et~al.}}]{Arsene:2005kd}%
  \BibitemOpen
  \bibfield  {author} {\bibinfo {author} {\bibfnamefont {I.}~\bibnamefont
  {Arsene}} \emph {et~al.} (\bibinfo {collaboration} {BRAHMS Collaboration}),\
  }\href@noop {} {\bibfield  {journal} {\bibinfo  {journal} {Nuclear Physics
  A}\ }\textbf {\bibinfo {volume} {757}},\ \bibinfo {pages} {1} (\bibinfo
  {year} {2005})}\BibitemShut {NoStop}%
\bibitem [{\citenamefont {Back}\ \emph {et~al.}(2005)\citenamefont {Back} \emph
  {et~al.}}]{Back:2005ck}%
  \BibitemOpen
  \bibfield  {author} {\bibinfo {author} {\bibfnamefont {B.~B.}\ \bibnamefont
  {Back}} \emph {et~al.} (\bibinfo {collaboration} {PHOBOS Collaboration}),\
  }\href@noop {} {\bibfield  {journal} {\bibinfo  {journal} {Nuclear Physics
  A}\ }\textbf {\bibinfo {volume} {757}},\ \bibinfo {pages} {28} (\bibinfo
  {year} {2005})}\BibitemShut {NoStop}%
\bibitem [{\citenamefont {Adams}\ \emph
  {et~al.}(2005{\natexlab{a}})\citenamefont {Adams} \emph
  {et~al.}}]{Adams:2005tm}%
  \BibitemOpen
  \bibfield  {author} {\bibinfo {author} {\bibfnamefont {J.}~\bibnamefont
  {Adams}} \emph {et~al.} (\bibinfo {collaboration} {STAR Collaboration}),\
  }\href@noop {} {\bibfield  {journal} {\bibinfo  {journal} {Phys. Rev. C}\
  }\textbf {\bibinfo {volume} {72}},\ \bibinfo {pages} {014904} (\bibinfo
  {year} {2005}{\natexlab{a}})}\BibitemShut {NoStop}%
\bibitem [{\citenamefont {Adcox}\ \emph {et~al.}(2005)\citenamefont {Adcox}
  \emph {et~al.}}]{Adcox:2005iw}%
  \BibitemOpen
  \bibfield  {author} {\bibinfo {author} {\bibfnamefont {K.}~\bibnamefont
  {Adcox}} \emph {et~al.} (\bibinfo {collaboration} {PHENIX Collaboration}),\
  }\href@noop {} {\bibfield  {journal} {\bibinfo  {journal} {Nuclear Physics
  A}\ }\textbf {\bibinfo {volume} {757}},\ \bibinfo {pages} {184} (\bibinfo
  {year} {2005})}\BibitemShut {NoStop}%
\bibitem [{\citenamefont {Huovinen}\ \emph {et~al.}(2001)\citenamefont
  {Huovinen}, \citenamefont {Kolb}, \citenamefont {Heinz}, \citenamefont
  {Ruuskanen},\ and\ \citenamefont {Voloshin}}]{Huovinen:483397}%
  \BibitemOpen
  \bibfield  {author} {\bibinfo {author} {\bibfnamefont {P.}~\bibnamefont
  {Huovinen}}, \bibinfo {author} {\bibfnamefont {P.~F.}\ \bibnamefont {Kolb}},
  \bibinfo {author} {\bibfnamefont {U.~W.}\ \bibnamefont {Heinz}}, \bibinfo
  {author} {\bibfnamefont {P.~V.}\ \bibnamefont {Ruuskanen}}, \ and\ \bibinfo
  {author} {\bibfnamefont {S.~A.}\ \bibnamefont {Voloshin}},\ }\href@noop {}
  {\bibfield  {journal} {\bibinfo  {journal} {Phys. Lett. B}\ }\textbf
  {\bibinfo {volume} {503}},\ \bibinfo {pages} {58} (\bibinfo {year}
  {2001})}\BibitemShut {NoStop}%
\bibitem [{\citenamefont {Kolb}\ and\ \citenamefont
  {Heinz}(2003)}]{Kolb:2003tx}%
  \BibitemOpen
  \bibfield  {author} {\bibinfo {author} {\bibfnamefont {P.~F.}\ \bibnamefont
  {Kolb}}\ and\ \bibinfo {author} {\bibfnamefont {U.}~\bibnamefont {Heinz}},\
  }\href@noop {} {\bibfield  {journal} {\bibinfo  {journal} {Arxiv preprint
  nucl-th/0305084}\ } (\bibinfo {year} {2003})}\BibitemShut {NoStop}%
\bibitem [{\citenamefont {Hirano}(2001)}]{Hirano:2001bn}%
  \BibitemOpen
  \bibfield  {author} {\bibinfo {author} {\bibfnamefont {T.}~\bibnamefont
  {Hirano}},\ }\href@noop {} {\bibfield  {journal} {\bibinfo  {journal} {Phys.
  Rev. C}\ }\textbf {\bibinfo {volume} {65}},\ \bibinfo {pages} {011901(R)}
  (\bibinfo {year} {2001})}\BibitemShut {NoStop}%
\bibitem [{\citenamefont {Bass}\ \emph {et~al.}(1998)\citenamefont {Bass} \emph
  {et~al.}}]{Bass:349359}%
  \BibitemOpen
  \bibfield  {author} {\bibinfo {author} {\bibfnamefont {S.~A.}\ \bibnamefont
  {Bass}} \emph {et~al.},\ }\href@noop {} {\bibfield  {journal} {\bibinfo
  {journal} {Prog. Part. Nucl. Phys.}\ }\textbf {\bibinfo {volume} {41}},\
  \bibinfo {pages} {225} (\bibinfo {year} {1998})}\BibitemShut {NoStop}%
\bibitem [{\citenamefont {Bleicher}\ \emph {et~al.}(1999)\citenamefont
  {Bleicher} \emph {et~al.}}]{Bleicher:1999wv}%
  \BibitemOpen
  \bibfield  {author} {\bibinfo {author} {\bibfnamefont {M.}~\bibnamefont
  {Bleicher}} \emph {et~al.},\ }\href@noop {} {\bibfield  {journal} {\bibinfo
  {journal} {J. Phys. G}\ }\textbf {\bibinfo {volume} {25}},\ \bibinfo {pages}
  {1859} (\bibinfo {year} {1999})}\BibitemShut {NoStop}%
\bibitem [{\citenamefont {Luzum}\ and\ \citenamefont
  {Romatschke}(2008)}]{Luzum:2008hz}%
  \BibitemOpen
  \bibfield  {author} {\bibinfo {author} {\bibfnamefont {M.}~\bibnamefont
  {Luzum}}\ and\ \bibinfo {author} {\bibfnamefont {P.}~\bibnamefont
  {Romatschke}},\ }\href@noop {} {\bibfield  {journal} {\bibinfo  {journal}
  {Phys. Rev. C}\ }\textbf {\bibinfo {volume} {78}},\ \bibinfo {pages} {034915}
  (\bibinfo {year} {2008})}\BibitemShut {NoStop}%
\bibitem [{\citenamefont {Song}\ \emph
  {et~al.}(2011{\natexlab{a}})\citenamefont {Song}, \citenamefont {Bass},\ and\
  \citenamefont {Heinz}}]{Song:2011fb}%
  \BibitemOpen
  \bibfield  {author} {\bibinfo {author} {\bibfnamefont {H.}~\bibnamefont
  {Song}}, \bibinfo {author} {\bibfnamefont {S.}~\bibnamefont {Bass}}, \ and\
  \bibinfo {author} {\bibfnamefont {U.}~\bibnamefont {Heinz}},\ }\href@noop {}
  {\bibfield  {journal} {\bibinfo  {journal} {Phys. Rev. C}\ }\textbf {\bibinfo
  {volume} {83}},\ \bibinfo {pages} {024912} (\bibinfo {year}
  {2011}{\natexlab{a}})}\BibitemShut {NoStop}%
\bibitem [{\citenamefont {Kodama}\ \emph {et~al.}(2001)\citenamefont {Kodama},
  \citenamefont {Aguiar}, \citenamefont {Osada},\ and\ \citenamefont
  {Hama}}]{Kodama:2001qv}%
  \BibitemOpen
  \bibfield  {author} {\bibinfo {author} {\bibfnamefont {T.}~\bibnamefont
  {Kodama}}, \bibinfo {author} {\bibfnamefont {C.~E.}\ \bibnamefont {Aguiar}},
  \bibinfo {author} {\bibfnamefont {T.}~\bibnamefont {Osada}}, \ and\ \bibinfo
  {author} {\bibfnamefont {Y.}~\bibnamefont {Hama}},\ }\href@noop {} {\bibfield
   {journal} {\bibinfo  {journal} {J. Phys. G}\ }\textbf {\bibinfo {volume}
  {27}},\ \bibinfo {pages} {557} (\bibinfo {year} {2001})}\BibitemShut
  {NoStop}%
\bibitem [{\citenamefont {Schenke}\ \emph {et~al.}(2010)\citenamefont
  {Schenke}, \citenamefont {Jeon},\ and\ \citenamefont
  {Gale}}]{Schenke:2010di}%
  \BibitemOpen
  \bibfield  {author} {\bibinfo {author} {\bibfnamefont {B.}~\bibnamefont
  {Schenke}}, \bibinfo {author} {\bibfnamefont {S.}~\bibnamefont {Jeon}}, \
  and\ \bibinfo {author} {\bibfnamefont {C.}~\bibnamefont {Gale}},\ }\href@noop
  {} {\bibfield  {journal} {\bibinfo  {journal} {Phys. Rev. C}\ }\textbf
  {\bibinfo {volume} {82}},\ \bibinfo {pages} {014903} (\bibinfo {year}
  {2010})}\BibitemShut {NoStop}%
\bibitem [{\citenamefont {Qiu}\ and\ \citenamefont {Heinz}(2011)}]{Qiu:2011js}%
  \BibitemOpen
  \bibfield  {author} {\bibinfo {author} {\bibfnamefont {Z.}~\bibnamefont
  {Qiu}}\ and\ \bibinfo {author} {\bibfnamefont {U.}~\bibnamefont {Heinz}},\
  }\href@noop {} {\bibfield  {journal} {\bibinfo  {journal} {Phys. Rev. C}\
  }\textbf {\bibinfo {volume} {84}},\ \bibinfo {pages} {024911} (\bibinfo
  {year} {2011})}\BibitemShut {NoStop}%
\bibitem [{\citenamefont {Bazavov}\ \emph {et~al.}(2009)\citenamefont {Bazavov}
  \emph {et~al.}}]{Bazavov:2009ep}%
  \BibitemOpen
  \bibfield  {author} {\bibinfo {author} {\bibfnamefont {A.}~\bibnamefont
  {Bazavov}} \emph {et~al.} (\bibinfo {collaboration} {HotQCD Collaboration}),\
  }\href@noop {} {\bibfield  {journal} {\bibinfo  {journal} {Phys. Rev. D}\
  }\textbf {\bibinfo {volume} {80}},\ \bibinfo {pages} {014504} (\bibinfo
  {year} {2009})}\BibitemShut {NoStop}%
\bibitem [{\citenamefont {Bors{\'a}nyi}\ \emph {et~al.}(2010)\citenamefont
  {Bors{\'a}nyi} \emph {et~al.}}]{Borsanyi:2010gh}%
  \BibitemOpen
  \bibfield  {author} {\bibinfo {author} {\bibfnamefont {S.}~\bibnamefont
  {Bors{\'a}nyi}} \emph {et~al.},\ }\href@noop {} {\bibfield  {journal}
  {\bibinfo  {journal} {J. High Energ. Phys.}\ }\textbf {\bibinfo {volume}
  {2010}},\ \bibinfo {pages} {77} (\bibinfo {year} {2010})}\BibitemShut
  {NoStop}%
\bibitem [{\citenamefont {Song}\ \emph
  {et~al.}(2011{\natexlab{b}})\citenamefont {Song}, \citenamefont {Bass},
  \citenamefont {Heinz}, \citenamefont {Hirano},\ and\ \citenamefont
  {Shen}}]{Song:2010mg}%
  \BibitemOpen
  \bibfield  {author} {\bibinfo {author} {\bibfnamefont {H.}~\bibnamefont
  {Song}}, \bibinfo {author} {\bibfnamefont {S.~A.}\ \bibnamefont {Bass}},
  \bibinfo {author} {\bibfnamefont {U.}~\bibnamefont {Heinz}}, \bibinfo
  {author} {\bibfnamefont {T.}~\bibnamefont {Hirano}}, \ and\ \bibinfo {author}
  {\bibfnamefont {C.}~\bibnamefont {Shen}},\ }\href@noop {} {\bibfield
  {journal} {\bibinfo  {journal} {Phys. Rev. Lett.}\ }\textbf {\bibinfo
  {volume} {106}},\ \bibinfo {pages} {192301} (\bibinfo {year}
  {2011}{\natexlab{b}})}\BibitemShut {NoStop}%
\bibitem [{\citenamefont {Alver}\ \emph
  {et~al.}(2008{\natexlab{a}})\citenamefont {Alver} \emph
  {et~al.}}]{Alver:2008fk}%
  \BibitemOpen
  \bibfield  {author} {\bibinfo {author} {\bibfnamefont {B.}~\bibnamefont
  {Alver}} \emph {et~al.} (\bibinfo {collaboration} {PHOBOS Collaboration}),\
  }\href@noop {} {\bibfield  {journal} {\bibinfo  {journal} {Phys. Rev. C}\
  }\textbf {\bibinfo {volume} {77}},\ \bibinfo {pages} {014906} (\bibinfo
  {year} {2008}{\natexlab{a}})}\BibitemShut {NoStop}%
\bibitem [{\citenamefont {Vredevoogd}\ and\ \citenamefont
  {Pratt}(2009)}]{Vredevoogd:2009jt}%
  \BibitemOpen
  \bibfield  {author} {\bibinfo {author} {\bibfnamefont {J.}~\bibnamefont
  {Vredevoogd}}\ and\ \bibinfo {author} {\bibfnamefont {S.}~\bibnamefont
  {Pratt}},\ }\href@noop {} {\bibfield  {journal} {\bibinfo  {journal} {Phys.
  Rev. C}\ }\textbf {\bibinfo {volume} {79}} (\bibinfo {year}
  {2009})}\BibitemShut {NoStop}%
\bibitem [{\citenamefont {Hanbury~Brown}\ and\ \citenamefont
  {Twiss}(1954)}]{HanburyBrown:433988}%
  \BibitemOpen
  \bibfield  {author} {\bibinfo {author} {\bibfnamefont {R.}~\bibnamefont
  {Hanbury~Brown}}\ and\ \bibinfo {author} {\bibfnamefont {R.~Q.}\ \bibnamefont
  {Twiss}},\ }\href@noop {} {\bibfield  {journal} {\bibinfo  {journal} {Philos.
  Mag.}\ }\textbf {\bibinfo {volume} {45}},\ \bibinfo {pages} {663} (\bibinfo
  {year} {1954})}\BibitemShut {NoStop}%
\bibitem [{\citenamefont {Goldhaber}\ \emph {et~al.}(1960)\citenamefont
  {Goldhaber}, \citenamefont {Goldhaber}, \citenamefont {Lee},\ and\
  \citenamefont {Pais}}]{Goldhaber:1960vf}%
  \BibitemOpen
  \bibfield  {author} {\bibinfo {author} {\bibfnamefont {G.}~\bibnamefont
  {Goldhaber}}, \bibinfo {author} {\bibfnamefont {S.}~\bibnamefont
  {Goldhaber}}, \bibinfo {author} {\bibfnamefont {W.}~\bibnamefont {Lee}}, \
  and\ \bibinfo {author} {\bibfnamefont {A.}~\bibnamefont {Pais}},\ }\href@noop
  {} {\bibfield  {journal} {\bibinfo  {journal} {Phys. Rev.}\ }\textbf
  {\bibinfo {volume} {120}},\ \bibinfo {pages} {300} (\bibinfo {year}
  {1960})}\BibitemShut {NoStop}%
\bibitem [{\citenamefont {Pratt}\ \emph {et~al.}(1990)\citenamefont {Pratt},
  \citenamefont {Cs{\"o}rg{\H o}},\ and\ \citenamefont
  {Zim{\'a}nyi}}]{Pratt:1990kx}%
  \BibitemOpen
  \bibfield  {author} {\bibinfo {author} {\bibfnamefont {S.}~\bibnamefont
  {Pratt}}, \bibinfo {author} {\bibfnamefont {T.}~\bibnamefont {Cs{\"o}rg{\H
  o}}}, \ and\ \bibinfo {author} {\bibfnamefont {J.}~\bibnamefont
  {Zim{\'a}nyi}},\ }\href@noop {} {\bibfield  {journal} {\bibinfo  {journal}
  {Phys. Rev. C}\ }\textbf {\bibinfo {volume} {42}},\ \bibinfo {pages} {2646}
  (\bibinfo {year} {1990})}\BibitemShut {NoStop}%
\bibitem [{\citenamefont {Brown}\ \emph {et~al.}(2005)\citenamefont {Brown}
  \emph {et~al.}}]{Brown:2005bl}%
  \BibitemOpen
  \bibfield  {author} {\bibinfo {author} {\bibfnamefont {D.}~\bibnamefont
  {Brown}} \emph {et~al.},\ }\href@noop {} {\bibfield  {journal} {\bibinfo
  {journal} {Phys. Rev. C}\ }\textbf {\bibinfo {volume} {72}},\ \bibinfo
  {pages} {054902} (\bibinfo {year} {2005})}\BibitemShut {NoStop}%
\bibitem [{\citenamefont {Alver}\ \emph
  {et~al.}(2008{\natexlab{b}})\citenamefont {Alver}, \citenamefont {Baker},
  \citenamefont {Loizides},\ and\ \citenamefont {Steinberg}}]{Alver:2008wu}%
  \BibitemOpen
  \bibfield  {author} {\bibinfo {author} {\bibfnamefont {B.}~\bibnamefont
  {Alver}}, \bibinfo {author} {\bibfnamefont {M.}~\bibnamefont {Baker}},
  \bibinfo {author} {\bibfnamefont {C.}~\bibnamefont {Loizides}}, \ and\
  \bibinfo {author} {\bibfnamefont {P.}~\bibnamefont {Steinberg}},\ }\href@noop
  {} {\bibfield  {journal} {\bibinfo  {journal} {arXiv}\ }\textbf {\bibinfo
  {volume} {nucl-ex}} (\bibinfo {year} {2008}{\natexlab{b}})}\BibitemShut
  {NoStop}%
\bibitem [{\citenamefont {Kharzeev}\ \emph {et~al.}(2005)\citenamefont
  {Kharzeev}, \citenamefont {Levin},\ and\ \citenamefont
  {Nardi}}]{Kharzeev:2005ku}%
  \BibitemOpen
  \bibfield  {author} {\bibinfo {author} {\bibfnamefont {D.}~\bibnamefont
  {Kharzeev}}, \bibinfo {author} {\bibfnamefont {E.}~\bibnamefont {Levin}}, \
  and\ \bibinfo {author} {\bibfnamefont {M.}~\bibnamefont {Nardi}},\
  }\href@noop {} {\bibfield  {journal} {\bibinfo  {journal} {Phys. Rev. C}\
  }\textbf {\bibinfo {volume} {71}},\ \bibinfo {pages} {054903} (\bibinfo
  {year} {2005})}\BibitemShut {NoStop}%
\bibitem [{\citenamefont {Broniowski}\ \emph {et~al.}(2009)\citenamefont
  {Broniowski}, \citenamefont {Florkowski}, \citenamefont {Chojnacki},\ and\
  \citenamefont {Kisiel}}]{Broniowski:2009hp}%
  \BibitemOpen
  \bibfield  {author} {\bibinfo {author} {\bibfnamefont {W.}~\bibnamefont
  {Broniowski}}, \bibinfo {author} {\bibfnamefont {W.}~\bibnamefont
  {Florkowski}}, \bibinfo {author} {\bibfnamefont {M.}~\bibnamefont
  {Chojnacki}}, \ and\ \bibinfo {author} {\bibfnamefont {A.}~\bibnamefont
  {Kisiel}},\ }\href@noop {} {\bibfield  {journal} {\bibinfo  {journal} {Phys.
  Rev. C}\ }\textbf {\bibinfo {volume} {80}} (\bibinfo {year}
  {2009})}\BibitemShut {NoStop}%
\bibitem [{\citenamefont {Schenke}\ \emph
  {et~al.}(2012{\natexlab{a}})\citenamefont {Schenke}, \citenamefont
  {Tribedy},\ and\ \citenamefont {Venugopalan}}]{Schenke:2012dk}%
  \BibitemOpen
  \bibfield  {author} {\bibinfo {author} {\bibfnamefont {B.}~\bibnamefont
  {Schenke}}, \bibinfo {author} {\bibfnamefont {P.}~\bibnamefont {Tribedy}}, \
  and\ \bibinfo {author} {\bibfnamefont {R.}~\bibnamefont {Venugopalan}},\
  }\href@noop {} {\bibfield  {journal} {\bibinfo  {journal} {Phys. Rev. Lett.}\
  }\textbf {\bibinfo {volume} {108}},\ \bibinfo {pages} {252301} (\bibinfo
  {year} {2012}{\natexlab{a}})}\BibitemShut {NoStop}%
\bibitem [{\citenamefont {Laine}\ and\ \citenamefont
  {Schr{\"o}der}(2006)}]{Laine:2006fj}%
  \BibitemOpen
  \bibfield  {author} {\bibinfo {author} {\bibfnamefont {M.}~\bibnamefont
  {Laine}}\ and\ \bibinfo {author} {\bibfnamefont {Y.}~\bibnamefont
  {Schr{\"o}der}},\ }\href@noop {} {\bibfield  {journal} {\bibinfo  {journal}
  {Phys. Rev. D}\ }\textbf {\bibinfo {volume} {73}} (\bibinfo {year}
  {2006})}\BibitemShut {NoStop}%
\bibitem [{\citenamefont {Cooper}\ and\ \citenamefont
  {Frye}(1974)}]{Cooper:1974ug}%
  \BibitemOpen
  \bibfield  {author} {\bibinfo {author} {\bibfnamefont {F.}~\bibnamefont
  {Cooper}}\ and\ \bibinfo {author} {\bibfnamefont {G.}~\bibnamefont {Frye}},\
  }\href@noop {} {\bibfield  {journal} {\bibinfo  {journal} {Phys. Rev. D}\
  }\textbf {\bibinfo {volume} {10}},\ \bibinfo {pages} {186} (\bibinfo {year}
  {1974})}\BibitemShut {NoStop}%
\bibitem [{\citenamefont {Pratt}\ and\ \citenamefont
  {Torrieri}(2010)}]{Pratt:2010kg}%
  \BibitemOpen
  \bibfield  {author} {\bibinfo {author} {\bibfnamefont {S.}~\bibnamefont
  {Pratt}}\ and\ \bibinfo {author} {\bibfnamefont {G.}~\bibnamefont
  {Torrieri}},\ }\href@noop {} {\bibfield  {journal} {\bibinfo  {journal}
  {Phys. Rev. C}\ }\textbf {\bibinfo {volume} {82}},\ \bibinfo {pages} {044901}
  (\bibinfo {year} {2010})}\BibitemShut {NoStop}%
\bibitem [{OSC(1997)}]{OSCAR97}%
  \BibitemOpen
  \href@noop {} {}\bibinfo {howpublished}
  {{https://karman.physics.purdue.edu/OSCAR/index.php}} (\bibinfo {year}
  {1997})\BibitemShut {NoStop}%
\bibitem [{\citenamefont {Amsler}\ and\ \citenamefont
  {Grp}(2008)}]{Amsler:2008kq}%
  \BibitemOpen
  \bibfield  {author} {\bibinfo {author} {\bibfnamefont {C.}~\bibnamefont
  {Amsler}}\ and\ \bibinfo {author} {\bibfnamefont {P.~D.}\ \bibnamefont
  {Grp}},\ }\href@noop {} {\bibfield  {journal} {\bibinfo  {journal} {Phys.
  Lett. B}\ }\textbf {\bibinfo {volume} {667}},\ \bibinfo {pages} {1} (\bibinfo
  {year} {2008})}\BibitemShut {NoStop}%
\bibitem [{\citenamefont {Luzum}\ and\ \citenamefont
  {Romatschke}(2009)}]{Luzum:2009bs}%
  \BibitemOpen
  \bibfield  {author} {\bibinfo {author} {\bibfnamefont {M.}~\bibnamefont
  {Luzum}}\ and\ \bibinfo {author} {\bibfnamefont {P.}~\bibnamefont
  {Romatschke}},\ }\href@noop {} {\bibfield  {journal} {\bibinfo  {journal}
  {Phys. Rev. C}\ }\textbf {\bibinfo {volume} {79}},\ \bibinfo {pages} {039903}
  (\bibinfo {year} {2009})}\BibitemShut {NoStop}%
\bibitem [{\citenamefont {Adler}\ \emph
  {et~al.}(2004{\natexlab{a}})\citenamefont {Adler} \emph
  {et~al.}}]{Adler:2004ki}%
  \BibitemOpen
  \bibfield  {author} {\bibinfo {author} {\bibfnamefont {S.~S.}\ \bibnamefont
  {Adler}} \emph {et~al.} (\bibinfo {collaboration} {PHENIX Collaboration}),\
  }\href@noop {} {\bibfield  {journal} {\bibinfo  {journal} {Phys. Rev. C}\
  }\textbf {\bibinfo {volume} {69}},\ \bibinfo {pages} {034909} (\bibinfo
  {year} {2004}{\natexlab{a}})}\BibitemShut {NoStop}%
\bibitem [{\citenamefont {Hirano}\ and\ \citenamefont
  {Tsuda}(2002)}]{Hirano:2002eh}%
  \BibitemOpen
  \bibfield  {author} {\bibinfo {author} {\bibfnamefont {T.}~\bibnamefont
  {Hirano}}\ and\ \bibinfo {author} {\bibfnamefont {K.}~\bibnamefont {Tsuda}},\
  }\href@noop {} {\bibfield  {journal} {\bibinfo  {journal} {Phys. Rev. C}\
  }\textbf {\bibinfo {volume} {66}} (\bibinfo {year} {2002})}\BibitemShut
  {NoStop}%
\bibitem [{\citenamefont {Pratt}(2009)}]{Pratt:2009bk}%
  \BibitemOpen
  \bibfield  {author} {\bibinfo {author} {\bibfnamefont {S.}~\bibnamefont
  {Pratt}},\ }\href@noop {} {\bibfield  {journal} {\bibinfo  {journal} {Phys.
  Rev. Lett.}\ }\textbf {\bibinfo {volume} {102}},\ \bibinfo {pages} {232301}
  (\bibinfo {year} {2009})}\BibitemShut {NoStop}%
\bibitem [{\citenamefont {Ollitrault}\ \emph {et~al.}(2009)\citenamefont
  {Ollitrault}, \citenamefont {Poskanzer},\ and\ \citenamefont
  {Voloshin}}]{Ollitrault:2009gm}%
  \BibitemOpen
  \bibfield  {author} {\bibinfo {author} {\bibfnamefont {J.-Y.}\ \bibnamefont
  {Ollitrault}}, \bibinfo {author} {\bibfnamefont {A.}~\bibnamefont
  {Poskanzer}}, \ and\ \bibinfo {author} {\bibfnamefont {S.}~\bibnamefont
  {Voloshin}},\ }\href@noop {} {\bibfield  {journal} {\bibinfo  {journal}
  {Phys. Rev. C}\ }\textbf {\bibinfo {volume} {80}},\ \bibinfo {pages} {014904}
  (\bibinfo {year} {2009})}\BibitemShut {NoStop}%
\bibitem [{\citenamefont {Adams}\ \emph {et~al.}(2004)\citenamefont {Adams}
  \emph {et~al.}}]{Adams:2004dg}%
  \BibitemOpen
  \bibfield  {author} {\bibinfo {author} {\bibfnamefont {J.}~\bibnamefont
  {Adams}} \emph {et~al.} (\bibinfo {collaboration} {STAR Collaboration}),\
  }\href@noop {} {\bibfield  {journal} {\bibinfo  {journal} {Phys. Rev. Lett.}\
  }\textbf {\bibinfo {volume} {92}},\ \bibinfo {pages} {112301} (\bibinfo
  {year} {2004})}\BibitemShut {NoStop}%
\bibitem [{\citenamefont {Adcox}\ \emph {et~al.}(2002)\citenamefont {Adcox}
  \emph {et~al.}}]{Adcox:2002bo}%
  \BibitemOpen
  \bibfield  {author} {\bibinfo {author} {\bibfnamefont {K.}~\bibnamefont
  {Adcox}} \emph {et~al.} (\bibinfo {collaboration} {PHENIX Collaboration}),\
  }\href@noop {} {\bibfield  {journal} {\bibinfo  {journal} {Phys. Rev. Lett.}\
  }\textbf {\bibinfo {volume} {88}},\ \bibinfo {pages} {192302} (\bibinfo
  {year} {2002})}\BibitemShut {NoStop}%
\bibitem [{\citenamefont {Adams}\ \emph
  {et~al.}(2005{\natexlab{b}})\citenamefont {Adams} \emph
  {et~al.}}]{Adams:2005cx}%
  \BibitemOpen
  \bibfield  {author} {\bibinfo {author} {\bibfnamefont {J.}~\bibnamefont
  {Adams}} \emph {et~al.} (\bibinfo {collaboration} {STAR Collaboration}),\
  }\href@noop {} {\bibfield  {journal} {\bibinfo  {journal} {Phys. Rev. C}\
  }\textbf {\bibinfo {volume} {71}},\ \bibinfo {pages} {044906} (\bibinfo
  {year} {2005}{\natexlab{b}})}\BibitemShut {NoStop}%
\bibitem [{\citenamefont {Adler}\ \emph
  {et~al.}(2004{\natexlab{b}})\citenamefont {Adler} \emph
  {et~al.}}]{Adler:2004ii}%
  \BibitemOpen
  \bibfield  {author} {\bibinfo {author} {\bibfnamefont {S.}~\bibnamefont
  {Adler}} \emph {et~al.} (\bibinfo {collaboration} {PHENIX Collaboration}),\
  }\href@noop {} {\bibfield  {journal} {\bibinfo  {journal} {Phys. Rev. Lett.}\
  }\textbf {\bibinfo {volume} {93}} (\bibinfo {year}
  {2004}{\natexlab{b}})}\BibitemShut {NoStop}%
\bibitem [{\citenamefont {Adler}\ \emph {et~al.}(2003)\citenamefont {Adler}
  \emph {et~al.}}]{Adler:2003gs}%
  \BibitemOpen
  \bibfield  {author} {\bibinfo {author} {\bibfnamefont {S.}~\bibnamefont
  {Adler}} \emph {et~al.} (\bibinfo {collaboration} {PHENIX Collaboration}),\
  }\href@noop {} {\bibfield  {journal} {\bibinfo  {journal} {Phys. Rev. Lett.}\
  }\textbf {\bibinfo {volume} {91}},\ \bibinfo {pages} {182301} (\bibinfo
  {year} {2003})}\BibitemShut {NoStop}%
\bibitem [{\citenamefont {Adare}\ and\ \citenamefont
  {other}(2012)}]{Adare:2012iv}%
  \BibitemOpen
  \bibfield  {author} {\bibinfo {author} {\bibfnamefont {A.}~\bibnamefont
  {Adare}}\ and\ \bibinfo {author} {\bibnamefont {other}} (\bibinfo
  {collaboration} {PHENIX Collaboration}),\ }\href@noop {} {\bibfield
  {journal} {\bibinfo  {journal} {Phys. Rev. C}\ }\textbf {\bibinfo {volume}
  {85}},\ \bibinfo {pages} {064914} (\bibinfo {year} {2012})}\BibitemShut
  {NoStop}%
\bibitem [{\citenamefont {Adare}\ \emph {et~al.}(2008)\citenamefont {Adare}
  \emph {et~al.}}]{Adare:2008cs}%
  \BibitemOpen
  \bibfield  {author} {\bibinfo {author} {\bibfnamefont {A.}~\bibnamefont
  {Adare}} \emph {et~al.} (\bibinfo {collaboration} {PHENIX Collaboration}),\
  }\href@noop {} {\bibfield  {journal} {\bibinfo  {journal} {Phys. Rev. C}\
  }\textbf {\bibinfo {volume} {77}},\ \bibinfo {pages} {064907} (\bibinfo
  {year} {2008})}\BibitemShut {NoStop}%
\bibitem [{\citenamefont {Lisa}\ \emph {et~al.}(2005)\citenamefont {Lisa},
  \citenamefont {Pratt}, \citenamefont {Soltz},\ and\ \citenamefont
  {Wiedemann}}]{Lisa:2005cg}%
  \BibitemOpen
  \bibfield  {author} {\bibinfo {author} {\bibfnamefont {M.~A.}\ \bibnamefont
  {Lisa}}, \bibinfo {author} {\bibfnamefont {S.}~\bibnamefont {Pratt}},
  \bibinfo {author} {\bibfnamefont {R.}~\bibnamefont {Soltz}}, \ and\ \bibinfo
  {author} {\bibfnamefont {U.}~\bibnamefont {Wiedemann}},\ }\href@noop {}
  {\bibfield  {journal} {\bibinfo  {journal} {Annu. Rev. Nucl. Part. Sci.}\
  }\textbf {\bibinfo {volume} {55}},\ \bibinfo {pages} {357} (\bibinfo {year}
  {2005})}\BibitemShut {NoStop}%
\bibitem [{\citenamefont {Broniowski}\ \emph {et~al.}(2008)\citenamefont
  {Broniowski}, \citenamefont {Chojnacki}, \citenamefont {Florkowski},\ and\
  \citenamefont {Kisiel}}]{Broniowski:2008iq}%
  \BibitemOpen
  \bibfield  {author} {\bibinfo {author} {\bibfnamefont {W.}~\bibnamefont
  {Broniowski}}, \bibinfo {author} {\bibfnamefont {M.}~\bibnamefont
  {Chojnacki}}, \bibinfo {author} {\bibfnamefont {W.}~\bibnamefont
  {Florkowski}}, \ and\ \bibinfo {author} {\bibfnamefont {A.}~\bibnamefont
  {Kisiel}},\ }\href@noop {} {\bibfield  {journal} {\bibinfo  {journal} {Phys.
  Rev. Lett.}\ }\textbf {\bibinfo {volume} {101}},\ \bibinfo {pages} {022301}
  (\bibinfo {year} {2008})}\BibitemShut {NoStop}%
\bibitem [{\citenamefont {Shen}\ \emph {et~al.}(2011)\citenamefont {Shen} \emph
  {et~al.}}]{Shen:2011zc}%
  \BibitemOpen
  \bibfield  {author} {\bibinfo {author} {\bibfnamefont {C.}~\bibnamefont
  {Shen}} \emph {et~al.},\ }\href@noop {} {\bibfield  {journal} {\bibinfo
  {journal} {J. Phys. G}\ }\textbf {\bibinfo {volume} {38}},\ \bibinfo {pages}
  {124045} (\bibinfo {year} {2011})}\BibitemShut {NoStop}%
\bibitem [{\citenamefont {Schenke}\ \emph
  {et~al.}(2012{\natexlab{b}})\citenamefont {Schenke}, \citenamefont {Jeon},\
  and\ \citenamefont {Gale}}]{Schenke:2012jr}%
  \BibitemOpen
  \bibfield  {author} {\bibinfo {author} {\bibfnamefont {B.}~\bibnamefont
  {Schenke}}, \bibinfo {author} {\bibfnamefont {S.}~\bibnamefont {Jeon}}, \
  and\ \bibinfo {author} {\bibfnamefont {C.}~\bibnamefont {Gale}},\ }\href@noop
  {} {\bibfield  {journal} {\bibinfo  {journal} {Phys. Rev. C}\ }\textbf
  {\bibinfo {volume} {85}},\ \bibinfo {pages} {024901} (\bibinfo {year}
  {2012}{\natexlab{b}})}\BibitemShut {NoStop}%
\end{thebibliography}%
\end{document}